\documentclass[oneside,11pt,reqno]{amsart}

 \pdfoutput=1
 \usepackage{hyperref}
 \usepackage[margin=2cm]{geometry}                		
\usepackage{graphicx}			
\usepackage{amsmath}
\usepackage{amssymb}

\usepackage{color}
\usepackage{subcaption}
\usepackage{mathtools}
\usepackage{amsaddr}
\usepackage{xfrac}

\usepackage{tikz}
\usepackage{xfrac}
\usepackage{bm}
\usetikzlibrary{arrows}
\newcommand{\ket}[1]{\left| #1 \right>} 
\newcommand{\bra}[1]{\left< #1 \right|} 
 
 \newcommand{\CN}{\mathbb{C}}
  
\renewcommand{\Re}{\operatorname{Re}}
\renewcommand{\Im}{\operatorname{Im}}
 \newcommand{\rmi}{\mathrm{i}}
 \newcommand{\rmd}{\mathrm{d}}
 \newcommand{\rme}{\mathrm{e}}

  \newcommand{\zbar}{\overline{z}}
  \newcommand{\Log}{\mathrm{Log}}

\newcommand{\UN}{\mathrm{U}(N)}

\hypersetup{
    colorlinks=true,
    linkcolor=black,
    citecolor=black,
    filecolor=black,
    urlcolor=black,
}
\newtheorem{theorem}{Theorem}

\newtheorem{proposition}{Proposition}

\usepackage{cite}
\newtheorem{subtheorem}{Theorem}[theorem]

\newcommand{\normord}[1]{\vcentcolon\mathrel{#1}\vcentcolon}
\usetikzlibrary{decorations.pathmorphing}

\tikzset{snake it/.style={decorate, decoration=snake}}\usepackage[T1]{fontenc}

\usepackage{amsfonts,amssymb}

\date{}							
\title[]{Asymptotic Correlations in Gapped and Critical Topological Phases of 1D Quantum Systems}

\author{N. G. Jones}
\address{School of Mathematics, University of Bristol, Bristol BS8 1TW, UK\\
and the Heilbronn Institute for Mathematical Research, Bristol, UK}
\email{n.g.jones@bristol.ac.uk}
\author{R. Verresen}
\address{Department of Physics T42, Technical University of Munich, 85748 Garching, Germany\\
and
Max-Planck-Institute for the Physics of Complex Systems, 01187 Dresden, Germany}
\email{rubenverresen@gmail.com}

\begin{document}

\begin{abstract}Topological phases protected by symmetry can occur in gapped and---surprisingly---in critical systems. 
We consider non-interacting fermions in one dimension with spinless time-reversal symmetry.
It is known that the phases are classified by a topological invariant $\omega$ and a central charge $c$.
We investigate the correlations of string operators, giving insight into the interplay between topology and criticality.
In the gapped phases, these non-local string order parameters allow us to extract $\omega$.
Remarkably, ratios of correlation lengths are universal. In the critical phases, the scaling dimensions of these operators serve as an order parameter, encoding $\omega$ and $c$.
We derive exact asymptotics of these correlation functions using Toeplitz determinant theory.
We include physical discussion, e.g., relating lattice operators to the conformal field theory. Moreover, we discuss the dual spin chains. Using the aforementioned universality, the topological invariant of the spin chain can be obtained from correlations of local observables.
\end{abstract}
\maketitle

\setcounter{tocdepth}{2}
\tableofcontents
\setcounter{theorem}{1}
\section{Introduction}

Topological phases are fascinating examples of quantum matter. In one spatial dimension, they can be stabilised if the Hamiltonian has a symmetry group. Gapped topological phases have been classified for both non-interacting fermionic systems (dubbed topological insulators or superconductors) \cite{Fu2007,Kitaev09,Schnyder08,Altland97,Ryu10,Hasan10} as well as general fermionic and bosonic systems (dubbed symmetry protected topological (SPT) phases) \cite{Wen2009,Pollmann10,Fidkowski11class,Turner-2010,Chen10,Schuch11}. However, it has recently been realised that critical matter can also form distinct topological phases---even without gapped degrees of freedom in the bulk \cite{VJP}. 
As in the gapped case, the topology manifests itself physically: for example, through exponentially localised zero-energy modes at the physical edges. As long as a symmetry is preserved, a topological invariant can prevent two critical systems from being smoothly connected. Relatedly, there is a lot of recent interest in topological critical phases which do have additional gapped degrees of freedom \cite{Kestner11,Cheng11,Fidkowski11longrange,Sau11,Kraus13,Keselman15,Iemini15,Lang15,Montorsi17,Ruhman17,Scaffidi17,Jiang17_preprint,Zhang17,Parker}.

\begin{figure}
	\includegraphics[width=\linewidth]{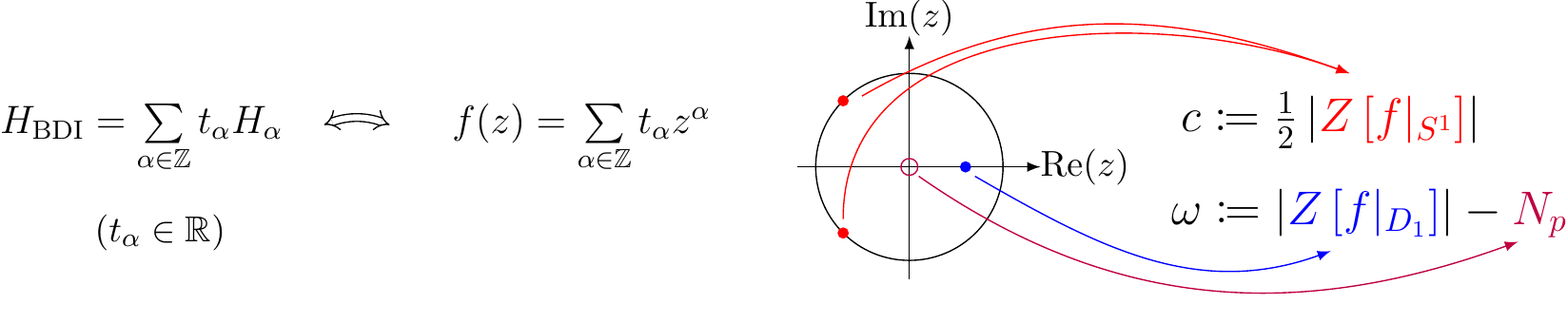}
	\caption{The Hamiltonians we consider can be expanded in a basis $H_\alpha$ (defined below equation \eqref{majoranaham}). The physics is encoded in the meromorphic function $f(z)$. The given definitions of $c$ and $\omega$ classify the phases of $H_{\mathrm{BDI}}$, where $Z[g]$ denotes the (multi)set of zeros of $g$ (with multiplicity) and $N_p$ the order of the pole at the origin. Physically, $c$ encodes the low-energy behaviour in the bulk, and $\omega$ the topological properties. \label{fig:zeros}}
\end{figure}

In a previous work, we extended the well-known classification of the gapped topological phases of quadratic fermionic Hamiltonians with spinless time-reversal symmetry \cite{Asboth15} (`BDI class' of Altland and Zirnbauer's tenfold way \cite{Altland97}) to gapless topological phases \cite{VJP}. These are labelled by a topological invariant $\omega$ ($\in \mathbb Z$) and the central charge $c$ ($\in \frac{1}{2} \mathbb Z_{\geq 0}$) of the conformal field theory (CFT) that describes the continuum limit if the model is critical. If the system is gapped, we say that $c=0$ and $\omega$ reduces to the well-known winding number of the BDI class \cite{Asboth15}. What allowed for a complete analysis was the fact that each Hamiltonian in this class can be efficiently encoded  into a holomorphic function $f(z)$ on the punctured complex plane $\mathbb C \setminus \{0\}$. Remarkably, $c$ and $\omega$ can then be obtained by counting zeros of $f(z)$ (see Figure \ref{fig:zeros}). This rephrasing allowed us to argue that two critical models in this class can be smoothly connected if and only if they have the same topological invariants and central charges.

What remained an open question is the extent to which the topological nature of these gapped and gapless phases is reflected in their correlation functions. Relatedly, it is natural to ask how the correlations are encoded in $f(z)$---especially since $c$ and $\omega$ are easily derived from its zeros.
Moreover, our earlier work left an uneasy tension: distinct critical phases could be distinguished by the topological invariant $\omega$, yet it was not clear to what extent this lattice quantity is related to the CFT in the continuum. Hence, bridging this gap in terms of a lattice-continuum correspondence is desirable. 
More generally, since these models are exactly solvable we can hope to obtain a lot of information, and perhaps uncover unexpected features. This is relevant also to the spin chains that are Jordan-Wigner dual to these fermionic chains: whilst the non-interacting classification is less natural there, the correlation functions we obtain contain useful physical information that can be related to an interacting classification.\par

The aim of this work is twofold: on the one hand, we focus on answering the aforementioned questions conceptually, linking universal properties of correlations to the function $f(z)$ and shedding light on the interplay of criticality and topology. On the other hand, since our models allow for a rigorous analysis, we give derivations of exact asymptotic expressions for important correlators. The method we use, Toeplitz determinant theory, has a long association with statistical mechanics (for a review, see \cite{DIK2}), and our analysis generalises the pioneering work of \cite{McI,BMc, FH} to a wider class of physical models.

\begin{figure}
	\includegraphics[width=\linewidth]{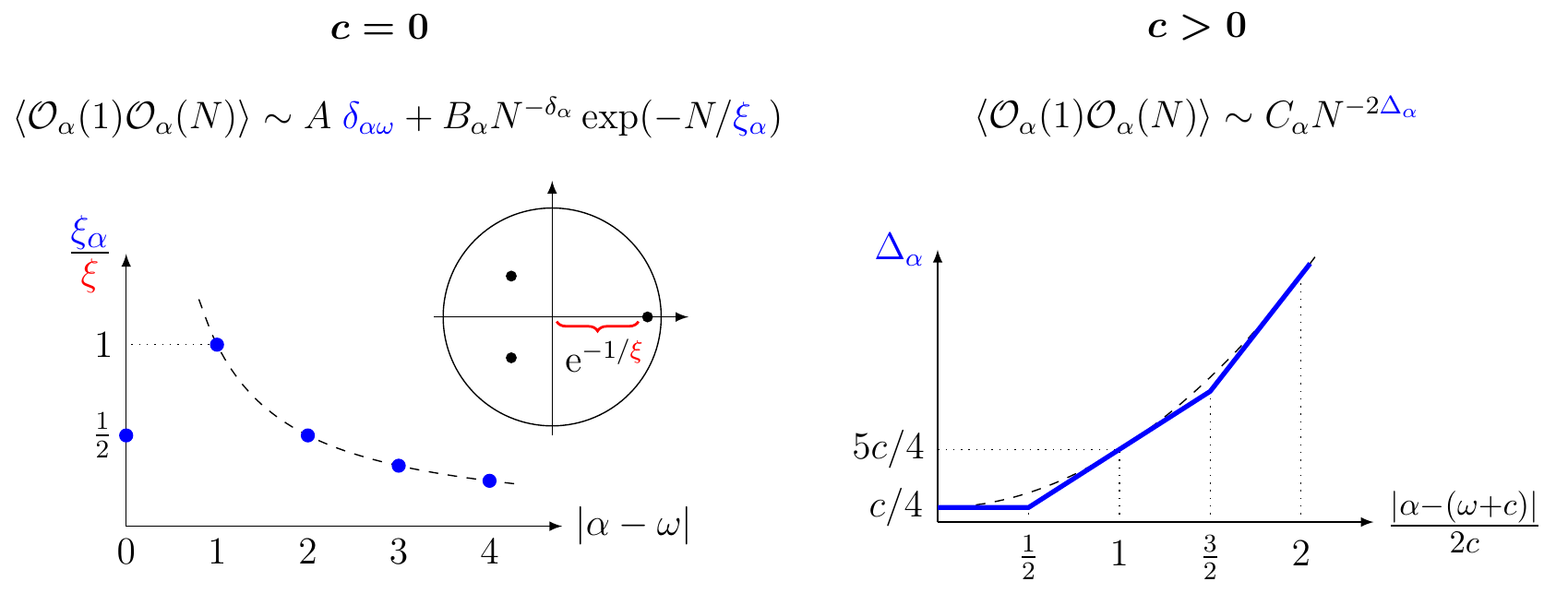}
	\caption{Universal asymptotics of the ground-state correlation functions considered in this work. If $c=0$ (i.e. the system is gapped), then $\mathcal O_\alpha$ has exponentially decaying correlations with correlation length $\xi_\alpha$. The ratios $\xi_\alpha/\xi_\beta$ are a universal function of $\omega$, with the global scale set by $1/\xi:=\min_{\zeta \in Z[f]} \lvert\log\lvert\zeta\rvert \rvert$; see the discussion before equation~\eqref{fzcanon}. (There is long-range order, i.e. $a_\alpha \neq 0$, if and only if $\alpha = \omega$; see Theorem \ref{orderparameter}.) If $c > 0$ and the zeros on the unit circle have multiplicity one (i.e. the bulk is described by a CFT with central charge $c$), then the correlation functions obey a power law with universal scaling dimension $\Delta_\alpha$; see Theorem \ref{scalingtheorem}. The dependence of $\Delta_\alpha$ on both $c$ and $\omega$ means that these parameters may be determined by measurements of scaling dimensions; see the discussion below Theorem \ref{scalingtheorem}. Note that there are exceptional cases that behave differently, as discussed in the text.\label{fig:correlation} }
\end{figure}

Since topological phases cannot be distinguished locally, in this work we study the correlations of so-called string-like objects $\mathcal O_\alpha$ (labelled by $\alpha \in \mathbb{Z}$), meaning that $\langle \mathcal{O}_\alpha(1) \mathcal{O}_\alpha(N) \rangle$ involves an extensive ($\sim N$) number of operators. Using Wick's theorem, these correlations reduce to $N\times N$ determinants. We calculate their asymptotic behaviour using the theory of Toeplitz determinants \cite{DIK2}, phrasing the answers in terms of the zeros of $f(z)$. Figure \ref{fig:correlation} summarises some of the main results. In the gapped case ($c=0$), it is well-known that SPT phases can be distinguished by string order parameters \cite{Kennedy1992,denNijs,PhysRevLett.100.167202,Haegeman,Else}, and we indeed prove that $\mathcal O_\alpha$ has long-range order \emph{if and only if} $\alpha = \omega$. More surprising is that the ratios of the correlation lengths of these operators are universal, i.e. they depend on $\omega$ only. Moreover, the largest correlation length has a universal relationship to the zero of $f(z)$ which is nearest to the unit circle. In the critical case ($c \neq 0$), all correlations are algebraically decaying and we obtain the corresponding scaling dimensions of $\mathcal O_\alpha$. It turns out that measuring these gives access to both $c$ and $\omega$. Moreover, we propose  a continuum-lattice correspondence for these operators. We expect that this correspondence will prove useful in exploring the effect of interactions on the phase diagram.
\par
Additionally, we discuss these correlators in the spin chain picture that is obtained after a Jordan-Wigner transformation. For odd $\alpha$, $\mathcal O_\alpha$ becomes a local spin operator, and long-range order in $\langle \mathcal{O}_\alpha(1) \mathcal{O}_\alpha(N) \rangle$ signals spontaneous symmetry breaking. However, for even $\alpha$ these correlators are string order parameters for spin chain SPTs. Whilst there is no natural notion of `non-interacting spin chains', our analysis may be helpful for determining the (interacting) classification of topologically non-trivial critical spin chains.\par

Since the physical consequences of our results can be understood without going into the mathematical details, we structure the paper as follows. First, in Section \ref{sec::summary}, we outline the model and state our main results. In Section \ref{sec::spin} we discuss the dual spin chain; then in Section \ref{previouswork} we discuss connections to previous works.
In Section \ref{sec::physics} we give further details of how our results fit into the broader physical context. In particular, we discuss general approaches to string order parameters and the consequences of universality in the gapped phases, give a CFT analysis of long-distance correlations and also show how our results allow us to deduce critical exponents. Only after this do we give the mathematical preliminaries in Sections \ref{sec::detcorr} and \ref{sec::Toeplitz}. The proofs of our results then follow in Sections \ref{sec::gap} and \ref{sec::gapless} for the gapped and critical cases respectively.  Finally, in Section \ref{sec::extensions}, we explain how our results may be extended in different directions.


\section{Statement of main results} \label{sec::summary}
\subsection{The model}
Consider a periodic chain where each site has a single spinless fermionic degree of freedom\footnote{Further details supporting this section are given in Appendix \ref{app::fermi}.} $\{c^\dagger_n, c_n; n = 1\dots L\}$. For convenience define the Majorana modes on each site:
\begin{align}
\gamma_n = c^\dagger_n + c_n, \qquad \tilde\gamma_n = \rmi (c^\dagger_n - c_n), \label{gammadef}
\end{align} 
where $\{\gamma_n,\gamma_m\} = 2 \delta_{nm}$ and $\{\gamma_n, \tilde \gamma_m\} = 0$.
Our class of interest---time-reversal symmetric, translation-invariant free fermions with finite-range couplings---has Hamiltonian \cite{deGottardi13,Niu12,VMP,VJP}
\begin{align}
H_{\mathrm{BDI}} = \frac{\rmi}{2}\sum_{\alpha=-\alpha_l}^{\alpha_r} \sum_{n \in \mathrm{sites}}  t_\alpha \tilde\gamma_n \gamma_{n+\alpha}, \qquad t_\alpha \in \mathbb{R}.\label{majoranaham}
\end{align}
This can be understood as an expansion in the basis $H_\alpha =\sfrac{\rmi}{2}\sum_{n \in \mathrm{sites}}   \tilde\gamma_n \gamma_{n+\alpha}$. The coupling between sites has maximum range $\alpha_{l/r}$ to the left and right. This model has an antiunitary symmetry, $T$, that acts as complex conjugation in the occupation number basis associated to the fermions $c_n$ and satisfies $T^2=1$. The Majorana operators $\gamma_n$ ($\tilde\gamma_n$) are called real (imaginary) since $T\gamma_nT = \gamma_n$ and  $T\tilde\gamma_nT =-\tilde \gamma_n$. This class of models is also invariant under parity symmetry $P = \rme^{\rmi\pi\sum_j c^\dagger_j c_j}$.
 We study the thermodynamic limit $L \rightarrow \infty$, with $\alpha_{l/r}$ finite but not fixed---i.e. we will consider models with differing maximum range. 
 The results given in this section are all for such finite-range chains, but we discuss the extension to long-range chains in Section \ref{sec::longrange}. This model was first analysed in its spin chain form in reference \cite{Suzuki71}. \par
 The coupling constants $t_\alpha$ establish a one-to-one correspondence between $H_{\mathrm{BDI}}$ and the complex functions 
\begin{align}
f(z) = \sum_\alpha t_\alpha z^{\alpha}. \label{fdef}
\end{align}
This is a holomorphic function away from a possible pole at the origin. By the fundamental theorem of algebra, $f(z)$ is specified by the degree of this pole and a multiset of zeros (up to an overall multiplicative constant). 
The basic relevance of $f(z)$ is that $|f(\rme^{\rmi k})|$ gives the one-particle energy of a mode with momentum $k$. The phase $\arg(f(\rme^{\rmi k}))$ is the angle required in the Boguliobov rotation that defines these quasiparticle modes \cite{KM,VJP}.
Remarkably, many other physical questions can be answered through simple properties of this function. Note that we will consistently abuse notation $f(k):=f(z=\rme^{\rmi k})$ whenever we restrict $z$ to the unit circle. 
\subsection{Phase diagram}
Our results characterise the correlations in the different phases of $H_{\mathrm{BDI}}$, hence we give them context by describing the phase diagram. First, phases of matter are defined as equivalence classes of ground states under smooth changes of the Hamiltonian, where two states are equivalent if they can be connected without a phase transition (i.e. a sharp change in physical behaviour\footnote{Transitions between gapped phases requires the closing of the gap. For two gapless phases a transition occurs when there is a change in the low-energy description, for example an increase in the central charge of the CFT.}). In particular, we define two \emph{critical} models to be in the same phase if physical quantities such as scaling dimensions vary smoothly. Smooth changes to the Hamiltonian (i.e. smooth changes to $t_\alpha$, including increasing the \emph{finite} range by tuning some $t_\alpha$ off zero) are equivalent to smooth motions of the zeros of $f(z)$, as we discuss in Appendix \ref{app::fermi2}.\par
$H_{\mathrm{BDI}}$ has two invariants that label both gapped and gapless phases (see also Figure \ref{fig:zeros}):
\begin{align}
c &= \frac{1}{2}\left(\textrm{\# zeros of}~ f(z)~\textrm{on the unit circle}\right) \\
\omega &=  N_z - N_p,
\end{align}
where $N_z$ is the number of zeros of $f(z)$ inside the unit disk and $N_p$ is the degree of the pole at the origin.
If $c=0$, the model is gapped. For gapless models, $c$ is the central charge of the low energy CFT when the zeros on the circle are non-degenerate\footnote{This is argued in Appendix \ref{app::fermi2}, see also Section \ref{sec::CFT}. If there are degeneracies then we have dynamical critical exponent greater then one---we will discuss this further below.}.
Note that $\omega$ is an invariant since it cannot change under smooth motion of the zeros without changing $c$. It is moreover topological: it cannot be probed locally, but distinguishes phases and manifests itself through protected edge modes \cite{VJP}.
That the pair $(c,\omega)$ specifies the phases of $H_{\mathrm{BDI}}$ was shown in reference \cite{VJP}. If in addition to the symmetries $P$ and $T$ that stabilise the aforementioned phases, one also enforces translation symmetry, then there are additional invariant signs, denoted $\Sigma$, that are discussed in Appendix \ref{app::fermi2}. \par
Note that the equivalence between $H_{\mathrm{BDI}}$ and $f(z)$ allows us to easily find a Hamiltonian within each phase: $H_\omega$ is a representative of the gapped phase with winding number $\omega$ and $H_\omega +H_{2c+\omega}$ is a representative of the gapless phase $(c,\omega)$.
\par

\subsection{String operators}
The above is already established in the literature. The results of the current work show that given $f(z)$, one can `read off' detailed information about two-point ground-state correlation functions of the operators $\mathcal{O}_\alpha(n)$:
\begin{align}\label{Odef}
\mathcal{O}_\alpha(n)=\begin{cases} \rmi^x \left(\prod_{m=1}^{n-1} \rmi\tilde\gamma_m\gamma_m\right)\gamma_{n}\gamma_{n+1}\dots\gamma_{n+\alpha-1} \quad& \alpha>0 \\
(-\rmi)^{x}\left(\prod_{m=1}^{n-1} \rmi\tilde\gamma_m\gamma_m\right)  (-\rmi\tilde\gamma_{n})\dots(-\rmi\tilde\gamma_{n+\lvert\alpha\rvert-1}) \quad &\alpha<0\\
\prod_{m=1}^{n-1} \rmi\tilde\gamma_m\gamma_m \quad &\alpha=0\end{cases}
\end{align}
where $x= |\alpha|/2$ for $\alpha$ even and $x=(|\alpha|-1)/2$ for $\alpha$ odd (the phase factors make $\mathcal{O}_\alpha$ hermitian). These operators are a cluster of $|\alpha|$ Majorana operators to the right of site $n$ multiplied by an operator giving the parity of the number of fermions to the left of $n$. Such operators appear naturally as we discuss in Section \ref{sec::stringorder}, see also \cite{BahriVishwanath}. There are two typical behaviours for these correlators. Let angle brackets denote ground-state expectation value, 
then in the gapped case we expect:
 \begin{align}
 \langle\mathcal{O}_\alpha(1)\mathcal{O}_\alpha(N+1)\rangle \sim A_\alpha + B_\alpha N^{-\delta_\alpha}\rme^{-N/\xi_\alpha}. \label{gapscale}
 \end{align}
 The constants $A_\alpha$ and $\delta_\alpha$ as well as the correlation length $\xi_\alpha$ do not depend on $N$. If $A_\alpha\neq 0$ we have long-range order.  $B_\alpha$ is $\Theta(1)$ and may include an oscillation with $N$. Note that, by translation invariance, we could equally well have considered the correlation function $\langle\mathcal{O}_\alpha(r)\mathcal{O}_\alpha(N+r)\rangle$ for any $r \in \mathbb{Z}$---we fix $r=1$ throughout for notational convenience. Note that $\langle\mathcal{O}_\alpha(1)\mathcal{O}_\beta(N+1)\rangle = 0$ if $\alpha \neq \beta$ as a simple consequence of the Majorana two-point functions given in Section \ref{sec::detcorr}. 
 \par
 We will see below that the ground state expectation value $\lim_{N\rightarrow\infty}\langle\mathcal{O}_\alpha(1)\mathcal{O}_\alpha(N+1)\rangle$ in the gapped phase with winding number $\omega$ is non-zero only when $\alpha = \omega$, and is hence an order parameter for that phase---this can be seen as an extension of the results of \cite{BahriVishwanath}.

 Because these correlators contain a string of fermionic operators of length of order $N$, these are called \emph{string order parameters} with value $A_\alpha$. Note that in the case that $\mathcal{O}_\alpha$ is local, (as happens in the spin picture given in Section \ref{sec::spin}), it is usual to call the one point function $\langle\mathcal{O}_\alpha(n)\rangle$ the order parameter. This is because in that case the ground state will spontaneous collapse such that $\sqrt{A_\alpha} = \langle\mathcal{O}_\alpha(n)\rangle$. In this work we prefer to use a single convention and always refer to the two point function as `the' order parameter. \par
  At critical points with a low-energy CFT description we expect:
 \begin{align}\label{critscale}
 \langle\mathcal{O}_\alpha(1)\mathcal{O}_\alpha(N+1)\rangle \sim C_\alpha  N^{-2\Delta_\alpha}
 \end{align}
 where $\Delta_\alpha$ is the smallest scaling dimension of a CFT operator that appears in the expansion of the continuum limit of $\mathcal{O}_\alpha$. The prefactor $C_\alpha$ may include spatial oscillations, and further details are given in Section \ref{sec::CFT}. Surprisingly, the set $\mathcal{O}_\alpha$ also act as order parameters for critical phases in a sense that we explain following Theorem \ref{scalingtheorem}.
 \par
 \subsection{Main results}\label{results}
 To fix notation, let us write:
 \begin{align}
f(z) = \frac{\rho}{z^{N_p}} \prod_{j=1}^{N_z} \left(z-z_j\right)\prod_{j=1}^{2c} \left(z-\rme^{\rmi k_j}\right)\prod_{j=1}^{N_Z} \left(z-Z_j\right). \label{fzcanon}
\end{align}
$N_p$ is the order of the pole at the origin, which is also the range of the longest non-zero coupling to the left. The number\footnote{We consider a multiset of zeros $\{\zeta_j\}$ and allow $\zeta_j$ with different index to coincide. This makes the counting unambiguous.} of zeros inside, on and outside the circle are denoted $N_z$, $2c$, $N_Z$ respectively, and $\rho$ is a real number. Since the $t_\alpha$ are real, all zeros are either real or come in complex conjugate pairs. 
\par
We first state results for the gapped case. Firstly we have that the correlators $\langle\mathcal{O}_\alpha(1)\mathcal{O}_\alpha(N+1)\rangle$ form a complete set of order parameters for the gapped phases of $H_{\mathrm{BDI}}$.

\begin{subtheorem}\label{phases}
 In the phase $(\omega, c=0, \Sigma)$ we have \begin{align}
\lim_{N\rightarrow\infty} |\langle\mathcal{O}_\alpha(1)\mathcal{O}_\alpha(N+1)\rangle| = \mathrm{const} \times \delta_{\omega\alpha}.
\end{align}
The non-zero constant is given in Theorem \ref{orderparameter}. The value of the sign $\Sigma$ may be inferred by the presence or absence of a $(-1)^N$ oscillation in this correlator. \end{subtheorem}

\begin{subtheorem}\label{orderparameter}
 In the phase $(\omega,c=0,\Sigma)$, the non-universal value of the order parameter is given by
\begin{align}
\lim_{N\rightarrow\infty}|\langle\mathcal{O}_\omega(1)\mathcal{O}_\omega(N+1)\rangle| = \left(\frac{ \prod_{i_1,i_2=1}^{N_z} (1-z_{i_1} z_{i_2})
	\prod_{j_1,j_2=1}^{N_Z} \left(1-\frac{1}{Z_{j_1} Z_{j_2}}\right)}{\prod_{i=1}^{N_z}\prod_{j=1}^{N_Z} \left (1-\frac{z_i}{ Z_j}\right)^2}\right)^{1/4}. 
\end{align}
\end{subtheorem}
Thus from the decomposition \eqref{fzcanon} we can read off $\omega=N_z-N_p$ and calculate the order parameter through the detailed values of the zeros. We discuss the mathematical form of the order parameter in Appendix \ref{app::multiplier}.\par
The next results show that the length scale in gapped phases is set by $\xi = 1/\lvert \log\lvert\zeta_\star\rvert \rvert$ where $\zeta_\star$ is any zero that maximises the right hand side of that equation (see Figure \ref{fig:correlation} for illustration). The set of $\zeta$ that are optimal in this way we call closest to the unit circle; we will always mean this logarithmic scale\footnote{That is, $1/\lvert\log|\zeta_i|\rvert$. This gives the natural length scale set by each zero, since the set of these lengths is invariant under spatial inversion $f(z)\rightarrow f(1/z)$.} when we talk about distance from the unit circle. The following results will be stated for `generic cases'---we argue that these cases are typical in Appendix \ref{app::fermi2}. \par
Now, in the phase $\omega$ we then have that  $\xi_\alpha$, as defined in \eqref{gapscale}, is equal to $\frac{\xi}{|\omega-\alpha|}$ (for $\alpha \neq \omega$)---this is a consequence of:
\begin{theorem}\label{lengththeorem}
If the system is in the phase $(\omega,c=0,\Sigma)$ then, in generic systems, we have the large $N$ asymptotics
\begin{align}
 \langle\mathcal{O}_\alpha(1)\mathcal{O}_\alpha(N+1)\rangle =  \det( M (N))\left(\lim_{M\rightarrow\infty}\lvert\langle\mathcal{O}_\omega(1)\mathcal{O}_\omega(M)\rangle\rvert\right)\rme^{-N|\omega-\alpha|/\xi}\rme^{\rmi \pi N m}\bigl(1 +o(1)\bigr) ;\end{align} 
$m\in\mathbb{Z}$ is a known constant and $M(N)$ is a known $|\omega-\alpha| \times|\omega-\alpha|$ matrix. The elements of this matrix have magnitudes that depend algebraically on $N$---in particular, $\det M(N) = \Theta(N^{-\delta})$ for some $\delta>0$. Generic systems are those where the nearest zero(s) to the unit circle is either a single real zero, or are a complex conjugate pair of zeros.
\end{theorem}
The analysis we give extends to exceptional cases---more than two closest zeros will almost always give the same $\xi_\alpha$, but if one has multiplicity, $\xi_\alpha$ \emph{may} be controlled by the next-closest zero. 
The $\xi_\alpha$ are always upper bounded by $\xi$, and in fact this bound is saturated in all exceptional cases except when there are mutually inverse closest zeros. See the discussion in Section \ref{sec::corrl} for full details. \par The form of $\det M(N)$ derived in Section \ref{proofcorrl} allows for some further general statements. Firstly, if there is one real zero nearest to the circle, then $\det M(N)$ is real and does not oscillate with $N$. The algebraic factor depends non-trivially on $|\omega- \alpha|$, as demonstrated in Table \ref{algscaling} for the case that $|\zeta_\star|<1$. If $|\zeta_\star|>1$ then the second and fourth columns of Table \ref{algscaling} should be interchanged (and the definitions of $\lambda$ and $\kappa$ change in the obvious way based on the formulae in Propositions \ref{prop1} and \ref{prop3}).
\begin{table}[tp]
\begin{center}\begin{tabular}{|c||c||c||c|}\hline $\omega-\alpha$ & $\det M(N)$  & $\omega-\alpha$ & $\det M(N)$  \\\hline\hline1 & $\kappa N^{-1/2}$ & $-1$ & $-\lambda N^{-3/2}$ \\\hline 2 & $-\frac{\kappa^2}{2}N^{-3}$ & $-2$ & $-\frac{3\lambda^2}{2}N^{-5}$ \\\hline 3 & $-\frac{3\kappa^3}{4}N^{-15/2}$ & $-3$ & $\frac{45\lambda^3}{4}N^{-21/2}$ \\\hline 4 & $\frac{135\kappa^4}{16}N^{-14}$ & $-4$ & $\frac{14175\lambda^4}{16}N^{-18}$ \\\hline \end{tabular} 
\vspace{0.1cm}
\caption{The value of $\det(M(N))$ in the case that there is one zero closest to the unit circle, and that zero is inside the circle. The constants $\kappa$ and $\lambda$ are independent of $N$ and defined in Propositions \ref{prop1} and \ref{prop3}. }
\label{algscaling}

\end{center}
\end{table}

If there are two complex zeros nearest to the unit circle then $\det M(N)$ is real but can contain $O(1)$ oscillatory terms such as $\sin(N\arg(\zeta_\star))$ (these oscillations may, however, not appear in the leading order term of $\det M(N)$). Moreover, if $|\zeta_\star|<1$ then $\det M(N) = \Theta(N^{-K|\omega-\alpha|})$, where $K=1/2$ for $\omega-\alpha > 0$ and $K=3/2$ for $\omega-\alpha <0$. The assignment of $K$ is reversed when $|\zeta_\star|>1$. \par
We complete our analysis of gapped models with a result for the asymptotic approach to the value of the order parameter. In particular, we prove that $\xi_\omega = \xi/2$, following from:
\begin{theorem}\label{orderlength}
In generic systems that are in the phase $(\omega, c=0,\Sigma)$, we have for large $N$ that
\begin{align}
 \langle\mathcal{O}_\omega(1)\mathcal{O}_\omega(N+1)\rangle  =\left(\rme^{\rmi \pi N m}\lim_{M\rightarrow\infty}\lvert\langle\mathcal{O}_\omega(1)\mathcal{O}_\omega(M)\rangle\rvert\right)\left(1+B_N\frac{ \rme^{-2N/\xi}}{N^2}\right)\left(1+o(1)\right). 
\end{align}
The factor $B_N$ is given implicitly in the proof and satisfies $|B_N| = O(1)$, $m\in \mathbb{Z}$ is a known constant. Generic systems are defined as in Theorem \ref{lengththeorem}.
\end{theorem}
The results of the discussion in Section \ref{sec::corrl} allow extension to non-generic systems. Given non-zero correlation lengths of $\mathcal{O}_\alpha$ for $\alpha\neq\omega$, the formula $1/\xi_\omega = 1/\xi_{\omega-1}+1/\xi_{\omega+1}$ holds. This agrees with Theorem \ref{orderlength} in the generic case where $ \xi_{\omega+1} =\xi_{\omega-1}$.\par
We now discuss results for the gapless phases. In critical chains the phases in the BDI class described in the bulk by a CFT and connected to a stack of translation invariant chains with arbitrary unit cell are classified by the semigroup $\mathbb{Z}_{\geq 0}\times\mathbb{Z}$: they are labelled by the
central charge $c\in \frac{1}{2} \mathbb{Z}_{\geq 0}$ and topological invariant $\omega \in \mathbb{Z}$.
The proof, using the $f(z)$ picture, is given in \cite{VJP}. Our present interest is confined to translation-invariant Hamiltonians that lie in one of these phases, and our next result gives the scaling dimension of the infinite class of operators $\mathcal{O}_\alpha$. A graphical representation of this theorem is given in Figure \ref{fig:correlation}.
  \begin{theorem}\label{scalingtheorem}
Consider a critical chain in the phase $(\omega,c>0,\Sigma)$ where the $2c$ zeros on the unit circle are non-degenerate. Let $\tilde \alpha = \alpha - (\omega + c)$. Then the operator $\mathcal{O}_\alpha$ has scaling dimension 
\begin{align}\Delta_\alpha(c,\omega) = c \left( \frac{1}{4} + x^2 - (x - [x])^2 \right) \Big|_{x = \tilde \alpha / 2c}\label{criticalscaling}
\end{align}
$[x]$ denotes the nearest integer to $x$.
\end{theorem}
Note that $\Delta_\alpha$ explicitly depends on the topological invariant $\omega$. Equation \eqref{criticalscaling} is independent of the choice in rounding half-integers, although for later notational convenience we define it to round upwards in that case. In Section \ref{sec::gapless} we prove Theorem \ref{scalingtheorem} on the way to the more detailed Theorem \ref{fullcritresult}. That theorem gives the full leading order term in the asymptotic expansion of $\langle \mathcal{O}_\alpha(1)\mathcal{O}_\alpha(N+1)\rangle$ at criticality, including nontrivial oscillatory factors that are helpful in identifying lattice operators with fields in the CFT description. 
 We give a discussion of this CFT description in Section \ref{sec::CFT}. A similar result holds when we have degenerate zeros on the unit circle, as long as every degeneracy is odd. We do not give results for the case that we have any zero of even degeneracy. \par
In the gapped case, Theorem \ref{phases} makes a simple link between measuring $\langle \mathcal{O}_\alpha(1)\mathcal{O}_\alpha(N+1)\rangle$ and learning $\omega$---one simply looks for the value of $\alpha$ with long-range order. It is not immediately obvious how to generalise this to critical models. Theorem \ref{scalingtheorem} shows that, as in the gapped phases, the behaviour of correlation functions allows us to see marked differences between different critical phases. In particular, the link between lattice operator and the operators that dominate its CFT description changes at a transition between critical phases (discussed in detail below).  In the critical case one can determine $c$ and $\omega$, but this requires information about more than one correlator. Inspecting the form of equation \eqref{criticalscaling}, displayed in Figure \ref{fig:correlation}, one concludes that it is not necessary to measure the scaling dimension of all $\mathcal{O}_\alpha$ in order to determine the phase. One method would be to measure the scaling dimensions of $\{\Delta_\alpha,\Delta_{\alpha\pm1},\Delta_{\alpha+2}\}$ for some convenient $\alpha$, and form the set of $\delta_\alpha = \Delta_{\alpha+1}-\Delta_\alpha$. This difference is equal to $[(\alpha - \omega)/2c-1/2]$---this means that $\delta_\alpha$ is a constant integer on plateaus of width\footnote{For clarity, by a plateau of width $2c$ we mean that $\delta_\alpha$ is constant for $2c$ consecutive values of $\alpha$.} $2c$, and that neighbouring plateaus differ in value by one. If the $\delta_\alpha$ are all different\footnote{We need three values of $\delta_j$ to check whether $c=1/2$ as $\delta_\alpha$ and $\delta_{\alpha+1}$ can be different if we happened to choose $\alpha$ at a kink in the scaling dimension plot---see Figure \ref{fig:correlation}.} then we must have $c=1/2$ and $\omega$ can be determined easily using $\omega = \alpha -\delta_\alpha $. Otherwise, one should then measure further scaling dimensions until the width of the constant plateau (equal to $2c$) is found. Once $c$ is known, $\omega$ may be determined: on the edge of the plateau we have $\omega = \alpha -2c \delta_\alpha$. Inferring the critical phase through these scaling dimensions is analogous to distinguishing the gapped phases through the string order parameter. If our model is taken to represent a spin chain then the $\mathcal{O}_\alpha$ are local for $\alpha$ odd. In Appendix \ref{app:delta} we show that it is possible to recover both $c$ and $\omega$ using scaling dimensions of local operators on the spin chain. 
Moreover, in gapped chains one can use the universality of the gapped correlations to similarly infer $\omega$ from knowing only two correlation lengths; this is explained in Section \ref{sec::stringorder}.\par
\subsection{The dual spin chain}\label{sec::spin}
Our results apply not only to $H_{\mathrm{BDI}}$ but also to certain spin-1/2 chains. We briefly review this correspondence so that the reader can have both pictures in mind, and to help us make links to the literature in the next section.
We write the Pauli operators as
\begin{align}\label{pauli}
X= \left(\begin{array}{cc}0~ & 1 \\1~ & 0\end{array}\right),\quad Y= \left(\begin{array}{cc}0 ~& -\rmi \\\rmi~ & 0\end{array}\right), \quad Z= \left(\begin{array}{cc}1 & 0 \\0 & -1\end{array}\right).
\end{align}
Define $X_n$, $Y_n$, $Z_n$ as the operators $X$, $Y$, $Z$ acting on the $n$th site (and tensored with identity on all other sites). A class of translation-invariant spin chains is given by Hamiltonians of the form
\begin{align}\label{hspin}
H_{\mathrm{spin}} &= \frac{t_0}{2} \sum_{n\in\mathrm{sites}} Z_n-\sum_{\alpha > 0} \frac{t_\alpha}{2} \left( \sum_{n\in\mathrm{sites}}X_n \left(\prod_{m=n+1}^{n+\alpha-1} Z_m\right)X_{n+\alpha}\right)-\nonumber\\&\hspace{1cm}\sum_{\alpha < 0} \frac{t_\alpha}{2}\left( \sum_{n\in\mathrm{sites}}Y_n \left(\prod_{m=n+1}^{n+\lvert\alpha\rvert-1} Z_m\right)Y_{n+\lvert\alpha\rvert}\right).\end{align} 
As before, we only allow a finite sum over $\alpha$, have $t_\alpha \in \mathbb{R}$ and take periodic boundary conditions. This is the class of generalised cluster models. Note that this includes the quantum Ising, XY and cluster models as special cases. 
In Appendix \ref{app::spin} we give details of the Jordan-Wigner transformation that relates $H_{\mathrm{spin}}$ to $H_{\mathrm{BDI}}$. The main point is that our results for the behaviour of $\langle\mathcal{O}_\alpha(1)\mathcal{O}_\alpha(N+1)\rangle$ apply equally well to the spin chain. The expressions for $\mathcal{O}_\alpha$ in terms of spin operators are given in Table \ref{tab::spinops}. Some of these operators appeared in the recent works \cite{Minami,Friedman}. Note that, as displayed in Table \ref{tab::spinops}, $\mathcal O_\alpha$ is local for odd $\alpha$ but remains a non-local string for even $\alpha$. One can easily see that for odd $\alpha$, $\langle \mathcal{O}_\alpha \rangle$ is zero in any symmetric state; hence Theorem \ref{phases} implies that for odd $\omega$ we have spontaneous symmetry breaking. For even $\omega$, however, $\mathcal O_\omega$ is a string
order parameter for the spin chain. As we discuss further in Section \ref{sec::stringorder}, this is indicative of the spin chain forming an interacting SPT phase. Note that the two phases can coexist: for $\alpha=3$, the order parameter is $\mathcal O_3(n) = X_{n+1}Y_{n+2}X_{n+3}$; hence $P = \prod_j Z_j$ and $T=K$ have been broken. However, $PT$ is preserved, and due to the similarity of $\mathcal O_3$ with the cluster model Hamiltonian---$\sum_j X_jZ_{j+1}X_{j+2}$---the ground state is \emph{also} an SPT phase protected by $PT$. Further details may be found in reference \cite{VMP}.\par
\begin{table}
\begin{center}
\begin{tabular}{|c||c|}\hline $\alpha$ & $\mathcal{O}_\alpha(n)$ \\\hline\hline Positive, odd & $X_{n}Y_{n+1}X_{n+2}Y_{n+3}\dots X_{n+\alpha-1}$  \\\hline Positive, even & $\left(\prod_{j=1}^{n-1} Z_j\right) Y_{n}X_{n+1}Y_{n+2}X_{n+3}\dots Y_{n+\alpha-2}X_{n+\alpha-1}$ \\\hline Zero & $ \prod_{j=1}^{n-1} Z_j$ \\\hline Negative, odd & $Y_{n}X_{n+1}Y_{n+2}X_{n+3}\dots Y_{n+|\alpha|-1}$ \\\hline Negative, even & $ \left(\prod_{j=1}^{n-1} Z_j\right) X_{n}Y_{n+1}X_{n+2}Y_{n+3}\dots X_{n+|\alpha|-2}Y_{n+|\alpha|-1} $ \\\hline \end{tabular}\vspace{0.1cm}
\caption{Spin operators that are the Jordan-Wigner dual of the fermionic operators $\mathcal{O}_\alpha $. }
\label{tab::spinops}   
\end{center}
\end{table}

\subsection{Relation to previous work}\label{previouswork}
In reference \cite{LSM}, Lieb, Schultz and Mattis set the stage for the analysis of determinantal correlations in free fermion models and related spin chains. The key reference related to our results for gapped models is the classic paper of Barouch and McCoy \cite{BMc}. There the authors study bulk correlations in the XY model which is the spin model equivalent to \eqref{majoranaham} with non-zero $t_0, t_1, t_{-1}$ only (and hence $f(z)$ depends on two zeros). The section of that paper on zero temperature correlations contains results for $\langle\mathcal{O}_\alpha(1)\mathcal{O}_\alpha(N+1)\rangle $ for $\alpha = 1,-1$ in the phases $\omega=0,1$ that include what one would obtain from our theorems. Beyond that, the paper \cite{clustermag} includes a calculation of the value of the order parameter for $\alpha = -1, 2$ in the special case that $f(z) = z^3 - \lambda$. Some portion of the phase diagram for $-2\leq\omega \leq 2$ is mapped out in reference \cite{Ohta} where order parameters are identified and calculated numerically. Several papers, for example \cite{son1,Nie}, study spin models with competing `large' cluster term and Ising term (i.e. non-zero $t_\alpha$, $t_{-1}$ and $t_0$). In these cases winding numbers are identified, but not order parameters or their values. Our computation giving Theorem \ref{orderparameter} is novel, extending previous calculations by addressing the full set of translation invariant models in the BDI class which require $f(z)$ with an arbitrary (finite) set of zeros. Moreover, this generality shows the robustness of these order parameters throughout the phase diagram.  \par
As mentioned, several papers have identified the form of the order parameters for $|\omega|\leq 4$ in the spin language. Equivalent fermionic order parameters are easily found using the Jordan-Wigner transformation and the paper \cite{BahriVishwanath} includes the fermionic $\mathcal{O}_{\alpha}$ for $|\alpha|\leq 2$ as well as discussing the general case.  In our work we prove that the intuitive general case holds by linking these order parameters to the generating function $f(z)$ and matching the winding number of $f(z)$ to the `unwinding number' of each correlator.\par
There are many works that study correlations in particular quantum phase transitions in our model. Again reference \cite{BMc} should be mentioned, along with \cite{Wu}, as seminal early works that derived critical behaviour for correlators $\langle\mathcal{O}_\alpha(1)\mathcal{O}_\alpha(N+1)\rangle $ with $|\alpha| \leq 1$ at the $c=1/2$ Ising transition. Transitions with higher $c$ include the $c=1$ XX model that is a standard model in physics \cite{Sachdev}, and the same correlators were analysed in reference \cite{ov} using the mathematical methods found below. We also mention the quantum inverse scattering method as a tool for calculating scaling dimensions in certain cases \cite{KBI} . \par
An isotropic spin chain is invariant under spin-rotation around the $Z$ axis. In our fermionic model $H_{\mathrm{BDI}}$, this manifests as invariance under spatial inversion $H_\alpha\leftrightarrow H_{-\alpha}$, and hence is a model for which $f(z) = f(1/z)$. This relation implies that $\omega=-c$. The correlators $\langle\mathcal{O}_\alpha(1)\mathcal{O}_\alpha(N+1)\rangle $ with $|\alpha| \leq 1$ in isotropic models with general $c$ and $\omega = -c$ were derived in references \cite{HKM,HJ} using the same methods as this paper. Our results go further by studying a wider class of models, including critical phases with general $(c,\omega)$, as well as a wider class of observables: $\langle\mathcal{O}_\alpha(1)\mathcal{O}_\alpha(N+1)\rangle $ for all $\alpha$. This allows us to observe that from knowledge of the scaling dimensions of these operators, one can identify the topological invariant $\omega$.\par
\section{Physical context and discussion}\label{sec::physics}
In this section we interpret our results and give them context. In Section \ref{sec::stringorder} we discuss universality and its implications for spin chains, as well as the relation to symmetry fractionalisation. In Section \ref{sec::majorana} we relate correlation length in the bulk to localisation length at the edge and in Section \ref{sec::critexp} we derive critical exponents from our results. In Section \ref{sec::CFT} we connect our lattice results to continuum (CFT) models and finally in Section \ref{sec::entanglementscaling} we discuss entanglement scaling approaching a transition. 
\subsection{SPT phases and string orders: universality and symmetry fractionalisation}\label{sec::stringorder}
In Section \ref{results}, we saw that for both gapped and critical systems in our class of models, one can measure the topological
invariant $\omega$ by looking at the string order parameters $\mathcal O_\alpha$. For gapped phases, one needs to find the value of $\alpha$ for which there is long-range order (Theorem \ref{phases}), whereas in the critical case, one uses the scaling dimensions (see Theorem \ref{scalingtheorem} and the discussion following it). The existence of topological string order parameters for critical phases is novel. However, even for the gapped phases that we consider, the string order parameters are unusual. This is for two reasons. Firstly, the usual justification for string orders relies on the concept of symmetry fractionalisation, which arises in the classification of \emph{interacting} SPT phases and is usually not employed in the classification of non-interacting topological insulators and superconductors. Secondly, even in the interacting case, phases which are protected by anti-unitary symmetries do not give rise to the kind of string order parameters we discuss in this work. Bridging this gap is the purpose of Section \ref{sec:interactions}. However, first we discuss the remarkable result that the correlations in the gapped phases exhibit universal properties.

\subsubsection{Universality} In the gapped phases, $\mathcal O_\alpha$ has correlation length $\xi_\alpha = \frac{\xi}{|\alpha-\omega|}$ (if $\alpha \neq \omega$), see Figure \ref{fig:correlation} (we assume the generic case for this discussion). This means that although $\xi$ depends on microscopic properties (like the position of the zeros of $f(z)$), the ratio ${\xi_\alpha}/{\xi_\beta}$ depends only on $\omega$ and is hence constant in each phase. This has interesting consequences. In principle, to determine the topological invariant of a gapped phase, one has to find an $\alpha$ such that $ \lvert\langle \mathcal O_\alpha(1) \mathcal O_\alpha(N)\rangle\rvert$ tends to a non-zero limit as $N \to \infty$. This requires going through an arbitrarily large set of observables. Surprisingly, it is sufficient to measure only, for example, \emph{two} correlation lengths $\xi_{\alpha_1}$ and $\xi_{\alpha_2}$ (for the observables $\mathcal O_{\alpha_1}$ and $\mathcal O_{\alpha_2}$) for any fixed choice of $\alpha_1$ and $\alpha_2$ satisfying $|\alpha_1 - \alpha_2| \in \{1,2 \}$. To see this, note that there are three cases. Firstly, if one finds long-range order for either $\alpha_1$ or $\alpha_2$, then $\omega$ is known. Secondly, if $\xi_{\alpha_1} = \xi_{\alpha_2}$, one knows that $\omega = \frac{\alpha_1 + \alpha_2}{2}$. In any other case, $\alpha_1$ and $\alpha_2$ will either both be larger or smaller than $\omega$, such that $\frac{\xi_{\alpha_1}}{\xi_{\alpha_2}} = \frac{\omega - \alpha_2}{\omega - \alpha_1}$. This can be uniquely solved, giving $\omega = \frac{\alpha_1 \xi_{\alpha_1} - \alpha_2 \xi_{\alpha_2}}{\xi_{\alpha_1} - \xi_{\alpha_2}}$.

The above shows that, using universality, one can replace an infinite number of observables by just two. However, it has an even more surprising consequence in the spin language. If we choose $\alpha_1 = 1$ and $\alpha_2 = -1$, then this corresponds to the correlation lengths $\xi_X$ and $\xi_Y$ of the local observables $X_n$ and $Y_n$. This fully determines the invariant
\begin{equation}
	\omega = \left\{
	\begin{array}{ccl}
		-1  & &\textrm{if }\lim_{N \to \infty}\langle Y_1 Y_N \rangle\neq 0\\
		1  & &\textrm{if }\lim_{N \to \infty}\langle X_1 X_N \rangle \neq 0\\
		0  &  &\textrm{otherwise and if } \xi_X = \xi_Y\\
		\frac{\xi_X+\xi_Y}{\xi_X-\xi_Y} & & \textrm{otherwise.}
	\end{array} \right.
\end{equation}
This means that one can distinguish, for example, the trivial paramagnetic phase from the topological cluster phase\footnote{I.e. models that can be connected to $H = -\sum_i X_i Z_{i+1}X_{i+2}$ without a phase transition.} by measuring the decay of correlation functions of \emph{local} observables. This is truly unusual and presumably an artifact of looking at spin models that are dual to non-interacting fermions. It would be interesting to investigate such ratios between correlation lengths in interacting models and determine whether this is a measure of the interaction strength between quasi-particles. 

\subsubsection{Symmetry fractionalisation} \label{sec:interactions} To contrast our analysis to the standard justification for string orders, we briefly repeat how string order parameters arise within the context of symmetry fractionalisation. It is worth emphasising that the known constructions for string order parameters of the type that we discuss are only for SPT phases which are protected by \emph{unitary} symmetries \cite{PollmannTurner,PhysRevLett.100.167202}.

\begin{figure}[h]
	\includegraphics[width=\linewidth]{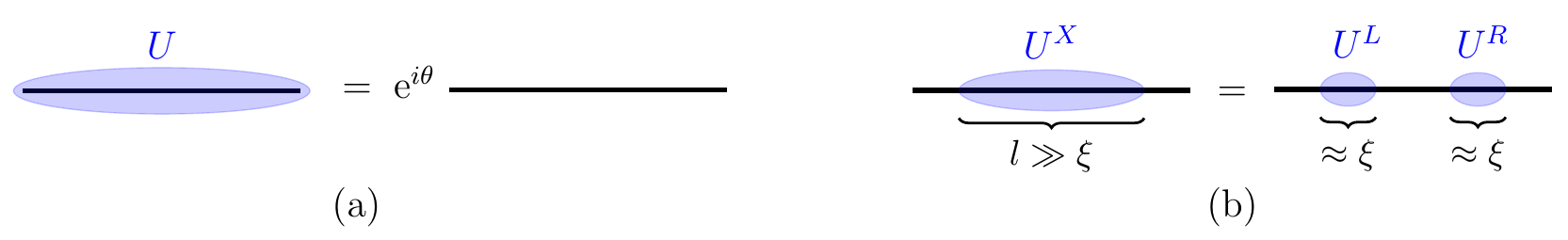}
	\caption{Consider a system which does not exhibit spontaneous symmetry breaking. (a) The ground state remains invariant when applying a global symmetry $U = \prod_n U_n$. (b) When acting with $U^X$ on a line segment $X$ of length $l \gg \xi$, then deep within that segment, the action is indistinguishable from the global symmetry operation $U$. Hence, the state can only be changed near the boundaries of $X$. In conclusion, effectively, $U \approx U^L U^R$. \label{fig:symfrac}}
\end{figure}

Let $U$ be some on-site unitary symmetry, i.e. $[U,H] = 0$ with $U = \prod_n U_n$. Consider the operator $U^X = \large \prod_{n\in X} U_n$ where $X$ is some large line segment of length $l$ (see Figure~\ref{fig:symfrac}). If $l \gg \xi$, then deep within $X$, $U^X$ looks like a bona fide symmetry operator. Hence, it is only near the edge of $X$ that $U^X$ can have a non-trivial effect. In other words, if $|\textrm{gs}\rangle$ is the ground state, then effectively $U^X |\textrm{gs}\rangle = U^L U^R |\textrm{gs}\rangle$, where $U^L$ and $U^R$ are operators that are exponentially localised near the boundary of the region $X$. This can be made rigorous using matrix product states \cite{Hastings,PhysRevLett.100.167202}. This phenomenon is known as \emph{symmetry fractionalisation} and is the essential insight that led to the classification of (interacting) SPT phases in 1D \cite{Fidkowski11class,Pollmann12,Chen10,Schuch11}. To illustrate this, consider the case with a symmetry group $\mathbb Z_2 \times \mathbb Z_2$, generated by global on-site unitary symmetries $U$ and $V$. Since $U_n$ and $V_n$ commute on every site, we have that $U^X V = V U^X$. Moreover, $U^X = U^L U^R$ (when acting on the ground state subspace), implying that $U^L$ and $V$ have to commute up to a complex phase. Using that $V^2 = 1$, we arrive at $U^L V = (-1)^{\frac{\omega}{2}} V U^L$ where $\omega \in \{0,2\}$. This defines a discrete invariant which allows to distinguish two symmetry-preserving phases. (One says that the phases are labelled by the inequivalent classes of projective representations of $\mathbb Z_2 \times \mathbb Z_2$.) In fact, one can show that in the phase where $\omega = 2$, the negative sign implies degeneracies both in the entanglement spectrum and, for open boundary conditions, in the energy spectrum \cite{Pollmann10}. Here we will not go into such details, referring the interested reader to the review in reference \cite{VMP}. Instead, we consider the effect on correlation functions.

One can consider the string correlation function $\bra{  \textrm{gs}} U^X \ket{ \textrm{gs}}  = \langle U^X \rangle = \langle U^L U^R \rangle$. Due to locality, we have that $\langle U^X \rangle \approx \langle U^L \rangle \langle U^R \rangle$ for $l \gg \xi$. Since SPT phases do not spontaneously break symmetries, we have that $\langle U^L \rangle = \langle U^L V \rangle = (-1)^\frac{\omega}{2} \langle V U^L \rangle = (-1)^\frac{\omega}{2} \langle U^L \rangle $. Hence, we conclude that the string correlation function $\langle U^X \rangle$ has to be zero if $\omega = 2$. Equivalently, measuring $\langle U^X \rangle \neq 0$ implies that $\omega = 0$, such that one calls this a string order parameter for the trivial phase ($\omega = 0$). Analogously, one can construct a string order parameter for the non-trivial phase: if $\mathcal V,\mathcal W$ are local operators anticommuting with $V$, then by repeating the above argument, we conclude that $\langle \mathcal V U^X \mathcal W \rangle \neq 0$ implies that $\omega = 2$ (with $\mathcal V$ ($\mathcal W$) localised near the left (right) of region $X$). Note that these string order parameters always work only in one direction: there is no information if one measures them to be zero. This is in striking contrast with the string order parameters we found for our non-interacting class of models.

Let us make this discussion more concrete with an example, where $U = P = \large \prod_n Z_n$ and $V = P_\textrm{odd} = \large \prod_{m\textrm{ odd}} Z_m$. Two models with this symmetry are $H_0 = \sum_n Z_n$ and $H_2 = - \sum_n X_{n-1} Z_n X_{n+1}$. One can calculate that their symmetry fractionalisations are $\omega =0$ and $\omega = 2$, respectively. The above tells us that $\prod_{n=1}^N Z_n$ is a string order parameter for the trivial phase. Similarly, taking $\mathcal V_{n,n+1} = Y_n X_{n+1}$---which indeed anticommutes with $P_\textrm{odd}$---then $\mathcal V_{1,2} \left( \prod_{n=1}^N Z_n \right) \mathcal V_{N+1,N+2}$, or equivalently $X_1 Y_2 \left( \prod_{n=3}^N Z_n \right) Y_{N+1} X_{N+2}$, is an order parameter for the topological phase $\omega = 2$. In this case, the string order parameters we have derived---with respect to $\mathbb Z_2\times\mathbb Z_2$--- for the trivial phase (connected to $H_0$) and the topological cluster phase (connected to $H_2$) happen to be the same as we encountered in the non-interacting case---with respect to the $P$ and $T$ symmetries---see Table~\ref{tab::spinops}.

Can we make a connection with our non-interacting classification and the concept of symmetry fractionalisation? For this it is easiest to work in the fermionic language. It is known that if one studies the fractionalisation of only the $P$ and $T$ symmetries, then there are only eight distinct phases \cite{Fidkowski10interaction}. However, since our model is non-interacting, the $P$ and $T$ symmetries imply an additional structural symmetry: the Hamiltonian can only contain terms which have an equal number of real and imaginary Majorana modes. This implies that if we have any operator which has a well-defined number of real minus imaginary Majorana operators (e.g. $\gamma_n \gamma_{n+1}$ would have `charge' two), then the Hamiltonian time evolution would conserve this. To see how this is useful, consider a fixed-point model $H_\alpha =  \sfrac{\rmi}{2}\sum_{n \in \mathrm{sites}}  \tilde\gamma_n \gamma_{n+\alpha}$. It is a simple exercise to check that for the symmetry fractionalisation of $P = P_L P_R$, we have that $P_L$ has charge $-\omega$ (and $P_R$ charge $\omega$). By the aforementioned argument and the concept of adiabatic connectivity, these charges of $P_L$ and $P_R$ should be stable throughout each gapped phase. It is easy to see that $\langle \mathcal V \rangle = 0$ for any operator whose charge is non-zero. Hence, in this way we are naturally led to consider $\gamma_1 \cdots \gamma_\alpha \left( \tilde \gamma_{\alpha+1} \gamma_{\alpha+1} \cdots \tilde \gamma_{N} \gamma_{N} \right) \tilde \gamma_{N+1} \cdots \tilde \gamma_{N+\alpha}$: this operator can only have long-range order if $\alpha = \omega$.\par
The power of symmetry fractionalisation is that it is not specific to the non-interacting case. Does that mean that if one considers interacting Hamiltonians preserving the aforementioned structural symmetry\footnote{That is, we allow any interaction term that contains the same number of real as imaginary Majorana modes.}, that we obtain an infinite number of distinct \emph{interacting} phases, labelled by $\mathbb{Z}$ and distinguished by the same string order parameters as the non-interacting case? This is not clear: the structural symmetry is not a conventional symmetry; in particular, the charge of this symmetry is not well-behaved under the multiplication of operators (which is usually essential to preserve charge under renormalization group flow). Hence, it is perhaps unlikely that this structural symmetry gives rise to non-trivial physics in the interacting case, but it might nevertheless be interesting to explore further.

\subsection{Majorana modes: localisation length versus correlation length}\label{sec::majorana}
In reference \cite{VJP}, we showed that if $\omega >0$, then the system has $\omega$ Majorana zero modes per edge. More precisely, to each of the $\omega$ largest zeros $\{z_i\}_{i=1,\cdots,\omega}$ of $f(z)$ within the unit disk, we associate a hermitian operator $\gamma_L^i$, all of them mutually commuting and squaring to the identity. Moreover, these operators commute with $T$ and are exponentially localised near the left edge, with respective localisation lengths $-1/\ln |z_i|$. (The same is true for the right edge, where they anticommute with $T$.) The crucial property which makes them so-called \emph{zero-energy} modes, is that they commute with the Hamiltonian (up to a finite-size error which is exponentially small in system size). Hence we have $\omega$ mutually anticommuting symmetries, from which one can show that to each edge we can associate a $\sqrt{2}^\omega$-fold degeneracy\footnote{Hence the system as a whole has a $2^\omega$-fold degeneracy.}. This is to be contrasted with the fact that the ground state is unique for periodic boundary conditions. This is a characteristic property of topological insulators (and more generally, symmetry-protected topological phases) in one spatial dimension. This is well-known for the gapped phases of the BDI class, but the proof in reference \cite{VJP} shows that this analysis goes through when the bulk is critical (i.e. $c\neq 0$).\par
Reference \cite{Niu12} notes the link between the localisation length of the Majorana edge mode and the behaviour of bulk spin correlations in a model equivalent to the XY model, and conjectured that this is a general phenomenon. Here we simply point out that the largest localisation length of a Majorana mode (if present) need not coincide with the bulk correlation length. Indeed, the latter is determined by the zero of $f(z)$ closest to the unit circle. In particular, if $\omega > 0$, the localisation lengths of the Majorana modes are determined by zeros \emph{within} the unit disk, whereas it could certainly be that a zero \emph{outside} the unit circle dominates the bulk correlation length. This disagreement between the two quantities is consistent with the observation in \cite{VJP} that one can tune a gapped phase to a critical point whilst (some of) the edge modes remain exponentially localised.   \par
This discussion for $\omega >0$ holds also for $\omega < -2c$ (if we interchange the words `left' and `right'), with the edge modes now being associated to zeros \emph{outside} the unit circle. As mentioned above, spatial left-right inversion corresponds to $f(z) \leftrightarrow f(1/z)$, which at the level the topological invariant corresponds to $\omega \leftrightarrow -\omega-2c$. For any other value of $\omega$, there are no edge modes \cite{VJP}.

\subsection{Critical exponents}\label{sec::critexp}
Critical exponents encode how different physical quantities diverge upon approaching a phase transition. In the classical case, the tuning parameter is usually the renormalised temperature $\tau = \frac{T-T_c}{T_c}$. In the current quantum setting, there is an equally natural\footnote{Moreover, one can check that this agrees with $\tau$ under the quantum-classical correspondence.} tuning parameter $\varepsilon \in \mathbb R$: if $f(z)$ represents a gapless Hamiltonian, then $f_\varepsilon(z) := f(z(1-\varepsilon))$ interpolates between two gapped phases. One can think of this as shrinking ($\varepsilon<0$) or expanding ($\varepsilon>0$) the radial component of the zeros of $f(z)$. We will work in the case where the system is gapless at $\varepsilon=0$ with $2c$ zeros on the unit circle, allowing for a multiplicity $m$ (which we take to be the same for every distinct zero). This means we have ${2c}/{m}$ \emph{distinct} zeros on the unit circle. We emphasise that the expressions in the remainder of this section are derived only in the case of uniform multiplicity. Our results allow for an analysis of the general case, but we do not wish to pursue this here. \par
We derive four critical exponents: the \emph{anomalous scaling dimension}, $\eta$, defined by the scaling of the order parameter, $\Psi$, \emph{at} the critical point ($\langle \Psi(1)\Psi(N) \rangle \sim 1/N^\eta$); $\nu$, which encodes the divergence of the correlation length ($\xi \sim |\varepsilon|^{-\nu}$); the \emph{dynamical critical exponent}, $z$, that relates how the correlation length $\xi$ diverges relative to the characteristic time scale defined by $\tau$, the inverse energy gap  ($\tau \sim \xi^z$); and $\beta$ which relates to the decay of the order parameter ($\Psi \sim |\varepsilon|^\beta$).

Exactly at the critical point, we have given explicit results above only for $m=1$. In particular, Theorem ~\ref{scalingtheorem} gives us that $\eta = {c}/{2}$. In the special case of $c={1}/{2}$, we recover the well-known result $\eta = 1/{4}$. For $m>1$ and \emph{odd}, we can use equation \eqref{sumrule1b} from Appendix \ref{app::degenerate} to easily derive that $\eta = mc/2$. The other three exponents are defined away from criticality where we can allow for any $m\geq 1$. The correlation length is determined by the nearest zero, such that $\xi \sim 1/|\ln|1+\varepsilon|| \sim 1/|\varepsilon|$. Hence, $\nu = 1$; independent of $c$ and $m$.
The energy gap, $\min_{z\in S^1} |f(z)|$, depends on the location of all zeros. However, close to criticality, we need to care only about the zeros describing the transition. Moreover, each of the distinct zeros has a local minimum, which all scale the same way. So without loss of generality, we can consider $f(z) = (z-(1+\varepsilon))^m$. The gap scales as $\sim |\varepsilon|^m$, such that $\tau \sim \xi^m$ with dynamical critical exponent $z=m$. This is consistent with $m=1$ being described by a CFT with central charge $c$, since that implies $z=1$. Lastly, we consider the order parameter given in Theorem \ref{orderparameter}. For each distinct zero, the order parameter has a factor $|1-(1+\varepsilon)^{-2}|^{m^2/8} \sim |\varepsilon|^{m^2/8}$; all other terms do not go to zero with $\varepsilon$. Combining $2c/m$ such factors, we have $\Psi \sim |\varepsilon|^{\frac{m^2}{8} \times \frac{2c}{m}}$. Hence, $\beta = mc/4$. In the special case of the Ising transition, $m=1$ and $c={1}/{2}$, this reduces to the well-known result $\beta = {1}/{8}$.\par
The above results are consistent with the well-known scaling relations \cite{cardy}. In particular, we can straightforwardly confirm that $2\beta = \nu (\eta+d-2)$, where our dimension is $d=2$. In fact, the aforementioned relationship implies that $\eta = {mc}/{2}$ for \emph{any} multiplicity $m$. Moreover, such scaling relations can be used to derive other critical exponents: such as $\gamma = \nu(2-\eta) = 2-mc/{2}$, $\alpha = 2-\nu d = 0$ and $\delta = {\nu d}/{\beta} -1 =  {8}/{mc}-1$. It is interesting to note that the critical exponent $\gamma$ changes sign for $mc = 4$. This data is summarised in Table \ref{tab:exp}.

\begin{table}[htp]
\begin{center}\begin{tabular}{|c||c|c|c|c|c|c|c|}\hline Critical exponent & $\alpha$ & $\beta$ & $\gamma$ & $\delta$  & $\eta$& $\nu$ & $z$   \\\hline\hline Value & $0$ & $mc/4$ & $2-mc/2$   & $8/mc-1$& $mc/2$& 1 & $m$ \\\hline \end{tabular} 
\caption{Summary of the critical exponents found in Section \ref{sec::critexp}. As above, $c$ is half the number of zeros on the unit circle and $m$ is the multiplicity of these zeros (taken to be uniform). If $m=1$, then $c$ is the central charge.}
\label{tab:exp}
\end{center}
\end{table}
\subsection{CFT and continuum-lattice correspondence}\label{sec::CFT}
We now explain how certain features of Theorem \ref{scalingtheorem} (and Theorem \ref{fullcritresult}) fit in to a CFT analysis of the critical point. This section is not rigorous, the aim is to complement the mathematical proofs with a perhaps more intuitive physical picture. We also use Theorem \ref{fullcritresult} to make claims about passing from the lattice to the continuum description of the operators $\mathcal{O}_\alpha$. \par
If our system has $2c$ non-degenerate zeros on the unit circle, then one can argue that the appropriate low-energy theory is a CFT built from $2c$ real, massless, relativistic free fermions \cite{shankar}. Briefly, one can linearise the dispersion relation $|f(k)|$ about all of its zeros on the circle (Fermi momenta), and each local linearised mode is such a fermion. Moreover, since $|f(k)| = |f(-k)|$ we can combine the real fermions from a pair of complex conjugate zeros to form a complex fermion with central charge one. This is helpful as complex fermions can be bosonised using the methods given in \cite{Haldane81,DS2,Sachdev}. In general, then, we have a low energy theory of an even number of complex fermions and either $0$, $1$ or $2$ real fermions (located at $k=0$ or $\pi$). The central charge is always equal to $c$. \par 
\subsubsection{High level analysis}
In reference \cite{BIR}, general CFT considerations are applied to integrable models to find asymptotic correlation functions of local operators  $\mathcal{A}_{\bm{n}}(x)$ that create fixed numbers, $n_j$, of quasiparticles of type $j$ (i.e. excitations near momentum $\pm k_j$). For equal times, these take the form
\begin{align}\label{integrableform}
\langle \mathcal{A}^\dagger_{\bm{n}}(0)\mathcal{A}_{\bm{n}}(x)\rangle = \sum_{\{m_j\}} C_{\mathcal{A}_{\bm{n}}}x^{-2\sum_j \Delta_{\bm{n}}^{(j)}} \rme^{ \rmi  x \sum_j m_j k_j}
\end{align}
for scaling dimensions $\Delta_{\bm{n}}^{(j)}$, Fermi momenta $k_j$ and the sum is over sets of $m_j \in 2\mathbb{Z}$ . The amplitude $C_{\mathcal{A}_n}$ depends on the appropriate form factor \cite{BIR}. The oscillatory term comes from a multiplicative factor $\rme^{-\rmi x \delta p}$ when $\mathcal{A}$ gives an intermediate excitation with momentum $\delta p$---these are non-zero when we have particle-hole excitations where the particles and holes are at different Fermi points. Relevant discussion is found in references \cite{Korepin2,BIR,KBI}. The $\mathcal{O}_\alpha$ that we consider are `square-roots' of local operators in the sense that the product $\mathcal{O}_{\alpha}(n)\mathcal{O}_{\alpha}(n+m)$ (for small $m$) is a local fermionic operator on the lattice and has an expansion dominated by such $\mathcal{A}$. One may then expect that correlations of $\mathcal{O}_{\alpha}$ will have the same form as \eqref{integrableform}, but with $m_j \in \mathbb{Z}$.
This is verified in Theorem \ref{fullcritresult}, apart from the possibility of an additional $(-1)^x$ oscillation. This is needed when the low energy degrees of freedom are modulated by this oscillation. The constant in Theorem \ref{fullcritresult} implicitly gives form factors of the relevant fields \cite{BIR,Smi}. \par
\subsubsection{CFT operator correspondence for one real fermion}
We now consider more details of the CFT description and the correspondence with canonical continuum operators. The following discussion will be in terms of the spin chain dual to the fermionic system, as discussed in Section \ref{sec::spin}.\par
If our model has one zero on the unit circle (which must be at $z=\pm1$, corresponding to $k=0, \pi$ respectively), then our continuum limit has $c=1/2$, and is hence described by the Ising CFT \cite{diF}. The operator content of this theory is well understood and, amongst the primary operators, there are exactly two with scaling dimension $1/8$. These are the `spin' and `disorder' operators, denoted by $\sigma$ and $\mu$ respectively. In the usual lattice Ising model, $\sigma$ is the field corresponding to the \emph{local} order parameter of the neighbouring ordered phase and $\mu$ is a \emph{non-local} string order parameter of the neighbouring paramagnetic phase. 
Moreover, $\sigma$ is odd under parity symmetry (realised on the lattice as $P = \prod_j Z_j$), while $\mu$ is even under $P$. 
Now, if we are in a model with $\omega =0$ and $c =1/2$, then we can continuously tune to a critical Ising chain $H = \sum_j X_jX_{j+1} \pm Z_j$ for which the correspondence is standard and discussed in \cite{diF}. In particular, we will have $X_n \rightarrow \sigma(n)$ and $\prod_{j=-\infty}^n Z_j  \rightarrow \mu(n)$; note that this is consistent with the aforementioned symmetry considerations. Moreover, models in our class with $\omega =-1$ and $c =1/2$ are in the same phase as a critical Ising chain $H = \sum_j Y_jY_{j+1} \pm Z_j$ where the analogous standard correspondence is $Y_n \rightarrow \sigma(n)$ and $\prod_{j=-\infty}^n Z_j  \rightarrow \mu(n)$, again consistent with locality and symmetry properties. \par
Our results on the lattice, in particular Theorem \ref{scalingtheorem}, allow us to extend this by showing that a system with $c=1/2$ and winding number $\omega$ has two operators with this dominant scaling dimension: $\mathcal{O}_\omega$ and $\mathcal{O}_{\omega+1}$.  By considering locality and parity symmetry of the operators, as well as scaling dimension, we identify $\mathcal{O}_\omega$ with $\sigma$ when $\omega$ is odd and $\mu$ when $\omega$ is even; $\mathcal{O}_{\omega+1}$ is then the other field. Importantly, we conclude that the operators on the lattice that have overlap with the dominant primary scaling fields depend on $\omega$. Indeed, along a path that connects a model with $c=1/2$ and winding number $\omega$ to a model with $c=1/2$ and winding number $\omega' \neq \omega$ then we must encounter a multicritical point with $c\geq 1$ where the pair of dominant operators in the $c=1/2$ models will have degenerate scaling dimension and the dominant pair changes as we go through this point---we give an example of this in Section \ref{examplecrittran}. Other operators $\mathcal{O}_\alpha$ for $\alpha$ neither $\omega$ nor $\omega+1$ will be dominated by descendants of $\sigma$ for $\alpha$ odd and of $\mu$ for $\alpha$ even (the lattice operators have dimension $1/8 + j$ for some $j\in \mathbb{Z}_+$, so we should take CFT descendants at level $j$). Note that in all correspondences we may have a $(-1)^n$ oscillatory factor.

\subsubsection{CFT operator correspondence for one complex fermion}
 Let us now consider the case $c=1$ with a complex conjugate pair of zeros, at $\rme^{\pm\rmi k_F}$, and a U(1) symmetry generated by $S^z_\mathrm{tot}= \sfrac{1}{2}\sum_i Z_i$ (the standard model in this class is the XX spin chain \cite{Sachdev}, which in our language corresponds to $H = H_{-1} + H_1$ with $f(z) =  z+1/z$). These are isotropic models with $f(z) = f(1/z)$, which implies that $\omega =-c$; we discuss other values of $\omega$ later.\par The fermionic Hamiltonian may be bosonised as described in \cite{Haldane81,FG,Affleck} and \cite[Chap. 20]{Sachdev}, passing also to a continuum limit. We denote the resulting bosonic fields by $\theta(x)$ and $\varphi(x)$, such that 
 \begin{align}
 [\partial_x\varphi(x),\theta(y)]=[\partial_x\theta(x),\varphi(y)] = \rmi\pi \delta(x-y) \label{CCR} \end{align}
where $\delta(x)$ is the Dirac delta function. We now recall some standard results for the isotropic model  \cite{Sachdev,polchinski}. Firstly, the vertex operator $\rme^{\rmi \theta(x)}$ creates a localised charge of the aforementioned U(1) symmetry, and $\partial_x \varphi(x)$ is a density fluctuation: the total density is $\rho(x) = (k_F + \partial_x \varphi(x))/\pi$.
Vertex operators of the form $\rme^{\rmi(\lambda_1\theta(x) + \lambda_2 \varphi(x))}$ have scaling dimension $(\lambda_1^2+ \lambda_2^2)/4$. When $\lambda_1 \in \mathbb{Z}$ and $\lambda_2 = 2k$ for $k \in \mathbb{Z}$, these operators are well-defined and mutually local. When $\lambda_1 \in \mathbb{Z}$ and $\lambda_2 = 2k +1$ for $k\in \mathbb{Z}$, these operators have a branch cut extending to infinity. As discussed below, on the lattice this branch cut is related to the non-local Jordan Wigner strings. Notice that this is an asymmetry in the role of the fields $\varphi$ and $\theta$.
\par
We now consider operator correspondences for $\mathcal{O}_\alpha$. Bosonisation will not fix constant coefficients of the operators, so we will usually suppress them in the following. Note, however, that hermiticity and symmetries constrain certain coefficients. We may also need to include an additional antiferromagnetic oscillation in order to correctly correspond to the lattice model and hence to our results.\par
 Firstly, following \cite{Haldane81,FG}, we note that $\rho(x)$ generates the U(1) symmetry. Hence, we can make the formal identification 
\begin{align}
   \rme^{-\rmi \frac{\chi}{2} \sum_{j} Z_j} = \rme^{\rmi {\chi} \sum_{j}(c^\dagger_jc_j-1/2)}  \rightarrow\rme^{\rmi {\chi} \int (\rho(x) -1/2) \rmd x}.\label{usub}
\end{align}  The Jordan-Wigner strings follow by setting $\chi = \pi$ and truncating the sum and integral respectively. We can then make the correspondence (see, for example, \cite{FG,Affleck} and \cite[Chap. 20]{Sachdev}):
\begin{align}\prod_{i=-\infty}^n Z_i \rightarrow a\rme^{\rmi (k_F n + \varphi(n))}+ \overline{a}\rme^{-\rmi (k_F n + \varphi(n))} \qquad (a \in \mathbb{C}).\label{zsub}\end{align} 
 The fermionic creation operator creates a U(1) charge and is multiplied by a Jordan-Wigner string, this leads to standard expressions for the dominant contributions to the right ($+$) and left ($-$) moving continuum fermion fields
\begin{align}
\psi_\pm(x) = \rme^{\rmi\theta(x) \pm \rmi \varphi(x)} + \dots \qquad\qquad \left( c_n \rightarrow \psi(n) \simeq \rme^{\rmi k_F n}\psi_+(n) + \rme^{-\rmi k_F n} \psi_-(n)  \right). 
\end{align}  
Time-reversal symmetry swaps these right and left moving fields (one may check the lattice expansion given in, for example, \cite{Affleck}) and so we have that $\varphi\rightarrow \varphi$ and $\theta \rightarrow -\theta$ under $T$. We can also confirm that the right-hand-side of \eqref{zsub} is hermitian and does not transform\footnote{Under a U(1) rotation through $\chi \in \mathbb{R}$, lattice operators are conjugated by $\exp({\rmi \chi \sum_n Z_n}/2)$ and fields transform as $\theta(x) \rightarrow \theta(x) + \chi$, $\varphi(x) \rightarrow \varphi(x)$.} under the U(1) or $T$, as required for the $Z$-string. The site at $-\infty$ may appear problematic in isolation, but we always consider two-point correlators and thus the infinite string to the left will drop out (similarly in the continuum one may take correlation functions to avoid considering the boundary). The meaningful correspondence here is:
\begin{align}
\left\langle \prod_{i=0}^n Z_i  \right\rangle \rightarrow \left\langle \left( \rme^{\rmi (k_F n + \varphi(n))}+ \rme^{-\rmi (k_F n + \varphi(n))}\right)\left( \rme^{\rmi \varphi(0)}+ \rme^{-\rmi \varphi(0)}\right)\right\rangle,
\end{align}
where we now suppress constant coefficients.\par
Now, by considering the U(1) action and time-reversal symmetry, we see that $\sigma^{\pm}_n \rightarrow \rme^{\pm\rmi \theta(n)}$ (with real coefficients). We then have that 
\begin{align}
X_n &\rightarrow \cos(\theta(n)) \nonumber\\
Y_n &\rightarrow \sin(\theta(n)).\label{spinsubs}
\end{align} 
Note that the relation between $\sigma^+$ and $\psi_{\pm}$ in the CFT corresponds on the lattice to \eqref{JWsigma}. 
\par
The above correspondences are well known \cite{FG,Affleck,Sachdev}. In our notation, \eqref{zsub} corresponds to $\mathcal{O}_0$ and \eqref{spinsubs} correspond to $\mathcal{O}_{\pm1}$. Operators $\mathcal{O}_\alpha$ for $|\alpha|>1$ are more involved. However, the correspondences that we have already established should allow us to find the continuum operator by taking operator products. We consider first the family of local spin operators, with $\alpha$ odd. In particular, let us analyse $\mathcal{O}_3(n) = X_nY_{n+1}X_{n+2}$.  From Theorem \ref{scalingtheorem}, we know that this operator has scaling dimension $9/4$. In the field theory the operators $\rme^{\pm3\rmi \varphi}$ and  $\rme^{\pm3\rmi \theta}$ share this scaling dimension, as well as operators that include derivatives, for example $(\partial^2 \theta)  \rme^{\rmi \theta}$. We should exclude terms that include $\rme^{\rmi (2k+1) \varphi}$ on the grounds of locality. We now show that all operators depending solely on $\theta$ and that have the correct scaling dimension may appear. Moreover, we give a method for determining the exact correspondence.  \par
First, note that any product of three neighbouring $X$ or $Y$ operators can be expanded as a linear combination of terms $\sigma_{n}^{\pm}\sigma^{\pm}_{n+1}\sigma^{\pm}_{n+2}$, with all possible sign combinations present. These products transform separately under U(1) with charge $m\in\{-3,-1,1,3\}$ equal to the sum of the signs. The dominant terms would be $\rme^{\pm \rmi \theta(n)}$, with subdominant contributions from $\rme^{\pm 3 \rmi \theta(n)}$ and products of $\rme^{\pm \rmi \theta(n)}$ and  $\rme^{\pm 3 \rmi \theta(n)}$ with derivatives of $\theta(n)$. It is possible that the coefficient of the dominant term and the first few subdominant terms vanish due to destructive interference---Theorem \ref{scalingtheorem} verifies that this occurs and as stated above $\mathcal{O}_{\pm3}(n)$ has the same scaling behaviour as $\rme^{\pm 3 \rmi \theta(n)}$. Details of a formal calculation in terms of the CFT operator product expansion are given\footnote{Parenthetically, this calculation indicates that other triples of neighbouring $X$ and $Y$ operators will scale as $\rme^{\rmi\theta}$; we have confirmed this in numerical simulations.} in Appendix \ref{appope}. Intuitively, $\mathcal{O}_{\pm3}$ is dominated by terms that create three charges, as well as the remnants of several terms that create one charge but (partially) destructively interfere with each other. \par By generalising this idea to all odd $\alpha$, we conjecture that $\mathcal{O}_\alpha (n)$ has an expansion of the form
\begin{align}\label{oddlabel}
\mathcal{O}_\alpha (n) \rightarrow \sum_{m=0}^{\lvert\alpha\rvert} \rme^{\rmi (\lvert\alpha\rvert - 2m)\theta(n)}\mathcal{D}_{(\alpha,m)}(\theta(n))+\dots \qquad\qquad(\alpha~ \mathrm{odd}).\end{align}
$\mathcal{D}_{\alpha,m}(\theta)$ is constant for $m = \pm\alpha$, and for other values of $m$ contains products of derivatives of $\theta$ such that the scaling dimensions of each term matches the extremal terms $\rme^{\pm\rmi \alpha \theta(n)}$. This is consistent with both Theorem \ref{fullcritresult} and calculations of the type in Appendix \ref{appope}. Note that this conjecture includes all operators in the field theory that depend only on $\theta$ and that have the correct scaling dimension; carefully taking operator products will give explicit expressions for the $\mathcal{D}$.
 \par
The case of $\alpha$ even follows the same pattern, except that we should always include a Jordan-Wigner string: the dominant term of which is given in \eqref{zsub}. Hence, we arrive at 
\begin{align}\label{evenlabel}
\mathcal{O}_\alpha (n) \rightarrow  \sum_{m=0}^{\lvert\alpha\rvert} \left(\rme^{\rmi (\lvert\alpha\rvert - 2m)\theta(n)+\rmi (k_F n + \varphi(n))}+\rme^{\rmi (\lvert\alpha\rvert - 2m)\theta(n)-\rmi (k_F n + \varphi(n))}\right)\mathcal{D}_{(\alpha,m)}(\theta(n))+\dots\qquad (\alpha ~\mathrm{even}),\end{align}
where, as above, $\mathcal{D}_{\alpha,m}(\theta)$ is constant for $m=\pm\alpha$ and contains appropriate numbers of derivatives such that all terms have the same scaling dimension. This is consistent with Theorem \ref{fullcritresult} and CFT calculations. In all cases we should allow multiplication by a global antiferromagnetic oscillation as discussed in the previous section.\par
The preceding analysis required the U(1) symmetry of isotropic models as a starting point. In reference \cite{VMP}, generalised Kramers-Wannier dualities were discussed that map between our models. One class of transformations swap models such that $f(z) \leftrightarrow z^n f(z)$ for some $m \in \mathbb{Z}$. If $f(z)$ is isotropic, then $z^nf(z)$ is not---this allows us to extend the preceding correspondence to anisotropic models. Anisotropic models that are dual to isotropic models in this way have an appropriately transformed U(1) symmetry. First, let us separate two cases: $m$ even and $m$ odd. When $m$ is even, the transformation will be local, and when $m$ is odd it will be non-local (this is related to whether the neighbouring gapped phases have a local order parameter). Now, in both cases we should take the correspondences \eqref{oddlabel} and \eqref{evenlabel} and shift the label on the left-hand-side $\mathcal{O}_\alpha \rightarrow \mathcal{O}_{\alpha +m}$ while \emph{not} shifting the right-hand-sides. In the case that $m$ is even, we take this as the correspondence. In the case that $m$ is odd, we alter the right-hand-side by swapping $\varphi$ and $\theta$.  
For example, in the transition $H=-\sum_n X_nZ_{n+1}X_{n+2} +\sum_n Z_n$ we identify $X_n$ with $\rme^{\rmi (k_F n + \theta(n))}+ \rme^{-\rmi (k_F n + \theta(n))}$. Notice that for odd $m$, the oscillatory factor $\rme^{\rmi k_F n}$ appears with the field $\theta(x)$. This is because $\partial_x \theta(x)$ is related to the density of the transformed U(1) symmetry. Moreover, this correspondence is consistent with the requirement that the non-local vertex operators with factors $\rme^{(2k+1)\varphi(x)}$ always correspond to non-local lattice operators.\par
The dualities discussed so far allow us to map an isotropic model to a representative of each phase $(c=1,\omega)$. As mentioned above, these representatives all have a U(1) symmetry and are thus not generic. It is then nontrivial to extend this analysis to general anisotropic models. Theorem \ref{fullcritresult} indicates, however, that this correspondence should continue to hold. Since the above argument does not make use of the fact that our underlying lattice model is non-interacting, we expect the correspondence to persist\footnote{Note that the scaling dimensions of the vertex operators will depend on the interactions, so the continuum operators in our expansions (including derivative terms) will not necessarily all have the same dimension. {The dominant operators should be identified as a subset of these. Note also that for sufficiently strong interactions, subdominant contributions that we ignore above can become important.}} in interacting models if the U(1) symmetry is preserved. However, if the U(1) symmetry is broken, we see no reason to expect it to continue to hold (in contrast to the non-interacting case).
 \par
\subsubsection{Example: transition between topologically distinct critical phases}\label{examplecrittran}
To complement the discussion so far, we consider the example \begin{align}f(z)& = (1-\lambda)z^2 + 2\lambda z - (1+\lambda)\nonumber
\\ &= (z-1)\bigl((1-\lambda)z + (1+\lambda)\bigr).
\end{align}
Tuning $-1\leq\lambda\leq1$ we interpolate between two critical phases, with a transition at $\lambda =0$. For $\lambda=1$ the system is the standard critical Ising chain with $H = -(\sum_j X_jX_{j+1}
+Z_j)$. Models with $\lambda>0$ are in the same phase: $(c=1/2,\omega=0)$.
For $\lambda =-1$ we have $H = -(\sum_jX_jZ_{j+1}X_{j+2} - X_jX_{j+1})$. Models with $\lambda<0$ can be smoothly tuned to this model and have $(c=1/2,\omega=1)$. For $\lambda = 0$, we have the $c=1$ transition between topologically distinct critical phases, with $H = -\frac{1}{2}(\sum_jX_jZ_{j+1}X_{j+2}+Z_j)$.
Table \ref{exampledims} gives the behaviour of the scaling dimensions of the most dominant $\mathcal{O}_\alpha$ as the system crosses the transition. We emphasise that while both sides of the transition are described by an Ising CFT, the scaling dimensions of the lattice operators change discontinuously. Note that the $c=1$ model has two real zeros so the CFT discussion above does not quite apply---a similar bosonisation scheme does work, as applied to a doubled Ising model in reference \cite{diF}.
\begin{table}
\begin{center}
\begin{tabular}{|c||c|c||c|c|c|c|c|}\hline   & $c$ & $\omega$ & $Y_n$&$\prod_{j=1}^n Z_j$ & $X_n$ & $\prod_{j=1}^{n} Z_j Y_{n+1}X_{n+2}$ & $X_nY_{n+1}X_{n+2}$ \\\hline\hline $\lambda>0$ & 1/2 & 0&9/8 & $\mathbf{1/8}$ --- $\mu$ & $\mathbf{1/8}$ --- $\sigma$ & 9/8 & 25/8 \\\hline $\lambda=0$ & 1 & 0 &5/4&$\mathbf{1/4}$ ---  $\cos (\varphi)$ & $\mathbf{1/4}$ --- $\rme^{\pm\rmi \theta}$ & $\mathbf{1/4}$ --- $\sin(\varphi)$ & 5/4 \\\hline $\lambda<0$ & 1/2 & 1&25/8 & 9/8 & $\mathbf{1/8}$ --- $\sigma$& $\mathbf{1/8}$ --- $\mu$ & 9/8 \\\hline \end{tabular}
\vspace{0.1cm}
\caption{Behaviour of scaling dimensions across a transition between critical phases---the model is $f(z)= (1-\lambda)z^2 + 2\lambda z - (1+\lambda)$. The dominant CFT fields associated to the dominant lattice operators are also given.}
\label{exampledims}
\end{center}
\end{table}
\subsubsection{CFT operator correspondence with $c$ complex fermions}
For higher values of $c$, we work formally with the linearised theory that consists of $2c$ real fermions. Note that it has been shown that spin models with $f(z) = \pm z^\omega(z^{2c}\pm1)$ are described at low energy by $so(2c)_1$ WZW models  \cite{son1,Ohta}, although a lattice-continuum operator correspondence has not been made. We can smoothly connect\footnote{That is, along a path where the CFT data varies smoothly.} any critical model in $H_{\mathrm{BDI}}$  to that subset of models \cite{VJP}.  \par Let us suppose for now that we have no real zeros and $c$ complex conjugate pairs of zeros (we order the zeros so that $k_i = -k_{2c-i}$).  Then we have $c$ canonical complex fermions which can each be bosonised as described above to give a set of fields $\theta_j(x)$ and $\varphi_j(x)$. The relevant vertex operators are of the form 
\begin{align}
\tau_{\bm{\mu},\bm{\nu}} (x)= \prod_{j=1}^c \rme^{\rmi \left((\mu_j + \nu_j) \theta_{j}(x) +(\mu_j - \nu_j) \varphi_{j}(x) \right)}\label{vertex}
\end{align}
where $\mu_i$ and $\nu_i$ are half-integer and have scaling dimension $\Delta_{\bm{\mu},\bm{\nu}} =\sum_j(\mu_j^2 +\nu_j^2)/2$ (we suppress Klein factors \cite{DS2}). The half-integer condition makes them twist operators for the linearised fermion field \cite{ginsparg}. They act nontrivially on all fermionic sectors, and by considering decoupled lattice models with Hamiltonian $H_{-m}+H_{m}$ (where the bosonisation in each sector is clear), their locality properties correctly reflect those of the operators $\mathcal{O}_\alpha$. Moreover, they have a minimal scaling dimension of $c/4$ (notice that this coincides with the smallest scaling dimension of the $\mathcal{O}_\alpha$, given in Theorem \ref{scalingtheorem}). 
As in the $c=1$ case, when we have a U(1) symmetry, then $\partial_x\varphi_j(x)$ or $\partial_x \theta_j(x)$ will correspond to fluctuations in the charge density. Hence, the appropriate field will be accompanied by $k_j x$---this is $\varphi(x)$ ($\theta(x)$) when the U(1) charge is local (nonlocal) on the lattice. Again, as above, these oscillatory factors persist away from these symmetric models.\par Suppose now that we are in an isotropic model, with U(1) symmetry generated by $S^z_\mathrm{tot}$.
The conjectured expansion of lattice operators $\mathcal{O}_\alpha$ goes through roughly as above. However, the identification of, say, $\sigma^+$ with a charge one operator is no longer so restrictive; this is because charges can have different signs in the different sectors and cancel. By considering the scaling dimensions, one concludes that the leading order terms have charge distributed evenly throughout the different sectors. For example, observe that the charge-two operator $\rme^{\rmi (\theta_1(x) + \theta_2(x))}$ with $\Delta = 1/2$ dominates the charge-two operator $\rme^{\rmi (3\theta_1(x) -\theta_2(x) )}$ with $\Delta =5/2$.  \par
More generally, as argued in \cite{HJ}, in isotropic models $\sigma^+=\left(\mathcal{O}_1 + \rmi\mathcal{O}_{-1}\right)/2$ will be dominated by operators $\tau_{\bm{\mu},\bm{\nu}}$ with $\sum_j (\mu_j+\nu_j) = 1$ (charge condition) and $\lvert\mu_i -\nu_j \rvert \leq 1$ (dominance condition). 
These conditions give a sum of terms that are products of $\rme^{\pm \rmi\theta_j(x)}$ or $\rme^{\pm \rmi(\varphi_j(x)+k_j x)}$ in each sector, and hence that can be distinguished by the presence or absence of oscillatory factors $\rme^{\pm \rmi k_j x}$. This is analogous to the sum over Fisher-Hartwig representations (see Section \ref{sec::Toeplitz}) that we derive in Section \ref{sec::scaling}, and the relevant oscillatory factors are confirmed in Theorem \ref{fullcritresult} (indeed, each such term is represented in the final result). Further, in \cite{HJ} it was argued that $\mathcal{O}_0$ will have $\sum_j( \mu_j+\nu_j) = 0$ (charge condition) and $\lvert\mu_i -\nu_j \rvert \leq 1$.
To extend this to operators with $\lvert \alpha \rvert > 1$, consider first the operators $\tau_{\bm{\mu},\bm{\nu}}$ with $\sum_j(\mu_j +\nu_j) =\lvert \alpha\rvert$ (a maximal charge condition) and $\lvert\mu_i -\nu_j \rvert \leq 1$ (dominance condition). This gives a set of operators that we expect to dominate the continuum limit of $\mathcal{O}_\alpha$. However, as in the $c=1$ case, we expect that we should include terms where maximally charged operators $\rme^{\rmi R \theta(x)}$ are substituted with  a series of terms $\rme^{\rmi (R-2m) \theta(x)}\lvert_{m=1,\dots ,R}$ multiplied by derivatives, such that each term has the same scaling dimension. The relevant scaling dimensions and oscillatory factors are confirmed in Theorem \ref{fullcritresult}.
\par 
To extend this to non-isotropic models, with $\omega = -c+m$, we note that there will again be two cases---$m$ even and $m$ odd. For $m$ even we simply shift the correspondences argued for the isotropic case. For $m$ odd we shift and further swap $\varphi_j(x)$ and $\theta_j(x)$. These correspondences may also be derived from $\sum_j(\mu_j +\nu_j) = \lvert c+\omega-\alpha \rvert$ (a maximal charge condition, when this is meaningful) and $\lvert\mu_i -\nu_j \rvert \leq 1$ (dominance condition)---although when $m$ is odd we should swap $\nu_j \rightarrow - \nu_j$ in these equations. Then, as mentioned above, we should include descendant operators with lower charge. These correspondences again agree with Theorem \ref{fullcritresult}.\par
 \textbf{Example:} Consider a model with $c=\omega =2$ (for example, $H= H_6+H_{2}$). By solving $\sum_{j=1}^2(\mu_j+\nu_j) = 3$ and then including descendants, we conjecture the correspondence:
 \begin{align}
 \mathcal{O}_{1}(x) \rightarrow \biggl(\rme^{\rmi(2 \theta_1(x)+\varphi_1(x) + k_1 x)}+\rme^{\rmi(2 \theta_1(x)-\varphi_1(x) -k_1 x)}+ \mathcal{D}(\theta_1)\left(\rme^{\rmi(\varphi_1(x) + k_1 x)}+\rme^{-\rmi(\varphi_1(x) +k_1 x)}\right)\nonumber\\ +\rme^{-\rmi(2 \theta_1(x)+\varphi_1(x) + k_1 x)}+\rme^{-\rmi(2 \theta_1(x)-\varphi_1(x) - k_1 x)}\biggr)\rme^{\rmi \theta_2(x)}+ \left(1\leftrightarrow 2\right) + \dots.
 \end{align} Note that all terms have $\Delta = 3/2$, and all oscillate as either $\rme^{\pm\rmi k_1 x}$ or $\rme^{\pm\rmi k_2 x}$ as expected. Furthermore, our conjecture gives the same expansion for $\mathcal{O}_{7}$, although the (suppressed) coefficients are not expected to be the same in general. Indeed, Theorem \ref{fullcritresult} indicates that the coefficients are different in the general case, since \eqref{multiplier} is not symmetric under taking sign-reversed Fisher-Hartwig representations.\par
In the case that $c\geq1$ and any zero is real, we do not conjecture the operator correspondence. We expect that similar arguments could work after bosonising a doubled model---this is performed for the $c=1/2$ case in \cite{Boyanovsky}. Note that Theorem \ref{fullcritresult} does not distinguish the case of two real zeros from two complex conjugate zeros at the level of scaling dimension.  \par

\subsection{Entanglement scaling}\label{sec::entanglementscaling}
The entanglement entropy of a subsystem is another physically important quantity. Let $\rho_A$ be the ground state reduced density matrix on sites $1$ up to $N$ and consider asymptotics in large $N$ after taking the length of the (periodic) chain to infinity. The most general results for isotropic critical chains in our class are given in \cite{KM,KM2}. Having identified the correlation length in gapped chains, derived from the nearest zero to the unit circle, it is interesting to consider the following theorem adapted from \cite{IMM}.
\begin{theorem}[Its, Mezzadri, Mo 2008]\label{IMM}
Consider a sequence of gapped chains (as defined in equation \eqref{majoranaham}) such that $2c$ of the zeros approach the unit circle, and that the limiting chain has no degenerate zeros on the circle. We label these approaching zeros by $\zeta_j$, noting that $\zeta_j$ can be either inside or outside the circle, and is either real or a member of a complex conjugate pair. Then the entanglement entropy of a subsystem of size $N$, in the limit $N \rightarrow \infty$, has the following expansion as $|\zeta_j| \rightarrow 1$: 
\begin{align}\label{IMMformula}
S[\rho_A]= -\frac{1}{6} \sum_{j=1}^{2c} \log \lvert \zeta_j - 1/\overline{\zeta}_j \rvert + O(1) .
\end{align}
\end{theorem}
Note that the $O(1)$ term is constant with respect to all the zeros that approach the unit circle (which are allowed to approach independently). Now, let us consider a sequence of models with a set of $2c$ zeros that approach the unit circle; for notational convenience let us fix them to be complex zeros outside the unit circle, other cases lead to the same result. Let this set of approaching zeros be specified by: $\{\rme^{\pm \rmi \phi_1} \rme^{1/\xi},\rme^{\pm \rmi \phi_2} \rme^{t_2/\xi^{r_2}},\dots \rme^{\pm \rmi \phi_c} \rme^{t_c/\xi^{r_c}}   \}$, where $\phi_i \neq \phi_j$ for $i \neq j$, $t_j>1$ and $r_j <1$, and we approach the circle letting $\xi \rightarrow \infty$. The conditions on $t_j$ and $r_j$ ensure that a closest zero is  $\rme^{ \rmi \phi_1} \rme^{1/\xi}$ for $\xi$ large enough. Inserting into Theorem \ref{IMM}, we get that:\par
\begin{align}
S[\rho_A] = \sum_{j=1}^c \frac{r_j}{3} \log(\xi) +O(1).
\end{align} Having different rates of approach to the circle means that we are necessarily approaching a multicritical point and we see crossover behaviour in the entanglement scaling. A simple example that allows this behaviour is the approach to the $c=1$ critical point with $H= \sum_i X_iX_{i+1} - Y_iY_{i+1}$ which is infinitesimally close to a $c=1/2$ line of transitions in the phase diagram of the XY model. \par
This is reminiscent of the Calabrese-Cardy formula \cite{CC}
that applies far more generally and gives asymptotics as the lattice spacing\footnote{Elsewhere in our work, units are fixed such that $a=1$.} $a \rightarrow 0$ 
\begin{align}
S[\rho_A] = \frac{c}{3} \log(\xi'/a) +O(1),  \label{CC}
\end{align} where $c$ is the central charge of the underlying CFT and $\xi'$ is the (fixed) correlation length of the system under consideration. This may also be interpreted as a scaling limit $\xi' \gg a$ \cite{calabrese}, and the formula was confirmed in this sense for the XY model in \cite{peschelEE}. Further relevant references are found in the review articles \cite{castro2009,calabrese}. We see that equation \eqref{IMMformula} is equivalent to formula \eqref{CC} in the vicinity of a regular critical point. At multicritical points the path approaching the transition is important, and the Calabrese-Cardy formula is expected to hold along renormalization group flow lines in parameter space.  
\par 

 \section{String correlators as determinants}\label{sec::detcorr}
We now begin the analysis necessary to prove the results given in Section \ref{results}.
 \subsection{Fermionic two point correlators}
 After defining $f(z)$ as in equation \eqref{fdef}, we have that:
\begin{align}
H = \sum_k \lvert f(k)\rvert \eta^\dagger_k \eta_k + \mathrm{const}
\end{align}
where the Boguliobov quasiparticles are found by rotating the Bloch sphere vector\footnote{The $c_k$ are the Fourier transform of the lattice fermions from which we built the $\gamma_n$ in equation \eqref{gammadef}.} $(c_{-k},c^\dagger_k)$ through an angle $f(k)/|f(k)|$ about the $x$-axis, giving
\begin{align}
\eta_k = \frac{1}{2}\left(1 +\frac{ {f(k)}}{|f(k)|} \right)c^\dagger_{k}+ \frac{1}{2}\left(1 -\frac{ {f(k)}}{|f(k)|} \right)c_{-k}.
\end{align}
 The sum over $k$ goes over momenta $k_n = 2\pi n/L$, although we always work in the limit where this sum becomes an integral from $0$ to $2\pi$. Details of this diagonalisation may be found in, for example, \cite{Suzuki71,KM,VJP}. The ground state, $\ket{\textrm{gs}}$, is the vacuum for the quasiparticles $\eta_k$, and from this we can easily calculate fermionic correlation functions---we refer the reader to \cite{KM} for details. We will use
\begin{align}
\langle -\rmi\tilde\gamma_n \gamma_m \rangle&:= \bra{\textrm{gs}} \left(-\rmi\tilde\gamma_n \gamma_m\right) \ket{\textrm{gs}}  = \frac{1}{2\pi} \int_0^{2\pi} \frac{ {f(k)}}{|f(k)|} \rme^{-\rmi(m-n) k}\rmd k \label{basiccorrelation}\\
\langle \tilde\gamma_n\tilde\gamma_m \rangle& = \langle \gamma_n\gamma_m \rangle = \delta_{nm}\nonumber
\end{align}
as elementary correlation functions in the rest of the paper, noting that it is $\arg(f(z))$, on the unit circle, that controls these correlations.\par
As an aside, note that for gapped chains the analysis of Section \ref{sec::corrl} allows us to find the large $N$ asymptotics of $\langle -\rmi\tilde\gamma_n \gamma_{n+N} \rangle$. As explained in the discussion around Section \ref{prooforder}, in generic cases this correlator will be $\Theta(N^{-K} \rme^{-N/\xi})$ where $K\in\{1/2,3/2\}$ is easily determined. For critical chains $f(k)$ has jump discontinuities. Decomposing as in equation \eqref{critde} and integrating by parts, we have that the fermionic two point function is $\Theta(1/N)$---this behaviour is as expected from the CFT description.

 \subsection{Wick's theorem}
Because the Hamiltonian \eqref{majoranaham} is quadratic, ground state expectation values have a Pfaffian structure. More precisely, suppose that we have $2N$ distinct and mutually anticommuting operators, $\mathcal{A}_n$, then:
\begin{align}
\langle &\mathcal{A}_1\cdots \mathcal{A}_{2N}\rangle = \sum_{\mathrm{all ~pairings}} (-1)^\sigma \prod_{\mathrm{all~ pairs~}(m,n)} \langle\mathcal{A}_m\mathcal{A}_n \rangle\\
&\bigg(=\langle\mathcal{A}_1\mathcal{A}_2 \rangle\langle\mathcal{A}_3\mathcal{A}_4 \rangle\cdots\langle\mathcal{A}_{2N-1}\mathcal{A}_{2N} \rangle -\langle\mathcal{A}_1\mathcal{A}_3 \rangle\langle\mathcal{A}_2\mathcal{A}_4 \rangle\cdots\langle\mathcal{A}_{2N-1}\mathcal{A}_{2N} \rangle +\dots \bigg).\nonumber
\end{align}
$(-1)^\sigma$ is the sign of the permutation that reorders the operators into each particular pairing. This expression is proportional to the Pfaffian of the antisymmetric matrix $\langle\mathcal{A}_m\mathcal{A}_n \rangle$, and is a form of Wick's theorem that is given in reference \cite{LSM}. 

\subsection{String correlation functions}
\label{sec::corr}
Consider the two point correlation function of $\mathcal{O}_\alpha$ for $\alpha>0$:
\begin{align}
\langle\mathcal{O}_\alpha(1) \mathcal{O}_\alpha(N+1) \rangle&= (-1)^x \langle  \gamma_1\dots \gamma_\alpha \left(\prod_{n=1}^{N} \rmi\tilde\gamma_n\gamma_n\right)\gamma_{N+1}\gamma_{N+2}\dots\gamma_{N+\alpha}\rangle\\ 
&= (-1)^x\langle (-\rmi\tilde\gamma_1)(-\rmi\tilde\gamma_2)\dots (-\rmi\tilde\gamma_\alpha) \left(\prod_{n=\alpha+1}^{N} \gamma_n(-\rmi\tilde\gamma_n)\right)\gamma_{N+1}\gamma_{N+2}\dots\gamma_{N+\alpha}\rangle \end{align}
We now transpose further terms to put unlike Majoranas as nearest neighbours
and apply Wick's theorem:
\begin{align}
\langle\mathcal{O}_\alpha(1) \mathcal{O}_\alpha(N+1)\rangle  &= (-1)^{N(\alpha-1)} \langle \prod_{n=1}^N(-\rmi\tilde\gamma_n \gamma_{n+\alpha})\rangle=(-1)^{N(\alpha-1)}\det (\langle -\rmi\tilde\gamma_n \gamma_{m+\alpha}\rangle)_{m,n=1}^{N}\\
&=(-1)^{N(\alpha-1)}\det \left( \frac{1}{2\pi} \int_0^{2\pi} \frac{ {f(k)}}{|f(k)|}\rme^{-\rmi\alpha k} \rme^{-\rmi(m-n) k}\rmd k\right)_{m,n=1}^{N}. \label{detcorr}\end{align}
For $\alpha<0$ an analogous calculation again leads to equation \eqref{detcorr}.
Table \ref{tab::spinops} gives the spin operators Jordan-Wigner dual to the fermionic operators $\mathcal{O}_\alpha(n)$ and Table \ref{tab::spincorrs} gives the equivalent spin correlators for all $\alpha$. A derivation is given in Appendix \ref{app::dual}. Notice that for odd $\alpha$ these operators and correlation functions are local in the spin variables, and for even $\alpha$ they are nonlocal; they are always nonlocal for the fermions. Understanding the asymptotic behaviour of the determinant $\eqref{detcorr}$ is the key to the results given in Section \ref{sec::summary}.
\section{Toeplitz determinants}\label{sec::Toeplitz}
Several theorems for the asymptotic behaviour of large Toeplitz determinants are required to prove our results, hence we use this section to review them in detail. 
This section is intended to not only state the results but to give an exposition of how to use them in practice. The reader already familiar with these ideas can hence skip this section and refer back to it where necessary. Note that we reformulate and simplify the statement of some theorems appropriately for our application, the most general statements are available in the given references. \par
 First, recall that an $N\times N$ Toeplitz matrix, $T$, takes the following `translation-invariant' form:
\begin{align}
(T)_{mn} = (t_{m-n}) = \left(\begin{array}{ccccc}t_0 & t_{-1} & t_{-2} & \hdots & t_{-(N-1)} \\t_{1} & t_0 & t_{-1} & \hdots &t_{-(N-2)} \\t_2 & t_{1} & t_0 & \hdots&t_{-(N-3)} \\\vdots & \vdots & \vdots  & \ddots&\vdots\\ t_{N-1} &t_{N-2} & t_{N-3} & \hdots & t_0 \end{array}\right).
\end{align}
This matrix can be thought of as the $N \times N$ truncation of an infinite matrix, with element $t_{-n}$ on the $n$th descending diagonal.
Consider a region of the complex plane, $U$, such that $S^1\subseteq U \subseteq \CN$. A function $t: U \rightarrow\CN$, integrable on the unit circle, generates a Toeplitz matrix through its Fourier coefficients:
\begin{align} \label{fourierc}
t_{n} = \frac{1}{2\pi} \int^{2 \pi}_{0} t\left(\rme^{\rmi k}\right) \rme^{- \rmi n k} \rmd k.
\end{align}
We refer to such $t(z)$ as the symbol of the Toeplitz matrix\footnote{We will always go in this direction: from symbol to matrix. The reverse is possible providing the $t_i$ decay fast enough.}
and denote the Toeplitz determinant of order $N$ that is generated by $t$ as
\begin{align}\label{Tdet}
D_{N}[t(z)] = \det (t_{m-n})_{m,n=1}^{N}, 
\end{align}
i.e. it is defined simply as the determinant of the $N\times N$ truncated matrix generated by $t$. It is the analytic properties of $t$ that govern the form of the asymptotics of this determinant as $N \rightarrow \infty$. By inspecting equation \eqref{detcorr}, we see that $\langle\mathcal{O}_\alpha(1)\mathcal{O}_\alpha(N+1)\rangle$ is, up to an oscillating sign, a Toeplitz determinant of order $N$ generated by \begin{align}
t(z)= z^{-\alpha} \frac{f(z)}{|f(z)|} .
\end{align}  \par
 To go further, we consider the symbol on the unit circle $z = \rme^{\rmi k}$ and attempt to factorise it as:
\begin{align}
t(z) = \rme^{V(z)}t_{\mathrm{singular}}(z).
\end{align}
Here $\rme^{V(z)} $ is called the smooth part of the symbol, which we take to mean that $V(z)$ is analytic on the unit circle\footnote{This smoothness requirement has been weakened by many authors; a strong result is given in \cite{DIK}, to which we refer the interested reader. For our purposes the strong condition of analyticity is acceptable.}. This implies that the winding number of $\exp\left(V\left(\rme^{\rmi k}\right)\right)$ is equal to zero.
 The Fourier coefficients of  $V\left(\rme^{\rmi k}\right)$ are\begin{align}
V_{n} = \frac{1}{2\pi} \int^{2 \pi}_{0} V\left(\rme^{\rmi k}\right) \rme^{- \rmi n k } \rmd k
\end{align}
and we define the Wiener-Hopf factorisation
 of $\rme^{V(z)}$ as:
\begin{align}
\rme^{V(z)} =b_+(z){\rme^{V_0}}b_-(z), \qquad b_+(z)=\rme^{\sum_{n=1}^\infty V_n z^n},\qquad b_-(z)=\rme^{\sum_{n=1}^\infty V_{-n} z^{-n}}.\label{WH}
\end{align}\par
In our work, we will have three families of symbol to consider. The first case is $t_{\mathrm{singular}}(z) = 1$, which works when our symbol $t(z)$ is smooth enough that its logarithm gives us an appropriate $V(z)$. The second case is $t_{\mathrm{singular}}(z) = z^\omega$, this is needed to represent symbols $t(z)$ that have an integral winding number $\omega$. Finally, the third case represents symbols $t(z)$ with sign-change jump discontinuities. Let $\zeta = \rme^{\rmi\theta}$ and consider the function on the unit circle:
\begin{align}\label{jumpdef}
g_{\zeta,\beta} (z) = 
\begin{cases}
\rme^{\rmi \pi \beta}, & 0 \leq \mathrm{arg} z < \theta, \qquad \\
\rme^{-\rmi \pi \beta}, & \theta\leq \mathrm{arg} z < 2 \pi. 
\end{cases}
\end{align}
For $\beta$ half-integer this is piecewise proportional to $\rmi$, with a sign change at $z=\zeta$ and at $z=1$. To represent a sign-change only at $z=\zeta$, we put $t_{\mathrm{singular}}^{\zeta,\beta}(z) = z^\beta g_{\zeta,\beta} (z) $, removing the jump at $z=1$. Conversely a jump only at $z=1$ would be represented simply by $z^\beta$.
Notice that \emph{any} half-integer $\beta$ represents the sign-change through $g_{\zeta,\beta}$, but the power of $\beta$ that appears distinguishes the $t_{\mathrm{singular}}^{\zeta,\beta}(z)$. The singular part of a function with several sign-change jump discontinuities can be decomposed as a product \begin{align}t_{\mathrm{singular}}(z)=\prod_jt_{\mathrm{singular}}^{\zeta_j,\beta_j}(z) = \prod_jz^{\beta_j} g_{\zeta_j,\beta_j} (z) = z^{\sum_j \beta_j} \prod_j  g_{\zeta_j,\beta_j} (z) ,\end{align} where all $\beta_j$ are half-integer, but note that now only the total $\sum_j \beta_j$ is fixed by the symbol we wish to represent---this redundancy has important consequences. As an example, consider the symbol $s(k) = \mathrm{sign}(\cos(k))$, this has jump discontinuities at $z=\pm \rmi$. Hence we should represent it by two $\beta$ half-integer singularities, and the fact that there is no overall winding implies that $\beta_1 = -\beta_{2}$. This gives a family of representations \begin{align}
s(z) = \mathrm{const}\times g_{\rmi,\frac{2n+1}{2}}(z)g_{-\rmi,-\frac{2n+1}{2}}(z)\end{align} where $n\in \mathbb{Z}$ and the constant fixes the correct overall sign at $z=1$. \par
With these ideas in place, notice that all three families of $t_{\mathrm{singular}}(z)$ can be represented in the same way. If we use  $t_{\mathrm{singular}}(z) = z^\beta g_{\zeta,\beta} (z) $ as a building block, then $t_{\mathrm{singular}}(z) =1$ is the case $\zeta=1, \beta =0$ and $t_{\mathrm{singular}}(z) =z^\omega$  is the case $\zeta=1, \beta =\omega$. Motivated by this discussion, we write down the canonical form of reference \cite{DIK} for a symbol that is non-vanishing on the unit circle and has sign-change jump discontinuities:
\begin{align}\label{canonform}
t_{\mathrm{canon}}(z) = \rme^{V(z)} z^{\sum_{j=1}^m \beta_j}\prod_{j=1}^m g_{z_j,\beta_j}(z) z_j^{-\beta_j}, \qquad z=\rme^{\rmi k}, \qquad k \in [0, 2\pi);
\end{align}
where  for $j=1, \ldots, m$ and $0 \leq k_{1} < \ldots < k_{m} < 2 \pi$, we have $z_{j} = \rme^{\rmi k_{j}}$, $\beta_{j} \in \frac{1}{2}\mathbb{Z}$ and the function $V(z)$ must be smooth as above. The factor $\prod_j z_j^{-\beta_j}$ is just a multiplicative constant and is there to align notation with \cite{DIK}. Any $\beta_j$ in this expression must be nontrivial, hence the symbol has $m$ jump discontinuities. Note that we allow $m=0$ when the symbol is simply $\exp(V(z))$, and the edge case $z_1=1$ has $g_{1,\beta_1} = \exp(-\rmi\pi\beta_1)$. Our notation deviates slightly from reference \cite{DIK}, where a $\beta_0$ is associated to $z=1$ even if there is no singularity there---this does not affect the adapted theorems we quote below.\par

Now, as explained above, for a symbol $t(z)$ with multiple jump discontinuities, there is an infinite class of different $t_{\mathrm{canon}}(z)$ to which it is equal. In fact, if we find a single representation with a set of $\{\beta_j\}$, we can find another representation by shifting each $\beta_j \rightarrow \beta_j +n_j$ such that $\sum n_j = 0$; however, we may have to amend our choice of $V(z)$ to include an additional multiplicative constant. We are interested in representations where $\sum_j  \beta_j^2$ is minimal---these will contribute to the leading-order asymptotics and so we refer to them as dominant. \par Following \cite{DIK}, in order to write down the dominant asymptotics, it is helpful to introduce the notion of \emph{FH-representations}. Given a symbol $t(z)$ written in canonical form \eqref{canonform}, replace all $\beta_j$ by $\tilde\beta_j=\beta_j+n_j$ such that $\sum_j n_j = 0$. This new function is the FH-representation $t(z; n_1, \dots,n_m)$, defined relative to the representation $t(z; 0, \dots,0)=t(z)$. We then have the equality:
\begin{align}
t(z;n_0,\dots,n_m) = \prod_{j=1}^m z_j^{n_j} t(z),
\end{align}
this means that, in general, the FH-representation differs from a canonical form for the symbol by a multiplicative constant. We illustrate this by example in Appendix \ref{app::fh}. An algorithm is given in \cite{DIK} to find the finite number of dominant FH-representations, where it is shown that \emph{all of these} contribute to the leading asymptotics of the determinant \eqref{Tdet}. For our purposes, finding a dominant representation will be simple; and given one dominant FH-representation of $f(z)$ for which we define $n_j=0$, all other dominant FH-representations have $n_j \in \{1,-1,0\}$.
\par
We now recall theorems relevant to the three cases introduced above. {Szeg\H{o}'s strong limit theorem} \cite{Szego} gives the dominant asymptotics for matrices generated by smooth symbols with no winding, i.e. the case $m=0$. We use a form adapted from reference \cite{DIK}:
\begin{theorem}[Szeg\H{o} 1952]  \label{szego}
Let $t(z) = \exp(V(z))$ be a symbol, with $V(z)$ smooth as explained above and such that $\sum_{n= -\infty}^\infty |n| |V_n|^2<\infty$. As the matrix dimension, $N$, goes to infinity:
\begin{align} \label{szego2}
D_{N}[t(z)] = \exp \left(N V_0 + \sum_{n=1}^{\infty} nV_nV_{-n}\right)(1+o(1)).
\end{align}
 \end{theorem}

 If we have a symbol with an integral winding number, i.e. $m=1$, $k_1=0$, $\beta_1 \in \mathbb{Z}$, the next theorem, adapted from a result of Fisher and Hartwig \cite{Hartwig1969}, allows us to reduce it to the product of a determinant that can be evaluated by Szeg\H{o}'s theorem and another small determinant.
 \begin{theorem}[Fisher, Hartwig 1969]\label{FHlength} 
 Let $t(z)=\rme^{V(z)}z^{-\nu}$, where $V(z)$ satisfies the conditions for Theorem \ref{szego}. Given $b_\pm(z)$ as defined in \eqref{WH}, define the auxilliary functions: \begin{align}
 l(z) = \frac{b_-(z)}{b_+(z)} \qquad m(z) = \frac{b_+(1/z)}{b_-(1/z)},\end{align}
 with associated Fourier coefficients\footnote{These exist by the Wiener-L\'evy theorem \cite{Hartwig1969}.} $l_k,m_k$. \\For $\nu >0$ we have:
\begin{align}
D_N[z^{-\nu}\rme^{V(z)}] = (-1)^{N\nu} D_{N+|\nu|}[\rme^{V(z)}]\times \det\left( \begin{array}{cccc}d_{N} & d_{N-1} & \dots & d_{N-\nu+1} \\d_{N+1} & d_{N} &  & d_{N-\nu+2} \\\vdots &  &  & \vdots \\d_{N+\nu-1} & d_{N+\nu-2} &  \dots & d_{N}\end{array}\right)  \label{FH2}
\end{align}
where $d_k = l_{k} + \delta_k^+$. For $\nu <0$ we instead have $d_k = m_k + \delta_k^-$; \eqref{FH2} is otherwise unchanged. \par
General estimates for the error terms $\delta_k^\pm$ are given in \cite{Hartwig1969}---the only case we need is as follows. Suppose that the large Fourier coefficients of $h(z)=\rme^{V(z)}$ behave as $\lvert h_n \rvert = O(\rho^n)$ and $\lvert h_{-n} \rvert = O(\sigma^n)$ then for large $k$, $\delta_k^+ = O(\rho^{2k}\sigma^k)$ and $\delta_k^- = O(\rho^{k}\sigma^{2k})$.
\end{theorem}
Given the definitions in the above theorem, we can also state a formula from \cite{Hartwig1969} for the leading order correction to Theorem \ref{szego}. 
\begin{theorem}[Fisher, Hartwig 1969]\label{szegoerror} Let $t(z) = \rme^{V(z)}$ satisfy the conditions for Theorem \ref{szego}. Then we can write
\begin{align} 
\log D_{N}[t(z)] = N V_0 + \sum_{n=1}^{\infty} nV_nV_{-n} + E^{(1)}_N + E^{(2)}_N,
\end{align} 
where, for $l(z)$ and $m(z)$ defined above, we have $E^{(1)}_N = - \sum_{n=1}^\infty n l_{N+n}m_{N+n}$. The error term  $E^{(2)}_N$ is subdominant---see \cite{Hartwig1969} for general estimates. For the case relevant to us, with $\rho$ and $\sigma$ defined in Theorem \ref{FHlength}, we have $E^{(1)}_N = O(\rho^N \sigma^N)$ and $E^{(2)}_N = O(\rho^{2N} \sigma^{2N})$.
\end{theorem} 
The final theorem we need is the \emph{generalised Fisher-Hartwig conjecture}. The asymptotics for symbols with fractional jump discontinuities was initially conjectured by Fisher and Hartwig in \cite{FH}; this conjecture was then generalised to the class of symbols that we need  by Basor and Tracy \cite{basor1991}. This generalised case was proved by Deift, Its and Krasovsky in \cite{DIK}, and we give a simplified form of their result relevant to our work.
\par

\begin{theorem}[Deift, Its, Krasovsky 2011]\label{FHT}
Consider a Toeplitz matrix generated by $t(z)$ in the canonical form \eqref{canonform}. Suppose $\beta_j \not\in \mathbb{Z}$ for all $j$. Then, as the matrix dimension, $N$, goes to infinity:
\begin{align}
D_{N}[t(z)]=\sum_{\substack{\mathrm{Dominant~}\\
                              \mathrm{FH-reps:}~\{n_j\}}}\left(\prod_{j=1}^m z_j^{n_jN}\right)
\mathcal{R}(t(z;\{n_j\})(1+o(1)).\end{align}
Where:\begin{align}\label{FHformula}
\mathcal{R}(t(z;\{n_j\}))&=N^{-\sum_{j=1}^m \tilde\beta_j^2}\exp\left(NV_0 + \sum_{n=1}^\infty nV_nV_{-n}\right) \nonumber
\prod_{1\le i<j\le m}
|z_i-z_j|^{2\tilde\beta_i\tilde\beta_j}\nonumber\\
&\hspace{2.5cm}\times \prod_{j=1}^m b_+(z_j)^{\tilde\beta_j}b_-(z_j)^{-\tilde\beta_j}\prod_{j=1}^mG(1+\tilde\beta_j) G(1-\tilde\beta_j).\\
&\hspace{4cm}(\mathrm{Recalling~that}~~~\tilde\beta_j=\beta_j+n_j.) \nonumber
\end{align} \end{theorem}
The $V_n$ are unaltered when passing between FH-reps. Branches of $b_{\pm}(z_j)^{\tilde\beta_j}$ are determined by $b_\pm(z_j)^{\tilde\beta_j} = \rme^{\tilde\beta_j\sum_{n=1}^\infty V_{\pm n} z_j^{\pm n}}$. $G(z)$ is the Barnes $G$-function \cite[\textsection 5.17]{NIST:DLMF}; given as a Weierstrass product by \begin{align} 
\mathop{G\/}\nolimits\!\left(z+1\right)=(2\pi)^{z/2}\rme^{-\frac{1}{2}z(z+1)-\frac{1}{2}\gamma_E z^{2}}\*\prod_{j=1}^{\infty%
}\left(\left(1+\frac{z}{j}\right)^{j}\mathop{\exp\/}\nolimits\!\left(-z+\frac{%
z^{2}}{2j}\right)\right),\label{barnes}
\end{align}
where $\gamma_E$ is the Euler-Mascheroni constant. 
It is clear from equation \eqref{barnes} that $G$ vanishes whenever the argument is a negative integer. Hence if $\beta_j \in \mathbb{Z}$, the RHS of \eqref{FHformula} vanishes and this is not the first term of the asymptotic expansion (instead we should use Theorem \ref{FHlength}). 
\section{Gapped chains - analysis}\label{sec::gap}

\subsection{Closed form for the order parameter---Proof of Theorem \ref{orderparameter}}\label{sec::orderparameter} 
Since $c=0$, the complex function is given by \begin{align}
f(z)& = \rho\frac{1}{z^{N_p}} \prod_{i=1}^{N_z} (z-z_i) \prod_{j=1}^{N_Z} (z-Z_j)\\
&= \left( \rho \prod_{j=1}^{N_Z} (-Z_j) \right) z^\omega \prod_{i=1}^{N_z} (1-z_i/z) \prod_{j=1}^{N_Z} (1-z/Z_j)
=:  \rho' z^\omega f_0(z).
\end{align}
$\rho' = \rho\prod_{j=1}^{N_Z} (-Z_j)$, and it is only the sign of this real number that is important; moreover, since the $Z_j$ come in complex conjugate pairs, the sign only depends on $N_{Z}^{+}$, the number of zeros on the positive real axis and outside the unit circle. For bookkeeping purposes, define 
\begin{align}\label{bookkeep}
s =\mathrm{sign}(\rho)\times(-1)^{N_{Z}^+}.
\end{align}
If we consider $(-1)^{N(\omega-1)}\langle\mathcal{O}_\omega(1)\mathcal{O}_\omega(N)\rangle$, then this is generated by $t(z) =  \rme^{V(z)}$, where 
\begin{align}
V(z) - V_0& = \frac{1}{2} \left(\log f_0(z) - \log f_0(\overline{z})\right) 
\end{align}
for a continuous logarithm that could be found by integrating the logarithmic derivative of $f$. We instead jump to the following solution:
\begin{align}
V(z) - V_0&= \frac{1}{2} \sum_{i,j} \Log(1-z_i/z)-\Log(1-z_i/\overline{z})+ \Log(1-z/Z_{j})-\Log(1-\zbar/{Z_j})\\
&= -\frac{1}{2} \sum_{n=1}^\infty \frac{1}{n}\left(\sum_{i=1}^{N_z} \left(\frac{z_i}{z}\right)^n-\left(\frac{z_i}{\zbar}\right)^n+ \sum_{j=1}^{N_Z}\left(\frac{z}{Z_j}\right)^n-\left(\frac{\zbar}{Z_j}\right)^n\right),\label{Vexpansion}
\end{align}
where the function $\Log(z)$ is the principal branch of the complex logarithm---this is clearly smooth and recovers $f(z)$ when we take the exponential. Note that we used that the zeros are either real, or occur in complex conjugate pairs. On the unit circle $z=\rme^{\rmi k}$ we can put $\overline{z}=1/z$ into \eqref{Vexpansion}. 

This gives us an honest $V(z)$ from which one can read off the Fourier coefficients:
\begin{align}
V_0 &= \log s = 0, \rmi \pi\\
V_n &=\begin{cases}
\frac{1}{2n}\left( \sum_iz_i^{n} - \sum_jZ_j^{-n}\right) \quad &n>0\\
 -\frac{1}{2n}\left(\sum_i z_i^{n} - \sum_jZ_j^{-n}\right)\quad &n<0 . 
\end{cases}\label{fourierV}
\end{align}
Inserting into Theorem \ref{szego} we reach:
\begin{align}
\det[t(z)] &= s^N \exp\left( \sum_{n=1}^\infty - \frac{1}{4n}{\left( \sum_{i=1}^{N_z}z_i^{n} -\sum_{j=1}^{N_Z}Z_j^{-n}\right)}^{\!\!2}~ \right) .
\end{align}
On expanding the square and interchanging the finite sums with the sum over $n$ in the exponent, we can then perform the sum over $n$ leading to Theorem \ref{orderparameter}. The term under the fourth root is always a positive real, and the principal logarithm implies that we take the positive root. For completeness, note that the oscillatory factor multiplying the order parameter is given by $\rme^{\rmi \pi N(\omega-1) + N\log(s)}$.\par
Note that with the Fourier coefficients of $V$ in hand, we can find the Wiener-Hopf decomposition \eqref{WH} of our symbol when $z$ is on the unit circle.
\begin{align}\label{WHb}
b_+(z) &= \rme^{\sum_{n=1}^\infty \frac{1}{2n}\left( \sum_iz_i^{n} - \sum_jZ_j^{-n}\right) z^n} =\prod_{i=1}^{N_z}\rme^{-\frac{1}{2} \Log(1-zz_i)}\prod_{j=1}^{N_Z}\rme^{\frac{1}{2} \Log(1-z/Z_j)}\\
b_-(z) &= \rme^{-\sum_{n=1}^\infty\frac{1}{2n}\left(\sum_i z_i^{n} - \sum_jZ_j^{-n}\right)z^{-n}} =\prod_{i=1}^{N_z}\rme^{\frac{1}{2} \Log(1-z_i/z)}\prod_{j=1}^{N_Z}\rme^{-\frac{1}{2} \Log(1-1/(zZ_j))}.\nonumber
\end{align}
Note that $1/b_+(z)=b_{-}(1/z)$. Moreover, 
\begin{align}
b_+(z) = \sqrt{\frac{\prod_{j=1}^{N_Z}(1-z Z_j^{-1})}{\prod_{i=1}^{N_z}(1-zz_i)}}
\end{align}
where the square-root is continuous on the unit circle and the branch is fixed as the positive root of a positive real at $z=1$.
\subsection{Correlation lengths}\label{sec::corrl}
We now use Theorem \ref{FHlength} to find the behaviour of the correlation function $\langle\mathcal{O}_\alpha(1)\mathcal{O}_\alpha(N+1)\rangle$ in the gapped phase $\omega$. For definiteness, let us label the zeros by proximity to the unit circle: $|Z_i| \leq |Z_j|$ and $|z_i| \geq |z_j|$ for $i<j$.
\subsubsection{Asymptotics of $l(z),m(z)$}
The key ingredient that we need are the asymptotically large Fourier coefficients of the auxilliary functions  
\begin{align}
l(z) &= \frac{b_-(z)}{b_+(z)} =\sqrt{\frac{\prod_{i=1}^{N_z}(1-zz_i)(1-z_i/z)}{\prod_{j=1}^{N_Z}(1-zZ_j^{-1}))(1-Z_j^{-1}/z)}} \\ m(z) &= 1/l(z).
\end{align}
Note that so far $l$ and $m$ are defined only on the unit circle and with the principal branch of the square-root (in fact, due to the complex-conjugate pairs of roots, the arguments of the square-root are strictly positive). For the purposes of this calculation, we assume the generic problem where the branch points in $R=\{z_i,z_i^{-1},Z_j,Z_j^{-1}\}$ are all distinct, we will comment later on the effect of multiplicity.
We need the dominant asymptotic term of the $n$th Fourier coefficient of $l(k)$ for large $n$:
\begin{align}
l_n& = \frac{1}{2\pi} \int_0^{2\pi} l(k) \exp(-\rmi n k) \rmd k\nonumber\\
&=\frac{1}{2\pi \rmi} \int_{S^1}\left(\frac{\prod_{i=1}^{N_z}(1-sz_i)(1-z_i/s)}{\prod_{j=1}^{N_Z}(1-sZ_j^{-1})(1-Z_j^{-1}/s)}\right)^{\!\!1/2}~ s^{-(n+1)}\rmd s. \label{nfourier}
\end{align}
We analytically continue $l(k)$ off the unit circle into the complex $s$-plane. The idea is to move the contour of integration out to infinity, where the $s^{-n}$ term in the integrand will cause the integral to vanish there. The integrand has branch cuts on which the contour gets snagged, and the dominant contribution will come from the nearest branch points outside the unit circle---this is the Darboux principle \cite{Dingle}.\par

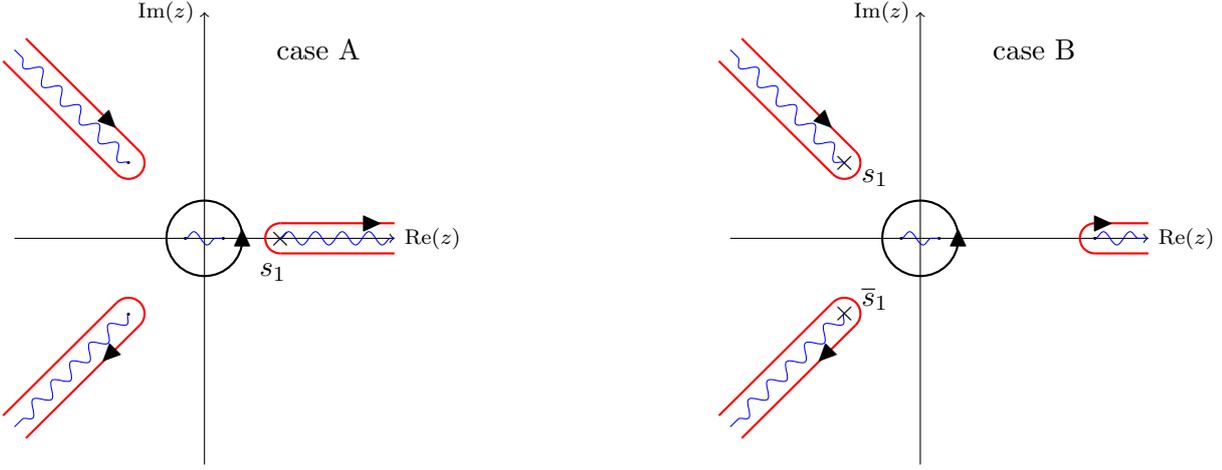
\begin{figure}
\centering
\begin{minipage}{.5\textwidth}
\begin{tikzpicture}
 \draw[font=\scriptsize] [->] (-2.5,0) -- (2.5,0) node [right]  {$\Re(z)$};
    \draw[font=\scriptsize] [->] (0,-3) -- (0,3) node [left] {$\Im(z)$};
\draw (1.5,2.5) node {$\mathrm{case ~A}$};

%\draw (0,0) node [rotate=45] {$+$};
\draw (1/4,0) node {$.$};
\draw (-1/4,-0) node {$.$};
\draw (1,-0) node {$\times$};
\draw (.9,-0.2) node [below] {$s_1$};

  \path [draw=blue,snake it]
    (1,0) -- (2.5,0);

\draw (-1,1) node {$.$};
\draw (-1,-1) node  {$.$};
\draw[thick]  [draw=red]
(1,.2) -- (2.5,.2);
\draw[thick]  [draw=red]
(1,-.2) -- (2.5,-.2);

\draw[thick]  [draw=red]
(-1.15,0.85) -- (-2.65,2.35);

\draw[thick]  [draw=red]
(-.85,1.15) -- (-2.35,2.65);

\draw[thick]  [draw=red]
(-1.15,-0.85) -- (-2.65,-2.35);

\draw[thick]  [draw=red]
(-.85,-1.15) -- (-2.35,-2.65);
\draw[thick] [draw=red] (-.85,1.15) arc (45:-135:0.212cm);
\draw[thick] [draw=red] (-.85,-1.15) arc (-45:135:0.212cm);\draw[thick]  [draw=red](1,0.2) arc (90:270:0.2cm);

  \path [draw=blue,snake it] %inside circle
    (-1/4,0) -- (1/4,0);
     \path [draw=blue,snake it]
    (-1,1) -- (-2.5,2.5);
        \path [draw=blue,snake it]
    (-1,-1) -- (-2.5,-2.5);
\draw[thick](0,0) circle (0.5);
%\draw[dashed](0,0) circle (1.15);
\draw (2.2,0.2) node  [scale=1, rotate=-90] {$\blacktriangle$};
\draw (-1.25,1.55) node  [scale=1, rotate=-135] {$\blacktriangle$};
\draw (-1.25,-1.55) node  [scale=1, rotate=135] {$\blacktriangle$};
\draw (.5,0) node  [scale=1] {$\blacktriangle$};
\end{tikzpicture}
\end{minipage}%END OF FIRST FIG
\begin{minipage}{.5\textwidth}
  \centering
\begin{tikzpicture}
 \draw[font=\scriptsize] [->] (-2.5,0) -- (3,0) node [right]  {$\Re(z)$};
    \draw[font=\scriptsize] [->] (0,-3) -- (0,3) node [left] {$\Im(z)$};
\draw (1.5,2.5) node {$\mathrm{case ~B}$};

%\draw (0,0) node [rotate=45] {$+$};
\draw (1/4,0) node {$.$};
\draw (-1/4,-0) node {$.$};
\draw (2.3,-0) node {$.$};
\draw (-.6,.8) node {$s_1$};
\draw (-.6,-.8) node  {$\overline{s}_1$};

  \path [draw=blue,snake it]
    (2.3,0) -- (3,0);

\draw (-1,1) node {$\times$};
\draw (-1,-1) node  {$\times$};
\draw[thick]  [draw=red]
(2.3,.2) -- (3,.2);
\draw[thick]  [draw=red]
(2.3,-.2) -- (3,-.2);

\draw[thick]  [draw=red]
(-1.15,0.85) -- (-2.65,2.35);

\draw[thick]  [draw=red]
(-.85,1.15) -- (-2.35,2.65);

\draw[thick]  [draw=red]
(-1.15,-0.85) -- (-2.65,-2.35);

\draw[thick]  [draw=red]
(-.85,-1.15) -- (-2.35,-2.65);
\draw[thick] [draw=red] (-.85,1.15) arc (45:-135:0.212cm);
\draw[thick] [draw=red] (-.85,-1.15) arc (-45:135:0.212cm);
\draw[thick]  [draw=red](2.3,0.2) arc (90:270:0.2cm);

  \path [draw=blue,snake it] %inside circle
    (-1/4,0) -- (1/4,0);
    
  \path [draw=blue,snake it]
    (-1,1) -- (-2.5,2.5);
        \path [draw=blue,snake it]
    (-1,-1) -- (-2.5,-2.5);
\draw[thick](0,0) circle (0.5);
%\draw[dashed](0,0) circle (1.5);
\draw (2.4,0.2) node  [scale=1, rotate=-90] {$\blacktriangle$};
\draw (-1.25,1.55) node  [scale=1, rotate=-135] {$\blacktriangle$};
\draw (-1.25,-1.55) node  [scale=1, rotate=135] {$\blacktriangle$};

\draw (.5,0) node  [scale=1] {$\blacktriangle$};
\end{tikzpicture} 
\end{minipage}
\caption{Schematic for the two generic cases of the computation \eqref{nfourier}. Blue (wavy) lines indicate branch cuts in the integrand. The black curve is the initial integration contour $S^1$, and the red (lighter) curve is the deformed contour. $\times$ indicates the closest branch cuts to the unit circle that are outside the circle.}
\label{contourfig}
\end{figure}

By inspection we have either a square-root or inverse square-root branch point at every element of $R$. If there are an odd number of such points inside (and therefore, by symmetry, outside) the unit circle, then zero and infinity are also branch points---hence there are always an even number of branch points both inside and outside the unit circle.  We choose any branch cut pattern inside the unit circle (where no cut crosses the circle). Outside the unit circle we order the branch points by radial distance from the origin. In generic circumstances there will be either one real branch point (case A), or a complex-conjugate pair of branch points (case B), closest to the origin. Choose the cuts to be leaving all branch points radially. An example for each of the two cases is depicted in Figure \ref{contourfig}---we call the nearest branch point(s) $s_1$ (and $\overline{s}_1$), for $\arg(s_1)\in[0,\pi]$. We connect up the radial cuts outside a circle of large radius, the precise choice is unimportant. \par

In case A we consider the Hankel contour connecting infinity to the nearest real zero and back ---this is exactly the relevant part of the snagged contour. After parameterising $s=s_1\rme^t$ for $t \in \mathbb{R}_+$ and where $\arg(t) =0$ below the axis and $\arg(t) = -2\pi$ above the axis, this integral obeys the conditions for \emph{Watson's lemma for loop integrals}---see, for example, \cite[\S15.6.1]{Temme} and \cite{Olver}. This gives us an asymptotic series of which we need only the first term. Recall that we have ordered our zeros so that $s_1$ is either $1/z_1$ or $Z_1$---then we have
\begin{proposition} Suppose there is a single real root closest to the unit circle. Then, \label{prop1}
\begin{align}
l_n = \begin{cases}
- z_1^n\frac{1}{n^{3/2}}\underbrace{\frac{1}{2\sqrt{\pi}}\left(\frac{(1-z_1^2)\prod_{i=2}^{N_z}(1-z_i/z_1)(1-z_1z_i)}{\prod_{j=1}^{N_Z}(1-Z_j^{-1}/z_1)(1-z_1Z_j^{-1})}\right)^{\!\!1/2}}_{\lambda}\left(1+O(1/n)\right) \qquad &s_1=1/z_1\\\\
Z_1^{-n}\frac{1}{n^{1/2}}\frac{1}{\sqrt{\pi}}\left(\frac{\prod_{i=1}^{N_z}(1-Z_1z_i)(1-z_i/Z_1)}{(1-Z_1^{-2})\prod_{j=2}^{N_Z}(1-Z_1Z_j^{-1})(1-(Z_1Z_j)^{-1})}\right)^{\!\!1/2}\left(1+O(1/n)\right)\qquad &s_1=Z_1\end{cases}
\end{align}
where the square-root is principal (with positive real argument). 
\end{proposition}
This follows from the above discussion after using the same method to estimate the contribution of all other snagged contours---these are bounded above by $|z_*|^{-n}$ where $|z_*|>s_1$, and are thus exponentially subdominant.\par
In case B we use the same method but now sum over the dominant contributions coming from the two branch points. This leads to
\begin{proposition}\label{prop2}
\begin{align}
l_n = \begin{cases}
\frac{1}{\sqrt{\pi}}\Im(c_1 z_1^n )\frac{1}{n^{3/2}}\left(1+O(1/n)\right) \qquad &s_1=1/z_1\\\\
\frac{2}{\sqrt{\pi}}\Im(c_2 Z_1^{-n})\frac{1}{n^{1/2}}\left(1+O(1/n)\right)\qquad &s_1=Z_1\end{cases}
\end{align}
for
\begin{align}\label{taylor0}
c_1 = &-\left(-\frac{(1-z_1^2)(1-z_1\overline{z}_1)(1-\overline{z}_1/z_1)\prod_{i=2}^{N_z}(1-z_i/z_1)(1-z_1z_i)}{\prod_{j=1}^{N_Z}(1-Z_j^{-1}/z_1)(1-z_1Z_j^{-1})}\right)^{\!\!1/2}\\
c_2 =& -\left(-\frac{\prod_{i=1}^{N_z}(1-Z_1z_i)(1-z_i/Z_1)}{(1-Z_1^{-2})(1-Z_1/\overline{Z}_1)(1-(Z_1\overline{Z}_1)^{-1})\prod_{j=2}^{N_Z}(1-Z_1Z_j^{-1})(1-(Z_1Z_j)^{-1})}\right)^{\!\!1/2}.\nonumber
\end{align}
This constant is the first term of the Taylor series of the regular part of the integrand at the branch point, and the square root is continuously connected to the principal branch on the real axis. \end{proposition}
Note that in the case where we have only two roots, and they form a conjugate pair (as happens in the XY model), the constants are evaluated with the principal square root.\par
The exceptional cases where Propositions \ref{prop1} and \ref{prop2} do not apply are when $f(z)$ has zeros with multiplicity, more than a pair of zeros closest to the unit circle, or both. We discuss these cases below.\par
We also need the asymptotic behaviour of $m_n$. Fortunately no further analysis is needed: $m(z)$ and $l(z)$ share the same structure but are mutually inverse. Hence we have:
\begin{proposition}\label{prop3}
In the case of a nearest singularity $s_1$ on the real axis we have:
\begin{align}
m_n = \begin{cases}
- Z_1^{-n}\frac{1}{n^{3/2}}\frac{1}{2\sqrt{\pi}}\left(\frac{(1-Z_1^{-2})\prod_{j=2}^{N_Z}(1-Z_1Z_j^{-1})(1-(Z_1Z_j)^{-1})}{\prod_{i=1}^{N_z}(1-Z_1z_i)(1-z_i/Z_1)}\right)^{\!\!1/2}\left(1+O(1/n)\right) \qquad &s_1=Z_1\\\\
z_1^{n}\frac{1}{n^{1/2}}\underbrace{\frac{1}{\sqrt{\pi}}\left(\frac{\prod_{j=1}^{N_Z}(1-Z_j^{-1}/z_1)(1-z_1Z_j^{-1})}{(1-z_1^2)\prod_{i=2}^{N_z}(1-z_i/z_1)(1-z_1z_i)}\right)^{\!\!1/2}}_{\kappa}\left(1+O(1/n)\right)\qquad &s_1=1/z_1.\end{cases}
\end{align}
For a complex conjugate pair of nearest singularities we have:
\begin{align}
m_n = \begin{cases}
\frac{1}{\sqrt{\pi}}\Im(c_2^{-1} Z_1^{-n} )\frac{1}{n^{3/2}}\left(1+O(1/n)\right) \qquad &s_1=Z_1\\\\
\frac{2}{\sqrt{\pi}}\Im(c_1^{-1} z_1^{n})\frac{1}{n^{1/2}}\left(1+O(1/n)\right)\qquad &s_1=1/z_1,\end{cases}
\end{align}
where the $c_i$ are defined in \eqref{taylor0}. \end{proposition}
 
Now, if a zero has multiplicity two then we get either a simple pole of $l(z)$ (and hence a zero of $m(z)$) or a zero of $l(z)$ (and hence a simple pole of $m(z)$). A simple pole will give an exponential decay $\rme^{-n/\xi}$, using Cauchy's theorem, with no algebraic prefactor (recall that $\xi = 1/|\log|\zeta_\star||$ where $\zeta_\star$ is (any) one of the zeros of $f(z)$ closest to the unit circle). A zero of $l(z)$ is not a singularity so our contour will not be snagged there---we must hence look at the next-nearest singularity to the unit circle. Higher order multiplicities will give branch points, higher-order poles or higher-order zeros, and the calculations similarly go through. Higher-order poles will never have a vanishing residue for all $n$, and in fact for large $n$ the dominant term in the residue will come from derivatives of $s^{-(n+1)}$ in \eqref{nfourier}. Importantly, even in these exceptional cases, the nearest zero \emph{always} sets the longest correlation length for the operators $\mathcal{O}_\alpha$. This is because, from the discussion above, either $l_n$ or $m_n$ has asymptotic decay controlled by the nearest zero (and hence there is an observable with correlation length $\xi$ which follows from the calculation below).
\par
Having more than two equidistant singularities requires summing over the contributions from each of them; this will give an $\rme^{- n/\xi}$ decay for zeros of multiplicity one (the coefficient must be calculated in each case, and for higher multiplicity one sums the contributions outlined above)---there may be destructive interference for certain values of $n$. This can include equidistant singularities coming from zeros both inside and outside the unit circle. Another exceptional case of this type is two closest zeros both on the real axis (i.e. at $a$ and $-a$). Again we sum over the contributions which are given explicitly by the formulae in Propositions \ref{prop1} and \ref{prop3}. 
\par The final exceptional case is where we have degenerate closest zeros which are \emph{mutually inverse}. For example, if the closest zeros are at $a$ and at $1/a\in\mathbb{R}$. This is the only case where $\xi$ defined in terms of one of these closest zeros is not realised as the longest correlation length (although it is still an upper bound)---the contribution of the mutually inverse zeros cancels in the definition of $b_\pm(z)$ and so they do not contribute to the asymptotics of \emph{any} $\mathcal{O}_\alpha$. In such a case, the longest correlation length is set by the closest zero of $f(z)$ whose inverse is not a zero. The starkest examples of this behaviour are in isotropic models, where $b_\pm(z) = 1$ and the correlation length is zero for all observables! This also follows from the observation that the ground state of a gapped isotropic model in our class is always a product state.

\par
In summary, we have, in generic cases, that $l_n$ and $m_n$ decay exponentially with correlation length $\xi$. In exceptional cases their decay is at least this fast. Generically, if the nearest zero is inside the circle, we have an algebraically decaying prefactor $n^{-3/2}$ for $l_n$ and $n^{-1/2}$ for $m_n$, this assignment is reversed if the nearest zero is outside. 
 Moreover, if the nearest zero is complex then $l_n$ and $m_n$ have an oscillatory prefactor. \par

\subsubsection{Error terms in Theorem \ref{FHlength}}\label{sec::errors}
In order to use Theorem \ref{FHlength} we need to estimate the errors $\delta_N^\pm$. To do so, we need to find $\rho$ and $\sigma$ such that for $h(z) = \rme^{V(z)}$, $\lvert h_n \rvert = O(\rho^n)$ and $\lvert h_{-n}\rvert = O(\sigma^n)$ for large $n$. Recall that the relevant $\rme^{V(z)} = e^{V_0}b_+(z) b_{-}(z)$---this has exactly the same singularities as $l(z)$ and $m(z)$ up to exchanging square-roots with inverse square-roots. The analysis above goes through and we see that, in all non-degenerate cases, $\rho = \sigma = 1/|s_1|$. We thus have that either $d_n = l_n + O(l_{2n}m_n)$ or $d_n = m_n + O(l_{n}m_{2n})$. For $n$ large this means that we can replace the matrix elements $d_n$ of the small determinant in Theorem \ref{FHlength} with either $l_n$ or $m_n$ without affecting the leading order behaviour.
\subsubsection{The asymptotics of the correlator $\langle\mathcal{O}_\alpha(1)\mathcal{O}_\alpha(N+1)\rangle$---Proof of Theorem \ref{lengththeorem}}\label{proofcorrl}
Suppose that we are in the phase $\omega$, then the generating function of the correlator is $s\rme^{V(z)}z^{\omega-\alpha}$. In the case $\omega-\alpha>0$, using Theorem \ref{FHlength} we have that
\begin{align}
\langle\mathcal{O}_\alpha(1)\mathcal{O}_\alpha(N+1)\rangle = (-1)^{N(\omega-1)} &D_{N+\omega-\alpha}(s \rme^{V(z)})\\\nonumber& \times\det\left( \begin{array}{cccc}m_{N} & m_{N-1} & \dots & m_{N-(\omega-\alpha)+1} \\m_{N+1} & m_{N} &  & m_{N-(\omega-\alpha)+2} \\\vdots &  &  & \vdots \\m_{N+(\omega-\alpha)-1} & m_{N+(\omega-\alpha)-2} &  \dots & m_{N}\end{array}\right).
\end{align}
The large determinant $D_{N+\omega-\alpha}(s \rme^{V(z)})$ is of Szeg\H{o} form, and is, to leading order, equal to the result of Theorem \ref{orderparameter}---i.e. the value of the order parameter. Inserting the dominant term of $m_N$ as found in the previous section, the second determinant may be evaluated directly to find the leading order term of the correlator.  \par We have almost proved Theorem \ref{lengththeorem}, but need to do some further analysis to isolate the exponential decay. This is the point where we specialise to generic situations, so that we are guaranteed that $m_N =\Theta(\rme^{-N/\xi})$. Then, in the position $(i,j)$ of the second matrix we have a factor of $\rme^{(-N+i -j)/\xi}$. The row and column index multiplicatively decouple, and so any individual term of the Laplace expansion of the determinant contains a factor of $\rme^{-N(\omega-\alpha)/\xi}$, hence we may factor this out and we have that:
\begin{align} \label{algdet}
\langle\mathcal{O}_\alpha(1)\mathcal{O}_\alpha(N+1)\rangle =  \rme^{\rmi \pi N(\omega-1)+N \log s}&\left(\lim_{R\rightarrow\infty}\lvert\langle\mathcal{O}_\omega(1)\mathcal{O}_\omega(R)\rangle\rvert\right) \\ &\times\rme^{-N(\omega-\alpha)/\xi}\underbrace{\det\left((N+i-j)^{-K} \alpha_{N+i-j}\right)_{i,j=1}^{\omega-\alpha}}_{\det M(N)}.\nonumber
\end{align}
The matrix elements of $M(N)$ are derived from the propositions above: i.e. $K = 1/2$ or $3/2$ and $\alpha_n$ are the coefficients that can oscillate with $n$. Hence, $\det M(N)$ will contribute an algebraic dependence on $N$ (and not affect the exponential scaling).
For $\omega-\alpha<0$ the same calculation goes through with $m_n$ replaced by $l_n$ (and the second matrix has dimension $|\omega-\alpha|$). We have hence proved Theorem \ref{lengththeorem}.\par
Now, putting together Theorems \ref{orderparameter} and \ref{lengththeorem} prove Theorem \ref{phases}. In particular, we have shown that the correlators $|\langle\mathcal{O}_\alpha(1)\mathcal{O}_\alpha(N+1)\rangle|$ do indeed form a set of order parameters that distinguish $\omega$. The sign of $f(z)$ (an invariant of our model) may be inferred by the presence or absence of $(-1)^N$ oscillation. As can be seen in \eqref{algdet}, this oscillation depends on both $\omega$ and $s$, as defined in \eqref{bookkeep} (one must also take into account oscillations coming from $\det M$).
\subsubsection{The correlation length of $\langle\mathcal{O}_\omega(1)\mathcal{O}_\omega(N)\rangle$ in the phase $\omega$---Proof of Theorem \ref{orderlength}} \label{prooforder}
The proof follows from Theorem \ref{szegoerror} and our calculations above. Firstly, using \ref{sec::errors} we have that $E^{(2)}_N$ is exponentially subdominant compared to $E^{(1)}_N$. We do not evaluate $E^{(1)}_N$ in closed form, but need that the first term in the sum $(-l_{n+1}m_{n+1})$ gives the dominant scaling, as claimed in \cite{FH}. Thus, in the generic case,  we have $E^{(1)}_N=O(|s_1|^{-2N}/N^2)$. To see this, one needs to consider the different orders in the \emph{full} asymptotic expansion of $l_n$ and $m_n$, as given by Watson's lemma \cite{Temme}.  In particular, one can factor out the dominant term from $|l_{n+1} m_{n+1}|$ and $E^{(1)}_N$ then becomes a sum of many convergent series multiplied by non-positive powers of $N$ (along with exponentially subdominant contributions coming from other singularities further from the circle than $s_1$). One of these convergent series is $O(1)$ and we denote it by $B_N$---this will oscillate with $N$ if we have oscillation in $l_N$ and $m_N$. Putting this together one reaches Theorem \ref{orderlength}. The constant $B_N$, along with further corrections, are evaluated in \cite{Wu,BMc} for correlators that are equivalent to $X$ and $Y$ correlators in the XY model.
\subsubsection{Possible alternative proof}
An alternative approach to proving Theorem \ref{lengththeorem} would be to use the Fisher-Hartwig conjecture. The idea of such a proof is given in \cite{FF}---one should expand the Fourier contour defining the Toeplitz matrix \eqref{fourierc} out to the nearest singularity in the generating function, and then rescale back to the unit circle. The deformed symbol is then singular on the unit circle (by construction), and, if it can be written in Fisher-Hartwig form, then Theorem \ref{FHT} can be used to derive the leading order asymptotics. This method is applied in \cite{ov} to $X$ and $Y$ correlation functions in the XY model. 
\section{Gapless chains - analysis}\label{sec::gapless}
\subsection{Scaling dimensions}\label{sec::scaling}
In this section we calculate the large $N$ asymptotics of $\langle\mathcal{O}_\alpha(1)\mathcal{O}_\alpha(N+1)\rangle$ for a system described by \eqref{fzcanon} with non-zero $c$.
This was solved for isotropic models (i.e. models where $f(z)=f(1/z)$) and for $\alpha \in\{-1,0,1\}$ in \cite{HJ}. We now explain how to use Theorem \ref{FHT} to find the answer in the general case. The idea is simple; take the symbol corresponding to $(-1)^{N(\alpha-1)}\langle\mathcal{O}_\alpha(1)\mathcal{O}_\alpha(N+1)\rangle$: $z^{-\alpha} f(z)/|f(z)|$ and find the dominant Fisher-Hartwig representations. This goes as follows: 
\begin{align}
z^{-\alpha} f(z)/|f(z)| &= s z^{-\alpha} \frac{z^{N_z}}{z^{N_p}} \prod_{j=1}^{N_z} \frac{(1-z_j/z)}{|1-z_j/z|}\prod_{j=1}^{2c} \frac{(z-\rme^{\rmi k_j})}{|z-\rme^{\rmi k_j}|}\prod_{j=1}^{N_Z} \frac{(1-z/Z_j)}{|1-z/Z_j|}\\
&=  \underbrace{ sC\prod_{j=1}^{N_z} \frac{(1-z_j/z)}{|1-z_j/z|}\prod_{j=1}^{N_Z} \frac{(1-z/Z_j)}{|1-z/Z_j|}}_{\rme^{V(z)}}
\underbrace{C^{-1}z^{\omega -\alpha}\prod_{j=1}^{2c} \frac{(z-\rme^{\rmi k_j})}{|z-\rme^{\rmi k_j}|}}_{\mathrm{singular}}.\end{align}
We reemphasise that in this analysis we pick the phase of the complex zeros such that $k_j \in [0,2\pi)$.
The smooth part $\rme^{V(z)}$ can be analysed as in the gapped case---in particular, the Fourier coefficients for $n\neq 0$ are given by \eqref{fourierV} (the phase factor $C$ is needed to put the singular part in canonical form, and this shifts $V_0$). Turning to the unit circle, $z= \rme^{\rmi k}$, the analysis of the singular part is split into three cases: a real zero at $k = 0,\pi$, a pair of complex conjugate zeros at $k = \phi$ and $k=2\pi -\phi$, or a set of zeros of multiplicity greater than one. The third (fine-tuned) case is discussed in Section \ref{FHdegen}, we ignore it for now. Note that we explicitly exclude such cases in the statement of Theorem \ref{scalingtheorem} where we limit ourselves to chains described at low energy by a CFT.
Now, for the real zero we have:
\begin{align}\label{realzero}
\frac{\exp(\rmi k)\pm1}{|\exp(\rmi k)\pm1|} = \exp(\rmi k/2) \times \begin{cases} \cos(k/2)/|\cos(k/2)| \\
\rmi\sin(k/2)/|\sin(k/2)| = \rmi.
\end{cases}
\end{align}
For a zero at $-1$ we have a sign-change discontinuity at $k = \pi$, and a zero at $1$ has a sign-change-type\footnote{I.e. we are in the limiting case where $g_{1,\beta}$ is actually constant, but the contribution to the Toeplitz determinant is \emph{as if it were} a sign-change singularity.} singularity at $k=0$.
For a complex conjugate pair of zeros at $\exp(\pm \rmi \phi)$ we have:
\begin{align}
\frac{(\rme^{\rmi k}-\rme^{\rmi\phi})(\rme^{\rmi k}-\rme^{-\rmi\phi})}{|\rme^{\rmi k}-\rme^{\rmi\phi}||\rme^{\rmi k}-\rme^{-\rmi\phi}|} = \exp(\rmi k) \times \mathrm{sign}(\cos(k)-\cos(\phi)) .
\end{align}
Since $\phi \neq 0$ or $\pi$, $\mathrm{sign}(\cos(k)-\cos(\phi))$ has sign-change discontinuities at $k = \phi$ and $k=2\pi-\phi$. We conclude that every zero contributes a factor $\exp(\rmi k/2)$ as well as a sign-change at the location of the zero, which we can represent with a $g_{k_j,\beta_j}(z)$ for $\beta_j$ any half-integer.\par
Putting this information back into the symbol we reach 
\begin{align}\label{critde}
z^{-\alpha} f(z)/|f(z)| &=  \underbrace{ s C\prod_{j=1}^{N_z} \frac{(1-z_j/z)}{|1-z_j/z|}\prod_{j=1}^{N_Z} \frac{(1-z/Z_j)}{|1-z/Z_j|}}_{\rme^{V(z)}}
\underbrace{z^{c+\omega -\alpha}\prod_{j=1}^{2c} g_{k_j,\beta_j}(z)\rme^{-\rmi\beta_jk_j}}_{\mathrm{singular}},\end{align}
where the $\beta_j$ are half-integer and \begin{align}\sum_{j=1}^{2c} \beta_j = c+\omega-\alpha. \label{sumrule1}\end{align} We need to fix the multiplicative constant, $C$, by noting that the singular factors, isolated above, usually jump between $\pm 1$, rather than $\pm \rmi$, and we also need to include $\prod \rme^{-\rmi\beta_jk_j}$ in the singular part of \eqref{canonform}. This leads to $C=\prod_{j=1}^{2c}\rme^{\rmi\beta_jk_j + \delta_{k_j,0}\rmi\pi/2}/g_{k_j,\beta_j}(1) $.
\par To find the asymptotics we need some minimal representation (a set of $\beta_j$ minimising $\sum_j \beta_j^2$), from which we can generate a set of minimal FH-reps to insert into Theorem \ref{FHT}. We find the solutions by first considering the cases $c+\omega-\alpha=(2m-1)c$, for $m\in\mathbb{Z}$, where the minimal solution is unique: $\beta_j = \frac{2m-1}{2}$ for all $j$. If we have $(2m-1)c<c+\omega-\alpha<(2m+1)c$ we form the set of minimal FH-reps by starting from $\tilde\beta_j = \frac{2m-1}{2}$ and sending $\tilde\beta_j \rightarrow \tilde\beta_j+1$ for \begin{align}c+\omega-\alpha-(2m-1)c = (1-m)2c+\omega-\alpha \in \mathbb{Z}\end{align} of the $\tilde\beta_j$. We will consider our starting FH-rep (with all $n_i=0$) to be the one where we shift the first $(1-m)2c+\omega-\alpha$ of the $\beta_j$. There are \begin{align}\binom{2c}{(1-m)2c+\omega-\alpha}= \frac{(2c)!}{((1-m)2c+\omega-\alpha)!(2mc-\omega+\alpha)!}\end{align} minimal FH-reps in total. Given the parameters $\omega$, $\alpha$, $c$ we get that $m = 1 + \lfloor \frac{\omega - \alpha}{2c} \rfloor$, where $\lfloor x \rfloor$ denotes the greatest integer less than or equal to $x$.
Theorem \ref{FHT} immediately gives the dominant scaling 
\begin{align}
|\langle\mathcal{O}_\alpha(1)\mathcal{O}_\alpha(N+1)\rangle| = \mathrm{const}\times N^{-2\Delta}(1+o(1)),
\end{align}
where 
\begin{align}
\Delta_\alpha(c,\omega) = \frac{1}{2}\left((2cm-\omega+\alpha)\frac{(2m-1)^2}{4}+(2c(1-m)+\omega-\alpha)\frac{(2m+1)^2}{4}\right).
\end{align}
This formula for $\Delta_\alpha$ obscures some features of this function, to bring them out define $\tilde\alpha = \alpha - (\omega+c)$, then one can show that
\begin{align}\Delta_\alpha(c,\omega) = c \left( \frac{1}{4} + x^2-(x - [x])^2 \right) \Big|_{x = \tilde \alpha / 2c}\label{criticalscale2}
\end{align}
where $[x]$ is the nearest integer to $x$. It is then clear that $\Delta_\alpha$ is symmetric under $\tilde\alpha \leftrightarrow -\tilde\alpha$ and that the minimal scaling dimension of our operators is $c/4$.
\subsection{The dominant asymptotic term}
We can go further with Theorem \ref{FHT} to get the first term in the asymptotic expansion of $\langle\mathcal{O}_\alpha(1)\mathcal{O}_\alpha(N+1)\rangle$ at large $N$. Firstly, note that:
\begin{align}
\rme^{V(z)} = \underbrace{s \prod_{j=1}^{2c}\rme^{\rmi\beta_jk_j}/g_{k_j,\beta_j}(1)}_{\rme^{V_0}}\underbrace{\prod_{j=1}^{N_z} \frac{(1-z_j/z)}{|1-z_j/z|}\prod_{j=1}^{N_Z} \frac{(1-z/Z_j)}{|1-z/Z_j|}}_{\rme^{V(z)-V_0}}.
\end{align}
The first factor is a pure phase and contributes to the asymptotics as $\rme^{N V_0}$. By inspection\footnote{The asymmetric second term in the sum is necessary because a singularity at $k_j =0$ is an edge case where $g_{0,\beta}(1)=\rme^{-\rmi\pi\beta}$. For all other $k$ we have $g_{k,\beta}(1)=\rme^{+\rmi\pi\beta}$.}
 \begin{align}
V_0 = \Log(s) + \rmi\sum_{j=1}^{2c}(\beta_j(k_j-\pi) + (2\beta_j\pi +\pi/2) \delta_{k_j,0}), \label{V0crit} \end{align}
it is important to emphasise that this is an imaginary number, and again recall that $k_j \in [0,2\pi)$.
The second factor contributes in two ways: firstly through $\rme^{\sum_n n V_nV_{-n}}$, exactly the quantity we calculated in Section \ref{sec::orderparameter}. Secondly, we need powers of the Wiener-Hopf factors that were derived at the end of Section \ref{sec::orderparameter}. Putting this all together we get: 
\begin{theorem}\label{fullcritresult}
\begin{align}\label{fullcritformula}
\langle\mathcal{O}_\alpha(1)\mathcal{O}_\alpha(N+1)\rangle = N^{-2\Delta_\alpha}\rme^{\rmi N K}\sum_{\{n_j\}}\left(\prod_{j=1}^{2c} \rme^{\rmi Nk_j n_j}\right)\mathcal{C}(\{\beta_j+n_j\})
(1+o(1)),\end{align}
where the sum is over all dominant FH-reps, these are parameterised by $\{n_j\}$ and defined in Section \ref{sec::scaling}. $\Delta_\alpha$ is $\Delta_\alpha(c,\omega)$ given in \eqref{criticalscale2} and $K$ is equal to $-\rmi V_0 +\pi(\alpha-1)$.
The representation dependent $O(1)$ multiplier is given by 
\begin{align}\label{multiplier}
\mathcal{C}(\{\tilde\beta_j\})&= \left(\frac{ \prod_{i_1,i_2=1}^{N_z} (1-z_{i_1} z_{i_2})
	\prod_{j_1,j_2=1}^{N_Z} \left(1-\frac{1}{Z_{j_1} Z_{j_2}}\right)}{\prod_{i=1}^{N_z}\prod_{j=1}^{N_Z} \left (1-\frac{z_i}{ Z_j}\right)^2}\right)^{1/4}\prod_{1\le i<j\le 2c}
|\rme^{\rmi k_i}-\rme^{\rmi k_j}|^{2\tilde\beta_i\tilde\beta_j}
\nonumber\\
&\hspace{1cm}\times \prod_{j=1}^{2c} \left(\frac{\prod_{l=1}^{N_Z}(1-\rme^{\rmi k_j} Z_l^{-1})(1-\rme^{-\rmi k_j}Z_l^{-1})}{\prod_{i=1}^{N_z}(1-\rme^{\rmi k_j}z_i)(1-\rme^{-\rmi k_j}z_i)}\right)^{\tilde\beta_j/2}\prod_{j=1}^{2c}G(1+\tilde\beta_j) G(1-\tilde\beta_j).\end{align}
\end{theorem}
In our construction of the dominant FH-reps we start from setting all $\tilde\beta_j$ to be equal and then add 1 to a fixed number of them. This means that the difference $\tilde\beta_i - \tilde\beta_j$ is either 0 , 1 or $-1$ in all dominant FH-reps. Hence, pairs of complex conjugate zeros $\rme^{\rmi k_i }=\rme^{-\rmi k_j}$ contribute $\rme^{\rmi n k_i N}$ to the oscillatory factor in \eqref{fullcritformula}, where $n \in \{0,\pm1\}$. We discuss the non-universal multiplier \eqref{multiplier} in Appendix \ref{app::multiplier}.
\subsection{Degenerate zeros on the unit circle}\label{FHdegen}
In the case that some of the zeros on the unit circle are degenerate, the analysis of the singular part given above follows through by raising to the power of the relevant multiplicity. Conjugate pairs of zeros must have the same multiplicity so contribute to the singular part as
\begin{align}
\left(\frac{(\rme^{\rmi k}-\rme^{\rmi\phi})(\rme^{\rmi k}-\rme^{-\rmi\phi})}{|\rme^{\rmi k}-\rme^{\rmi\phi}||\rme^{\rmi k}-\rme^{-\rmi\phi}|}\right)^m = \exp(\rmi m k) \times \biggl(\mathrm{sign}(\cos(k)-\cos(\phi))\biggr)^m .
\end{align}
Equation \eqref{realzero} is similarly raised to the power $m$. We see an important difference between odd and even multiplicity. For odd $m$ the degenerate zeros behave as above and we have a Fisher-Hartwig canonical form with half-integer $\beta$ singularities at $\rme^{\pm\rmi \phi}$. In the case that $m$ is odd for all zeros on the unit circle we can derive an analogue of Theorem \ref{fullcritresult}, the steps are given in Appendix \ref{app::degenerate}. For even $m$ at any zero, there is no jump discontinuity and we do not analyse this here. The multicritical point in the XY model, with $f(z) = (z+1)^2$ is an example of such a case.
\section{Extensions of our results}\label{sec::extensions}
\subsection{Long-range chains}
\label{sec::longrange}
In this section we discuss the effect of allowing our model \eqref{majoranaham} to have non-zero coupling constants between sites at `long-range'---i.e. that there is no finite constant beyond which all couplings vanish. \par
First, consider the case that the couplings decay with an exponential tail at large distances. This means that $f(z)$ has a $C^\infty$ smooth part, and a well defined winding number. We still have that $\omega = N_z - N_p$, but note that poles are no longer restricted to the origin. The theory of Section \ref{sec::Toeplitz}, with care, may still be used to reach the same broad conclusions as in the finite-range case. In addition, we need the main result of \cite{Ehrhardt}:
\begin{theorem}[Erhardt, Silbermann 1996]\label{Ehrhardt}
Take a symbol of the form $$f(z) = \exp(V(z)) z^\beta,$$ i.e. a symbol with a single Fisher-Hartwig jump singularity at $z=1$, and demand that $\exp(V(z))$ is a $C^\infty$ function. Then:
\begin{align}
D_{N}(f(z)) = \exp(N V_0) N^{-\beta^2}(E+o(1))
\end{align} 
where $E$ is the constant defined implicitly in \eqref{FHformula}.
\end{theorem}
For nonvanishing $E$, or equivalently $\beta \not \in \mathbb{Z}$, this is a special case of Theorem \ref{FHT}. However, for $\beta \in \mathbb{Z}$ this gives us a concrete asymptotic bound on the Toeplitz determinant in the case of a symbol with a $C^\infty$ smooth part.  \par
Szeg\H{o}'s theorem along with Theorem \ref{Ehrhardt} allows us to extend the classification of gapped phases via string order parameters to long-range chains with $C^\infty$ symbol. 
In particular, in the phase $\omega$ we have that $\langle\mathcal{O}_\omega(1)\mathcal{O}_\omega(N+1)\rangle$ tends to a non-zero value that can be calculated using Szeg\H{o}'s theorem. Moreover, Theorem \ref{Ehrhardt} proves that  $\langle\mathcal{O}_\alpha(1)\mathcal{O}_\alpha(N+1)\rangle$ for all $\alpha\neq \omega$ tends to zero at large $N$. This proves that Theorem \ref{phases} remains valid for long-range chains with exponentially decaying couplings. In fact, one can go further than Theorem \ref{Ehrhardt} and use the methods of \cite{Hartwig1969} to give an analogue of Theorem \ref{lengththeorem}. The $\alpha$th correlator in the phase $\omega$ is $O(\rme^{-N/\xi_\alpha})$, where $\xi_\alpha$ is defined as above and $\xi$ is derived from the singularity of the symbol closest to the circle (this singularity will come from \emph{either} a zero or a pole of $f(z)$).  \par
In critical chains, we may use the Fisher-Hartwig conjecture to derive the scaling dimensions exactly as in the finite-range case, on condition that there are finitely many zeros of $f(z)$ that are \emph{on} the unit circle, and that they remain well separated (this means that we may write our symbol in the canonical form \eqref{canonform}). Note that a study of the critical scaling of entanglement entropy for the isotropic subclass of such chains is included in \cite{KM}. While the winding number remains well defined, further analysis must be given to extend the results of \cite{VJP} to long-range chains.\par
Models where couplings have algebraic tails are also physically relevant, and of topical interest \cite{HGCA,Patrick}. In this case, $f(z)$ will no longer be analytic and so singularities occur in the symbol distinct from Fermi points (zeros on the circle) and winding number (discontinuities in the logarithm). As $f(z)$ is continuous, the winding number remains geometrically well-defined for gapped models. The theory of Toeplitz determinants may still be used in this case, and is deserving of a detailed analysis.
\subsection{Uniform asymptotics approaching transitions}
Our results give asymptotic correlations at particular points in the phase diagram. One may also be interested in how these correlations change along a path in parameter space, particularly where this path crosses a transition. This problem was studied analytically in reference \cite{Wu2} for the 2D classical Ising model (and hence the 1D quantum XY model). There are two cases where relatively recent `black-box' results in the literature can be applied to a broader class of models. Firstly, consider a generalised Ising transition where we begin in a general gapped phase and a single zero approaches the unit circle. The relevant Toeplitz determinant asymptotics are given in reference \cite{CIK}. Secondly, consider the case of two zeros $\rme^{\pm \rmi k}$ that come together. This is a generalisation of the approach to the multicritical point in the XY model along the isotropic critical line. The relevant Toeplitz determinant asymptotics are given in reference \cite{CK}. In both cases the crossover is controlled by a solution to the Painlev\'e V equation (althought a different one in each case). Due to the multiplicative nature of contributions to Toeplitz asymptotics, one would expect\footnote{We are grateful to the participants of the AIM workshop `Fisher-Harwig asymptotics, Szeg\H{o} expansions and statistical physics' for discussions on this point.} similar behaviour in more general transitions where, as well as the approaching zeros, there are additional `spectator' zeros on the unit circle.\par
\section{Conclusion}
Using Toeplitz determinant theory, we have investigated string-like correlation functions in a wide class of gapped and critical topological models. The salient features of their asymptotics can be deduced from the zeros of the associated complex function $f(z)$. For example, the location of the zeros in the complex plane allow us to deduce whether the system is gapped or critical, furthermore giving access to correlation lengths and universal scaling dimensions (as summarised in Figure \ref{fig:correlation}). Even detailed information, like the exact value of the order parameter, is a simple function of the zeros of $f(z)$. The generality of these results allowed us to derive lattice-continuum correspondences, critical exponents and order parameters for the topologically distinct gapless phases. We now mention a few interesting paths to explore.

One surprising result was the universality of the ratios between the correlation lengths $\xi_\alpha$---this allowed for the extraction of the topological invariant $\omega$. This was more striking for the dual spin chains, where local observables can be used to measure $\omega$. It would be interesting to explore what happens upon introducing interactions. One possible scenario is that ratios of distinct correlation lengths give a measure of the interaction strength between the quasi-particles created by the corresponding operators.

One of the motivations of this work was to study how the invariants $c$ and $\omega$ are reflected in physical correlations. The full classification of topological gapless phases within this symmetry class was obtained in the non-interacting case in reference \cite{VJP}. Since this relied on concepts that are well-defined only in the absence of interactions, it does not directly generalise\footnote{However, numerical simulations indicated the stability away from the non-interacting limit.}. However, correlation functions and their symmetries are much more general concepts, and having now characterised the topology in terms of them, a natural next step is to use this to extend the classification to the interacting case. \par
Lastly, as discussed in the previous section, the exact solvability and Toeplitz theory extend to cases with long-range couplings. This would certainly be interesting to explore, as removing constraints on $f(z)$ leads to new asymptotic behaviours of the correlation functions beyond those that we have analysed in this paper.
\subsection*{Acknowledgements}
We are grateful to Estelle Basor, John Cardy, Torsten Ehrhardt, Paul \mbox{Fendley}, Tarun Grover, Jon Keating, Francesco Mezzadri, Frank Pollmann, Tibor Rakovszky, Jonathan \mbox{Robbins}, Ryan Thorngren and Chris Turner for helpful discussions. RV thanks Roderich Moessner for discussions and collaboration on related topics. RV has been supported by the German Research Foundation (DFG) through the Collaborative Research Centre SFB 1143.

\bibliography{paper.bbl}{}

\appendix
\section{Details of $H_{\mathrm{BDI}}$ and the dual spin model}
In this appendix we give details of certain claims in the introductory sections.
\subsection{The model and its solution}\label{app::fermi}
Our model is a one-dimensional chain of $L$-sites, each hosting a spinless fermionic mode. In other words, we have operators $c_n$, labelled by a site index $n$, such that the fermionic anticommutation relations are satisfied:
\begin{align}
\{c_n^\dagger,c_m\} = \delta_{nm} \qquad \{c_n,c_m\} =0.
\end{align}
The Hilbert space is the Fock space built from these $L$ modes---i.e. $\mathcal{H} = \bigoplus_{n=0}^L \Lambda^n(\mathbb{C}^2)$, where the direct sum is over antisymmetric $n$-particle states. We take periodic boundary conditions, i.e. we identify sites $1$ and $L+1$, and reduce all site labels modulo $L$ when appropriate.
 The ordering of sites induces a notion of locality. We always work in a double scaling limit $N \rightarrow \infty, L \rightarrow \infty$ and $N/L \rightarrow 0$, where $L$ is the system size and $N$ is the scale at which we are studying correlations. A local operator at site $n$ should have support on a number of sites around $n$ that is independent of $N$ and $L$.\par
Our model $H_{\mathrm{BDI}}$ is defined in equation \eqref{majoranaham}. This may be rewritten in terms of the fermions $c_n$ as
\begin{align}
H_{\mathrm{BDI}} = \sum_{r=-R}^{R}\sum_{n \in \mathrm{sites}} a_r c^\dagger_n  c_{n+r} + \frac{b_r}{2} \left(c^\dagger_n  c^\dagger_{n+r}- c_n  c_{n+r} \right)\quad (+\mathrm{const}) \label{freefermi}
\end{align}
where:
\begin{align}
a_r = -\frac{t_r+t_{-r}}{2}\nonumber\\
b_r = -\frac{t_r-t_{-r}}{2}. \label{backandforth}
\end{align}
A general time-reversal symmetric, translation-invariant spinless free fermion Hamiltonian has a representation of the form \eqref{freefermi} with $a_r = a_{-r} \in \mathbb{R}$ and $b_r = -b_{-r} \in \mathbb{R}$---the first condition follows from $H=H^\dagger$ and the second from the anticommuting fermion algebra.
Through equation \eqref{backandforth}, this is in one-to-one correspondence with \eqref{majoranaham}, which is hence general as claimed.\par
\subsection{Further discussion of the phase diagram}\label{app::fermi2}
We first consider smooth changes to our model $H_\mathrm{BDI}$. The coefficients $t_\alpha$ are symmetric functions of the zeros $\zeta_j$, so vary continuously upon a continuous change of the $\zeta_j$; moreover the results of Harris and Martin show that, for a fixed degree polynomial, the zeros vary continuously with the coefficients \cite{HarrisMartin}. We allow an increase in the range of $f(z)$ by tuning $t_\alpha$ off $0$ for $\alpha<\alpha_L$ or $\alpha>\alpha_R$. In the first case we introduce a zero-pole pair at the origin, and in the second case we introduce a zero-pole pair at infinity---hence these should be allowed `smooth operations' when we want to classify phases by thinking of a Hamiltonian in terms of the zeros and pole of the corresponding $f(z)$. The reverse is also important: we can decrease the range by tuning $t_{\alpha_{R/L}}$ to zero, or deleting a zero-pole pair at the origin or infinity. \par
When we study gapless systems in this work, we usually focus on the case where the zeros on the unit circle are non-degenerate. In that case, each zero corresponds to a linear zero-energy crossing of the single-particle dispersion, which after linearisation contributes a single real fermionic field to the low-energy description. Hence, $c$, as defined in terms of the zeros, exactly coincides with the central charge of the bulk CFT---see also Section \ref{sec::CFT}.  If any zero has degeneracy greater than one, then the low-energy theory will not be a CFT and the scaling behaviour changes. One can see that under the allowed smooth operations, $\omega = N_z-N_p$ and $c$ are invariants of these phases. For $c>0$ these phases are always critical points between neighbouring gapped phases---we can continuously move all zeros off the unit circle either inside or outside to reach different $c=0$ phases.\par
We now consider what constitutes a generic model in our class. Since we have a finite number of zeros, fixed by the coupling range, by any reasonable distribution of zeros we expect either no zero or one independent zero at a particular radius. Since any complex zeros must come in conjugate pairs, we thus have either no zero, one real zero or two complex zeros at a particular radius. This means that typically gapless models will have $c=1/2$ or $c=1$. The theory extends easily to $2c$ nondegenerate zeros on the unit circle, and so we state our main results for this case. Such higher-$c$ models arise as multi-critical points in the phase diagram. Typical gapped models will have either a single, real, zero or a complex conjugate pair of zeros closest to the unit circle. For gapped models this will be the `generic case' that we refer to in some of our results. In the statements of our results we usually assume these generic cases, but discuss how the results are altered in other cases. For example, we do give results for some cases with degenerate zeros on the unit circle in Section \ref{FHdegen}. Then the dispersion relation $|f(k)|$ cannot be linearised and we do not have a CFT description. It is clear that even conditioning on having many zeros on the unit circle, these are rare points in parameter space.\par
Finally we mention the extra signs $\Sigma$. In the gapped case, the sign of $f(1)$ is invariant---it must be real so can only change by passing through zero and hence closing the gap. A gapped model in the phase $\omega$ can be smoothly connected to $f(z) =\pm z^\omega$. In reference \cite{VJP} we showed that there are two invariant signs when $c>0$ and the model is described by a CFT---in that case we can continuously connect any model to one with $f(z)= \pm z^\omega(z^{2c+\omega} \pm1)$, the two signs cannot be removed without a phase transition. We hence have a description of the phase diagram that labels both gapped and critical phases by the triple $(\omega,c;\Sigma)$ where $\Sigma \in \mathbb{Z}_2$ for $c=0$ and $\Sigma \in \mathbb{Z}_2\times\mathbb{Z}_2$ for $c>0$ gives the relevant signs. This sign information is easy to keep track of, so we classify phases including this sign---one is free to discard the extra information this gives.
\subsection{The spin model}\label{app::spin}
We now go into more detail related to Section \ref{sec::spin}. First a note on the Hilbert space of the spin chain. It is formally similar to that of the fermionic chains, as both are built from a set of two-dimensional Hilbert spaces. They differ in that the mathematical structure is simpler: $\bigotimes_{n=1}^M \mathcal{H}_n$, where the local Hilbert space $\mathcal{H}_n \simeq \mathbb{C}^2$---in contrast to the fermions, operators localised on distinct sites commute. \par 
We now define a Jordan-Wigner transformation that allows us to (almost) map $H_{\mathrm{BDI}}$ into $H_{\mathrm{spin}}$ and back. Let
\begin{align}
Z_n = \rmi \tilde\gamma_n \gamma_n\qquad
X_n = \prod_{m=1}^{n-1} \left(\rmi \tilde\gamma_m \gamma_m\right) \gamma_n \qquad
Y_n = \prod_{m=1}^{n-1} \left(\rmi \tilde\gamma_m \gamma_m\right) \tilde\gamma_n.
\end{align}
transform fermions into spins. 
The inverse transformation is given by
\begin{align}
\gamma_n = \prod_{m=1}^{n-1} Z_m X_n \qquad
\tilde \gamma_n = \prod_{m=1}^{n-1} Z_m Y_n.
\end{align}
Note that we also have the relationship 
\begin{align}
\sigma^+_n =  \prod_{m=1}^{n-1} \left(\rmi \tilde\gamma_m \gamma_m\right) c_n \qquad \sigma^-_n =  \prod_{m=1}^{n-1} \left(\rmi \tilde\gamma_m \gamma_m\right) c^\dagger_n.\label{JWsigma}
\end{align}
Applying this transformation to $H_{\mathrm{spin}}$ gives us \eqref{majoranaham}, except that for all couplings extending over the final bond between sites $L$ and $L+1\equiv 1$, we have a multiplicative factor of $(-1)^F$---the total fermionic parity. Since the Hamiltonian \eqref{majoranaham} is quadratic, it preserves the parity, and so we can solve \eqref{majoranaham} in two total parity sectors. Details may be found in \cite{Suzuki71}, where it is shown that we get two copies of \eqref{majoranaham}, one with periodic and one with antiperiodic boundary conditions. Since we will be interested in bulk correlation functions, which will be independent of boundary conditions in the $L \rightarrow \infty$ limit, we claim that simply using our results for the periodic fermion chain will be enough to understand these correlations in the periodic spin chain.

 \par 
 
 \begin{table}
 \begin{center}
\begin{tabular}{|c||c|}\hline $\alpha$ & $\langle \mathcal{O}_\alpha(1) \mathcal{O}_\alpha(N+1)\rangle $ \\\hline\hline Positive, odd & $\langle X_1Y_2 X_3Y_4\dots Y_{\alpha-1}X_\alpha~~ X_{N+1}Y_{N+2}X_{N+3}\dots Y_{N+\alpha-1}X_{N+\alpha}\rangle$  \\\hline Positive, even & $\langle X_1Y_2\cdots X_{\alpha-1} Y_{\alpha} \left( \prod_{j=\alpha+1}^N Z_j \right)Y_{N+1}X_{N+2}\cdots X_{N+\alpha}\rangle$ \\\hline Zero & $\langle \prod_{j=1}^N Z_j\rangle$ \\\hline Negative, odd & $\langle Y_1X_2 Y_3X_4\dots X_{|\alpha|-1}Y_{|\alpha|}~~~ Y_{N+1}X_{N+2}Y_{N+3}\dots X_{N+|\alpha|-1}Y_{N+|\alpha|} \rangle$ \\\hline Negative, even & $\langle Y_1X_2\cdots Y_{|\alpha|-1}X_{|\alpha|} \left( \prod_{j=|\alpha|+1}^N Z_j \right)X_{N+1}Y_{N+2}\cdots Y_{N+|\alpha|}\rangle$ \\\hline \end{tabular}\vspace{0.1cm}
\caption{Spin correlation functions that are the Jordan-Wigner dual of the fermionic string correlators $\langle \mathcal{O}_\alpha(1) \mathcal{O}_\alpha(N+1)\rangle $. }
\label{tab::spincorrs}   
\end{center}
\end{table}

 \subsection{$\mathcal{O}_\alpha$ as a spin operator}\label{app::dual}
In this section we explain how to derive the contents of Table \ref{tab::spinops}, from which Table \ref{tab::spincorrs} follows. The quickest way to proceed in all cases is to use the nearest-neighbour substitutions:
\begin{align}
X_nY_{n+1}&= \gamma_n \rmi\tilde\gamma_n\gamma_n \tilde\gamma_{n+1} = -\rmi \tilde\gamma_n\tilde\gamma_{n+1} \label{XYsub}\\
Y_nX_{n+1} &=\tilde\gamma_n \rmi\tilde\gamma_n\gamma_n \gamma_{n+1} = +\rmi \gamma_n\gamma_{n+1} \label{YXsub}.
\end{align}
First consider the operator $X_{n+1}Y_{n+2}X_{n+3}Y_{n+4}\dots X_{n+\alpha}$. By substituting with \eqref{YXsub} starting from the right and then inserting the Jordan-Wigner form of $X_n$, we reach:
\begin{align}
X_{n+1}Y_{n+2}X_{n+3}Y_{n+4}\dots X_{n+\alpha}&= X_{n+1}\rmi \gamma_{n+2}\gamma_{n+3}\dots \rmi \gamma_{n+\alpha-1}\gamma_{n+\alpha}\\
&= \prod_{m=1}^{n} \left(\rmi \tilde\gamma_m \gamma_m\right) \gamma_{n+1} \rmi \gamma_{n+2}\gamma_{n+3}\dots \rmi \gamma_{n+\alpha-1}\gamma_{n+\alpha}\\
&= \rmi^{(\alpha-1)/2}\prod_{m=1}^{n} \left(\rmi \tilde\gamma_m \gamma_m\right) \gamma_{n+1}  \gamma_{n+2}\gamma_{n+3}\dots  \gamma_{n+\alpha-1}\gamma_{n+\alpha}\\
&= \mathcal{O}_\alpha(n+1) \qquad \qquad\qquad(\alpha = 2m+1>0).
\end{align}
Using that $\prod_m Z_m= \prod_m  \tilde\gamma_m \gamma_m$ (i.e. the trivial correspondence for $\mathcal{O}_0$) and using \eqref{XYsub} and \eqref{YXsub}, the same reasoning leads to the other cases (including the correct phase factor). The $\alpha$ odd and $\alpha=0$ cases in Table \ref{tab::spincorrs} follow immediately. For $\alpha$ even, we put the operators on sites $1$ up to $\alpha$ together and then simplify using the Pauli algebra.

\section{Expansion of three neighbouring spin operators}\label{appope}
We wish to understand the CFT behaviour of lattice operators $P_{n+2} P_{n+1} P_n$ where $P_j=X_j$ or $\rmi Y_j$. Using the substitutions \eqref{spinsubs} we can write up to an overall multiplicative constant:
\begin{align}
P_{n+2} P_{n+1} P_n \rightarrow  \left(\normord{\rme^{\rmi\theta(n+2a)}} +s_2 \normord{\rme^{-\rmi\theta(n+2a)} }\right)\left(\normord{\rme^{\rmi\theta(n+a)}} +s_1 \normord{ \rme^{-\rmi\theta(n+a)} }\right)\left(\normord{\rme^{\rmi\theta(n)}} +s_0\normord{\rme^{-\rmi\theta(n)}} \right),
\end{align}
for $s_i=\pm1$, lattice spacing $a$ and colons indicate normal ordering (as defined in \cite{polchinski}). We then multiply out the brackets and use the normal ordering prescription to simplify. For all choices of $s_i$ apart from $(1,-1,1)$ and $(-1,1,-1)$, the dominant terms are proportional to $\rme^{\rmi\theta(n)}$ and $\rme^{-\rmi\theta(n)}$. For $s_i = (1,-1,1)$ we have

\begin{align}
\rmi X_{n+2} Y_{n+1} X_n \rightarrow&\normord{ \rme^{3 \rmi \theta(n)}} -\normord{ \rme^{-3 \rmi \theta(n)}} \nonumber\\&~~-  \sqrt{2}\rmi a^2 \normord{\theta''(x)\left(\rme^{\rmi\theta(n)} +\rme^{-\rmi\theta(n)}\right)}- \sqrt{8}a^2 \normord{\theta'(x)^2\left(\rme^{\rmi\theta(n)} -\rme^{-\rmi\theta(n)}\right)}+\dots
\end{align}
where the ellipsis indicates terms with subdominant scaling dimension. That these terms all have the same scaling dimension is a consequence of, for example, \cite[Eq. 2.4.19]{polchinski}. The number of derivatives in each term exactly balances the difference in scaling dimension of the vertex operators.\par
\section{Recovering $(c,\omega)$ from scaling dimensions}\label{app:delta}
We will show how to find $c$ and $\omega$ even when we have access to the scaling dimensions of $\mathcal{O}_\alpha$ only for $\alpha$ odd (i.e. for the spin chain we have access to correlation functions of local operators only). As explained in the main text, it is helpful to consider differences between scaling dimensions. In this restricted case we calculate $\delta'_\alpha:=\Delta_{\alpha+2}-\Delta_\alpha$. \par
First suppose that $c>2$, we are then guaranteed to see plateaus with repeated values of $\delta'_\alpha$. If these plateaus are constant width then this width gives us $c$---if not, then a `plateau' of width one implies the presence of a kink in $\Delta_\alpha(c,\omega)$ at the even value of $\alpha$ that is skipped over. We can then determine $c$, and hence $\omega$ as described in the main text. \par
For $c\leq 2$ we are not guaranteed to see these plateaus, however, we can still recover $c$ and $\omega$. By writing out $\delta'_\alpha$ using equation \eqref{criticalscaling}, we derive the formulae in Table \ref{tab:delta}. These can be easily distinguished by taking the next level of differences\footnote{Note that $c=2$ allows two possible patterns depending on the parity of $\omega$. For even $\omega$, starting at $\alpha=\omega+1$, we see the differences $\{0, 1, 2, 3, 4, \dots\}$, whereas for odd $\omega$, starting at $ \alpha=\omega$, we see $\{0,0,2,2,4,4, \dots\}$.}. Hence, given a finite set of $\Delta_\alpha$ that are derived from local observables in the spin chain, we can recover $(c,\omega)$. The size of the required set is of order $c$. 

\begin{table}\begin{center}
\begin{tabular}{|c||c|c|c|c|} \hline $c$ & $1/2$ & $1$ & $3/2$ & $2$ \\\hline\hline $\delta'_\alpha$ & $2(\alpha-\omega)+1$ & $\alpha - \omega$
 & $\{2\lfloor x \rfloor,\alpha-\omega-1- \lfloor x \rfloor \}$ &$\{2\lfloor x \rfloor,\alpha-\omega-2- 2\lfloor x \rfloor \}$  \\\hline
 $\delta'_{\alpha+2}-\delta'_\alpha$ & $4$ & $1$
 & $\{1,2\}$ & $\{0,2\}$ or $1$\\\hline
  \end{tabular}
   \vspace{0.1cm}
\caption{Differences in scaling dimension derived from equation \eqref{criticalscaling} for small $c$. As defined above, $x= (\alpha-(\omega+c))/2c$.}
\label{tab:delta}
\end{center}
\end{table}

\section{Example representations of a Fisher-Hartwig symbol}\label{app::fh}
\begin{table}[htp]
\begin{center}\begin{tabular}{|c||c|c|c||c||c|}\hline  & $k_1 = 0$ & $k_2 = \pi $&$V(z)$&Symbol $f_c(z)$:& \\\hline \hline Canonical form 1: & $\beta_1 = 1/2$ & $\beta_2 = -1/2$ &$\rmi \pi/2 $&$ \mathrm{sign}(\sin k)=f(z)$&$\times$\\\hline Canonical form 2: & $\beta_1= -1/2$ & $\beta_2 = 1/2$ & $-\rmi\pi/2$ &$ \mathrm{sign}(\sin k)=f(z)$&$\times$ \\\hline
 Canonical form 3: & $\beta_1 = -3/2$ & $\beta_2 = 3/2$ &$\rmi \pi/2 $&$ \mathrm{sign}(\sin k )=f(z)$&\\\hline 
FH-rep($n_0=0,n_1=0$)&$\beta_1 = 1/2$ & $\beta_2 = -1/2$&$\rmi \pi/2 $&$ \mathrm{sign}(\sin k)=f(z)$ &$\times$\\\hline&&&&&\\[-10pt] FH-rep($n_0=-1,n_1=1$)&$\beta_1 = -1/2$ & $\beta_2 = 1/2$&$\rmi \pi/2 $&$- \mathrm{sign}(\sin k) = \rme^{\rmi \sum k_j n_j}f(z)$&$\times$\\\hline &&&&&\\[-10pt] FH-rep($n_0=-2,n_1=2$)&$\beta_1 = -3/2$ & $\beta_2 = 3/2$&$\rmi \pi/2 $&$ \mathrm{sign}(\sin k) = \rme^{\rmi \sum k_j n_j}f(z)$&\\\hline \end{tabular} \vspace{0.1cm}
\caption{Example representations for the symbol $f(\rme^{\rmi k}) = \mathrm{sign}(\sin k)$. Note that the given parameters \{$k_i,\beta_i,V$\} fully specify the RHS of \eqref{canonform}. In the final column, $\times$ indicates a dominant representation.}\label{reptable}
\end{center}
\end{table}
In Table \ref{reptable}, we give some representations of the symbol $t(\rme^{\rmi k }) = \mathrm{sign}(\sin k)$. The aim is to illustrate the difference between canonical forms and FH-reps explained in Section \ref{sec::Toeplitz}.
\section{Discussion of nonuniversal factors}\label{app::multiplier}
It is interesting to note that the order parameter given in Theorem \ref{orderparameter} is a symmetric function in the variables $\{z_i\}$ and, separately, $\{1/Z_i\}$ (listing zeros with multiplicity as distinct symbols). Another way to see why this occurs is through noting that a Toeplitz determinant generated by $t(\rme^{\rmi k})$ is the same as the average of  $\tau(\rme^{\rmi k_j}):=\prod_j t(\rme^{\rmi k_j})$ over the group $\UN$ (with eigenangles labelled by $k_j$) \cite{itzkyson,KM}. From the analysis of Section \ref{sec::orderparameter}, for $t(z) = \rme^{V(z)}$ we have that $\tau$ is a symmetric function separately in the arguments $\{z_j\}$, $\{1/Z_j\}$ and $\rme^{\rmi k_j}$; so can be expanded in a basis of symmetric functions. Let us write $\tau(\{z_j,Z_j,\rme^{\rmi k_j}\}) = \sum_k a_k s^{(1)}_k(\{z_j\})s^{(2)}_k(\{1/Z_j\})s^{(3)}_k(\{\rme^{\rmi k_j}\})$ for some constants $a_k$ and symmetric functions $s^{(i)}_k$. When integrating over $\UN$, $s^{(1)}_k(\{z_j\})s^{(2)}_k(\{1/Z_j\})$ can be factored out for each $k$, leaving us with a result that is a sum over products of symmetric functions and so the determinant is a symmetric function in the appropriate variables. \par
In the critical case we can rewrite the $\Theta(1)$ multiplier \eqref{multiplier} in a way that gives a structure similar to the order parameter. In particular, when we have that all $|\beta_j| = 1/2$, then
\begin{align}
\mathcal{C}(\{\tilde\beta_j\}) &= \Biggl(\frac{ \prod_{i_1,i_2=1}^{N_z} (1-z_{i_1} z_{i_2})
	\prod_{j_1,j_2=1}^{N_Z} \left(1-\frac{1}{Z_{j_1} Z_{j_2}}\right)}{\prod_{i=1}^{N_z}\prod_{j=1}^{N_Z} \left (1-\frac{z_i}{ Z_j}\right)^2}\prod_{i=1}^{ 2c}\prod_{j\neq i}
(1-\rme^{\rmi k_j}\rme^{-\rmi k_i})^{\mathrm{sign}(\tilde\beta_i\tilde\beta_j)}
\nonumber\\
&\hspace{1cm}\times  \frac{\prod_{l=1}^{2c}\prod_{j=1}^{N_Z}\left((1-\rme^{\rmi k_l} Z_j^{-1})(1-\rme^{-\rmi k_l}Z_j^{-1})\right)^{\mathrm{sign}(\tilde\beta_l)}}{\prod_{l=1}^{2c}\prod_{i=1}^{N_z}\Bigl((1-\rme^{\rmi k_l}z_i)(1-\rme^{-\rmi k_l}z_i)\Bigr)^{\mathrm{sign}(\tilde\beta_l)}}\Biggr)^{1/4}\Big(G(3/2) G(1/2)\Big)^{2c}.\end{align}
Notice that, up to the normalising $G$-functions, this constant is built from terms of the form $(1-\zeta_i^{\pm1}\zeta_j^{\pm1})$ where the $\zeta_i$ are zeros of $f(z)$. The sign of $\tilde\beta_j$ somehow tells us whether the $j$th zero on the unit circle acts as if it is inside the unit circle, or acts as if it is outside. Indeed, we see that if the $i$th and $j$th zero are both inside or both outside, we get a positive power of the term $(1-\rme^{\rmi k_j}\rme^{-\rmi k_i})$ and if one is in and one is out it is a negative power---this mirrors the behaviour of the factors coming from the $z_i$ and $Z_j$. In the second line we have factors mixing zeros on the circle with zeros inside and outside the circle. Since all zeros on the circle come in complex conjugate pairs, terms of the form $(1-\rme^{\rmi k_l}z_j)$ appear twice---however depending on the relative sign of $\beta_j$ and $\beta_{j'}$ for this pair these factors can either cancel, or give a square. A similar squared term appears in the factor matching the $z_i$ to the $Z_j$ on the first line. For $|\beta_j|=n/2$ for $n>1$ we have a multiplicative effect where the contribution from the $j$th zero on the circle is counted $n$ times. This is reminiscent of the CFT description where operators that involve many excitations give multiplicative contributions from the same Fermi point (which is located at some momentum $k_j$).
\section{Critical models with degenerate roots on the unit circle}  \label{app::degenerate}
As explained in Section \ref{FHdegen}, we consider $f(z)$ with zeros of odd multiplicity $m_j$ at $\rme^{\rmi k_j}$. The index runs over $i = 1 \dots N_0$ and so the total number of zeros on the circle is given by $2c = \sum_{j=1}^{N_0} m_i$. Note that by symmetry we must have equal multiplicities at complex conjugate zeros.\par
The main difference in the analysis is that there is only one $\beta$ for each unique zero (i.e. the degenerate zeros at that point correspond to only one FH singularity, but contribute to the winding multiple times)---this alters the sum rule \eqref{sumrule1}, which becomes
\begin{align}\label{sumrule1b}
\sum_{j=1}^{N_0}  \beta_j = c + \omega - \alpha.
\end{align} 
A method for solving \eqref{sumrule1b} is to first solve \eqref{sumrule1} as in Section \ref{sec::scaling}: assigning a half-integer $\hat\beta_{\hat{j}}$ where $\hat{j}=1\dots 2c$. We then group these as:
\begin{align}\beta_j = \sum_{\hat{j}: ~z_{\hat{j}}=z_j} \hat\beta_{\hat{j}} \qquad j = 1 \dots N_0.\end{align}
As all multiplicities are odd, this will lead to a canonical form for the symbol, but not necessarily a dominant FH-rep---this is because we minimised $\sum_{\hat{j}=1}^{2c} \hat\beta_{\hat{j}}^2$  whereas we need to minimise $\sum_{j=1}^{N_0} \beta_j^2$. We proceed by adding one to the smallest $\beta_j$ and subtracting one from the largest $\beta_j$ until the distance between smallest and largest is equal to zero or one. With this set of $\beta$ we can construct a dominant canonical form as in Section \ref{sec::scaling}, noting that the sum in the definition of $V_0$ \eqref{V0crit} should now range over unique zeros only (i.e. goes from $1$ to $N_0$).
Moreover, if the $\beta_j$ are not all equal, we construct the other dominant FH-reps $\tilde\beta_j = \beta_j + n_j$ as described in Section \ref{sec::Toeplitz}.
We then have that $\tilde\beta_j \in \{\lfloor \frac{c+\omega -\alpha}{N_0} \rfloor-1/2, \lfloor \frac{c+\omega -\alpha}{N_0} \rfloor+1/2\}$, and the scaling dimension follows. Theorem \ref{FHT} leads again to a variant of Theorem \ref{fullcritresult}, where we sum over the dominant FH-reps just described, and where all products over the $k_j$ are over unique zeros only. 
\end{document}